\documentclass[a4paper,12pt]{article}
\usepackage{amsmath,amssymb}
\usepackage{hyperref}
\usepackage{graphicx}
\usepackage{mathrsfs}
\usepackage{float}
\usepackage{ulem}
\hypersetup{
    colorlinks,%
    citecolor=blue,%
    filecolor=blue,%
    linkcolor=blue,%
   urlcolor=blue,
   linktoc=page
}
\usepackage{cite}

\newcommand{\be}{\begin{equation}}
\newcommand{\ee}{\end{equation}}
\newcommand{\beq}{\begin{equation}}
\newcommand{\eeq}{\end{equation}}

\newcommand{\bea}{\begin{equation}\begin{aligned}}
\newcommand{\eea}{\end{aligned}\end{equation}}
\newcommand{\ba}{\begin{align}}
\newcommand{\ea}{\end{align}}

\begin{document}

\begin{titlepage}

\vspace{.4cm}
\begin{center}
\noindent{\Large \textbf{AdS$_2$ holography and the SYK model}}\\
\vspace{1cm}
G\'abor S\'arosi$^{a,b,}$\footnote{gsarosi@vub.ac.be}

\vspace{.5cm}
 {\it
 $^{a}$Theoretische Natuurkunde, Vrije Universiteit Brussels and \\ International Solvay Institutes,\\
Pleinlaan 2, Brussels, B-1050, Belgium\\
\vspace{0.2cm}
 }
 \vspace{.5cm}
  {\it
  $^{b}$David Rittenhouse Laboratory, University of Pennsylvania,\\
  Philadelphia, PA 19104, USA\\
\vspace{0.2cm}
 }

\end{center}


\begin{abstract}
These are lecture notes based on a series of lectures presented at the XIII Modave Summer School in Mathematical physics aimed at PhD students and young postdocs. The goal is to give an introduction to some of the recent developments in understanding holography in two bulk dimensions, and its connection to microscopics of near extremal black holes. The first part reviews the motivation to study, and the problems (and their interpretations) with holography for AdS$_2$ spaces. The second part is about the Jackiw-Teitelboim theory and nearly-AdS$_2$ spaces. The third part introduces the Sachdev-Ye-Kitaev model, reviews some of the basic calculations and discusses what features make the model exciting. 
\end{abstract}

\end{titlepage}

\tableofcontents
\pagebreak

\section{Foreword}

These notes are an extended summary of lectures that I gave at the XIII Modave Summer School in Mathematical Physics in September, 2017. There are three main parts. The first one aims to give an introduction to what we understand about holography in AdS$_2$ and its connection to black hole physics, with the primary upshot that in the deep infrared the dynamics is largely universal and described by the so called Jackiw-Teitelboim model. The second one is an overview of gravitational dynamics and the coupling to matter in this model. The third part is an introduction into the Sachdev-Ye-Kitaev (SYK) model. This is a quantum mechanical model of $N$ Majorana fermions with all-to-all random couplings, gaining considerable attention recently along with many of its variants. This last section is mostly readable independently. For the reader only interested in this, the purpose of the first two parts is mostly to explain the reason why the high energy community is excited about this model, namely that it shares some common features with the physics of the near horizon region of near extremal black holes.

Beyond the usual background assumed in a high energy theory graduate school, in the first two parts we do assume familiarity with many aspects of the AdS/CFT dictionary in higher dimensions. There are many reviews available online where this can be picked up, here we point to \cite{Aharony:1999ti} as a nice example. The third part on the SYK model is mostly readable without this knowledge.


I hope that beyond the participants of the school, these notes can be useful for those who have not followed these developments but wish to gain some basic familiarity with them. There are certainly many errors (hopefully mostly typos), so comments and corrections are warmly welcome.

\section{Motivation}
\label{sec:motiv}

The first step before talking about dynamics in AdS$_2$ spaces is of course getting a bit familiar with AdS$_2$ spaces, which is the main purpose of this section. We first discuss how AdS$_2$ appears and why it is interesting in the context of black hole physics, and while doing so, we review its causal structure and different coordinate systems. Then we move on to discuss the problem with backreaction with the use of a family of two dimensional dilaton-gravity models which describe a large class of near extremal black holes. We will interpret this problem using the holographic renormalization group and show that the dynamics of these models is largely universal in the IR.

Most of this section is based on \cite{Maldacena:1998uz}.

\subsection{Near horizon region of extremal black holes}

As a sufficiently simple example, consider the magnetically charged Reisner-Nordstr\"om solution in four dimensions. The metric and the electromagnetic field are given by
\bea
\label{eq:RN}
ds^2 &=-\frac{(r-r^+)(r-r^-)}{r^2}dt^2 + \frac{r^2}{(r-r^+)(r-r^-)}dr^2 + r^2 d\Omega^2_2, \\
F &= Q \sin \theta d\phi \wedge d\theta,\\
r_{\pm} &= Q\ell_P +E \ell_P^2 \pm \sqrt{2QE\ell_P^3+E^2 \ell_P^4}.
\eea
Here, $d\Omega^2_2$ is the usual line element on the two-sphere, $Q$ is the magnetic charge, and
\beq
E=M-\frac{Q}{\ell_P},
\eeq
is the excitation energy above extremality. The mass of the black hole is $M$ and the Planck length is $\ell_P=\sqrt{G_N}$. These are the only dimensionful parameters.\footnote{Of course, we are setting all unit conversion parameters $c,k_B,...$ to one throughout these lectures, so everything is measured in units of some power of length.} 

\begin{figure}[h!]
\centering
\includegraphics[width=0.4\textwidth]{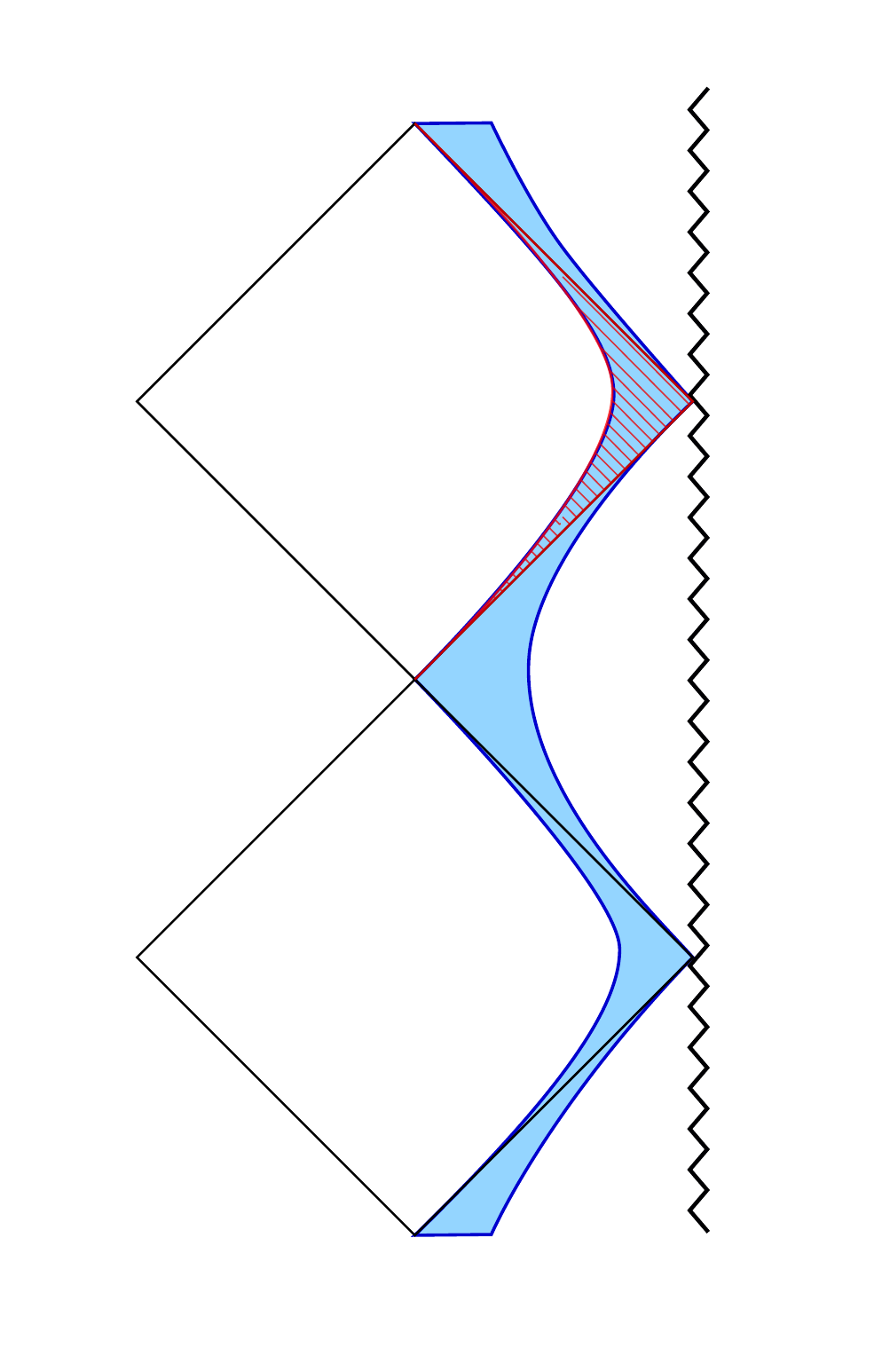} \includegraphics[width=0.4\textwidth]{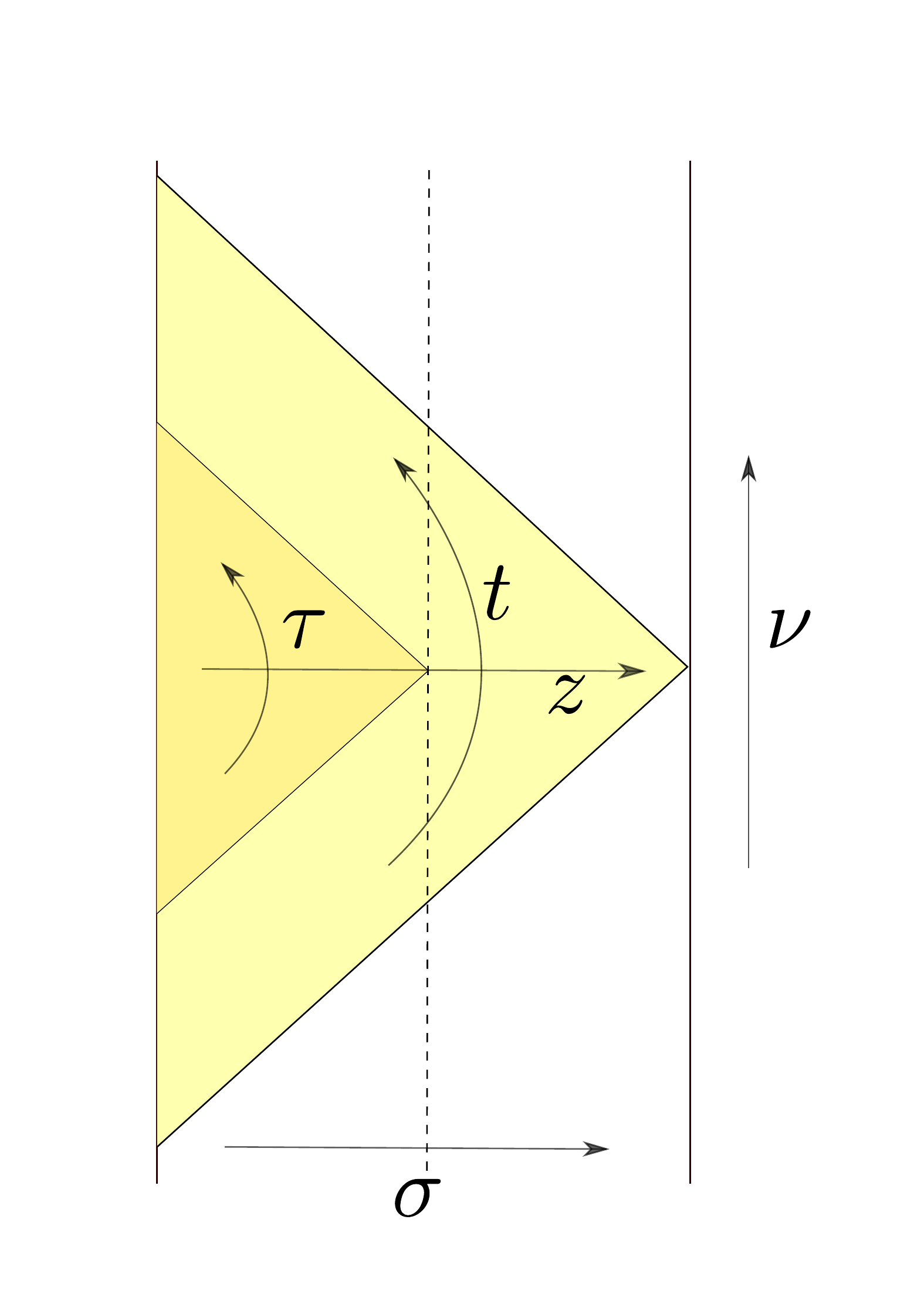}
\caption{Left: Maximally extended Penrose diagram of the extremal Reisner-Nordstr\"om solution. The blue region zigg-zagging up is the AdS$_2$ near-horizon region. The red dashed region is the patch covered by Poincar\'e coordinates. Right: Penrose diagram and coordinates of global AdS$_2$. Notice that there are two boundaries. The Poincar\'e patch is the light yellow region, while the dark yellow region is the Rindler patch.}
\label{fig:1}
\end{figure}

Let us now examine how the near horizon region looks like when the black holes is extremal, i.e. $E=0$. In this case only $\ell_P$ carries dimension so this is the parameter that defines what ``near" means in a near horizon limit. So we define a new coordinate
\beq
z=\frac{Q^2 \ell_P^2}{r-r_+},
\eeq
and zoom in to $r_+$ by taking $\ell_P \rightarrow 0$ while holding $z$ fixed. The resulting metric is 
\beq
\label{eq:nhlimit}
ds^2\approx \ell_P^2 Q^2 \left( \frac{-dt^2+dz^2}{z^2}+d\Omega_2^2\right),
\eeq
which is the product space AdS$_2\times S^2$. The AdS$_2$ metric is given in so called Poincar\'e coordinates
\beq
 ds^2=\ell_{AdS}^2\frac{-dt^2+dz^2}{z^2},
\eeq
where $\ell_{AdS}$ is the curvature radius of the space. We depict the Penrose diagram\footnote{The Penrose diagram is a projection to two dimensions of a conformal compactification of the spacetime. Since the Lorentzian ``angles" are left invariant by this compactification, we can use this diagram to depict the causal structure of a spacetime: lines of 45 degrees are light rays, any line more horizontal than that is spacelike, while more vertical lines are timelike.} of the extremal Reisner-Nordstr\"om spacetime on the left panel of Fig. \ref{fig:1}.

The coordinates $(t,r)$ that we have started with in \eqref{eq:RN} (with $r_+=r_-$) cover one exterior diamond and one connecting interior triangle. The $(t,z)$ coordinates of \eqref{eq:nhlimit} cover the region dashed with red lines. The complete diagram shows the structure of the geodesically completed extremal Reisner-Nordstr\"om spacetime, consisting of an infinite number of exterior and interior regions on top of each other. The blue stripe zigg-zagging up along the horizon is the geodesically completed version of AdS$_2$. We can pick coordinates that cover this maximally extended AdS$_2$. We will call such coordinates global coordinates and we will primarily use the following forms of the metric
\bea
\label{eq:ad2coords}
 ds^2 & =\ell_{\rm AdS}^2\frac{-dt^2+dz^2}{z^2} && \text{Poincar\'e}, \\
 &= \frac{-4\ell_{\rm AdS}^2 du^+ du^-}{\sin^2(u^+-u_-)} && u^\pm =\arctan (t \pm z), \;\; \text{Global (light cone)}, \\
  &= \ell_{\rm AdS}^2\frac{- d\nu^2 + d\sigma^2}{\sin^2 \sigma} && u^\pm =\frac{\nu \pm \sigma}{2}, \;\; \text{Global} .
\eea
We basically obtain the extension of the spacetime by decompactifying the timelike coordinate $\nu$, which appears compact when written in terms of Poincar\'e coordinates. The spacelike coordinate is still confined to $\sigma \in [0,\pi]$. 

By dropping the conformal factor $\ell_{\rm AdS}^2\frac{1}{\sin^2 \sigma}$, we see that the causal structure of AdS$_2$ is that of a strip in 2d Minkowski space. The Penrose diagram along with the above coordinate systems is shown in the right panel of Fig. \ref{fig:1}. The region covered by the $(t,z)$ coordinates is called the Poincar\'e patch and is the light yellow region on the figure. Notice that unlike in the case of higher dimensional AdS spaces, there are two distinct boundaries. The left boundary is at $z=0$. There is an additional darker region on the figure, which will be useful later for finite temperature discussions, called the Rindler patch. The coordinates of the Rindler and the Poincar\'e patches depicted on Fig. \ref{fig:1} are related as\footnote{For further reference, we note that the Poincar\'e metric in \eqref{eq:ad2coords} is invariant under M\"obius transformations $z\pm t \mapsto \frac{a (z\pm t)+b}{c(z\pm t)+d}$, $ad-bc=1$. Different Rindler patches are related to \eqref{eq:Rindler} by such transformations, e.g. a patch which shares the future horizon with the Poincar\'e coordinates is given by $t=\coth \rho e^\tau$, $z=(\sinh \rho)^{-1} e^\tau$.}
\beq
\label{eq:Rindler}
z \pm t = \frac{(1\pm \cosh \rho) e^{\tau/2}-\sinh \rho e^{-\tau/2}}{(1\pm \cosh \rho) e^{\tau/2}+\sinh \rho e^{-\tau/2}}.
\eeq
The metric in these coordinates is
\beq
ds^2=\ell_{\rm AdS}^2 (d\rho^2-\sinh^2 \rho d\tau^2),
\eeq
and the boundary is located at $\rho \rightarrow \infty$.

\subsection{The backreaction problem}

Now we are going to review what makes asymptotically AdS$_2$ spaces qualitatively different than their higher dimensional cousins. In a nutshell, we will see that the backreaction from any excitation destroys the asymptotic AdS$_2$ geometry. We will begin with a physical argument why this must be the case by analysing our previous example of magnetically charged Reisner-N\"ordstrom black hole. We then argue that the near horizon regions of a large class of extremal black holes are described by a family of two dimensional dilaton-gravity models (discussed in great detail in \cite{Almheiri:2014cka}), and review why all these models suffer from the backreaction problem.

\subsubsection{Black hole gap}

The Hawking temperature of the black hole \eqref{eq:RN} near extremality is given as\footnote{The Hawking temperature is basically the surface acceleration of the black hole, see e.g. \cite{Townsend:1997ku} for details.}
\bea
T_H &= \frac{r_+-r_-}{4\pi r_+^2} \\
&= \frac{1}{2\pi} \left(\frac{2}{\ell_P Q^3}E\right)^{\frac{1}{2}} + O(E^{3/2}),
\eea
where in the second line we expanded for small excitation energy $E$. As a consequence, we have the energy-temperature relation
\beq
E=2\pi^2 Q^3 \ell_P T_H^2.
\eeq
We see that there is no way to take a near horizon limit $\ell_P \rightarrow 0$ while keeping both $E,Q$ and $T_H$ fixed. A $T_H \rightarrow \infty$ limit is problematic to describe within general relativity, because this amounts to the black hole becoming very small and its horizon eventually getting within Planckian distance to the curvature singularity. So there are two options: either we fix $Q$ and describe only the ground states, i.e. the microstates of the extremal black hole, or we take $Q\rightarrow \infty$, which amounts to a large $N$ limit in AdS/CFT terminology, and consists of free supergravity with all backreaction suppressed by some powers of $1/Q$ (see \cite{Maldacena:1997ih,Maldacena:1998uz} for more on this limit). We should contrast this with a planar black $p$-brane, where we have a (IR-regulated) transverse spatial volume $V_p$ of the brane worldsheet, which can step into the place of $\ell_P$ and provide the correct dimensions for the excitation energy
\beq
E \sim V_p T_H^{p+1},
\eeq
which makes it possible to take an $\ell_P$ (or in AdS/CFT more commonly $\ell_{\rm string}$) $\rightarrow 0$ limit with keeping the charges of the state fixed.

We can understand what goes wrong with the extremal black hole by thinking about when we expect the semiclassical description to break down. Describing a black hole in terms of general relativity leads to exact thermodynamical laws, which suggests that this description only applies in a thermodynamic limit. Now a nonextremal black hole Hawking radiates, and a typical Hawking quantum has energy $\sim T_H$. Therefore, a thermodynamic description requires $E \gg T_H$, i.e. that emission of a Hawking quantum can be considered a quasi-equilibrium process. This description must break down when $E \sim T_H$, i.e.
\beq
\label{eq:gap}
E \sim \frac{1}{\ell_P Q^3}.
\eeq
This energy is called the black hole gap, and it is expected to be the rough magnitude of the energy gap above the ground states in the microscopic spectrum of the black hole.\footnote{String theory provides quantum descriptions for some near-extremal black holes as composite objects of branes wrapping compact dimensions. For these descriptions, the gap in the spectrum is indeed of the order \eqref{eq:gap} \cite{Maldacena:1996ds} (keeping in mind that the actual powers depend on the spacetime dimensionality, here we have $d=4$).} We see that the gap goes to infinity as $\ell_P \rightarrow 0$, so that all excitations of the black hole are lifted from the spectrum in this limit, and we are left with the ground states. It is useful to contrast this situation again with that of the planar $p$-branes. There, the spectrum is quantized only because of box-quantization, so the gap is
\beq
E_{gap} \sim V_p^{-\frac{1}{p-1}},
\eeq
where $V_p$ is again the transverse spatial volume. We see that the problem with near-extremal black holes is basically that a zero dimensional object cannot have any spatial volume.

\subsubsection{Dilaton-gravity models}
\label{sec:backreact}
Having an idea of how excitations above extremality are lifted from the spectrum in a near horizon limit, let us now review the concrete manifestation of this problem in terms of gravity in two dimensions. 

To get a handle on what these two dimensional models have to do with the previous discussion focused on the Reisner-Nordstr\"om black hole, we first do a quick exercise. The RN solution \eqref{eq:RN} extremizes the action
\beq
S_{\rm Einstein-Maxwell} \sim {\frac{1}{\ell_P^2}}\int d^4 x \sqrt{-g} \left( R_g-\frac{{\ell_P^2}}{4}F_{\mu \nu}F^{\mu \nu}\right),
\eeq
where $R_g$ is the Ricci-scalar of the metric $g$ and $F_{\mu \nu}$ the field strength tensor of the Maxwell field. Looking for static, spherically symmetric solutions we may dimensionally reduce to the $r$-$t$ plane by considering the ansatz
\bea
\label{eq:warpedprod}
ds^2 &= h_{ij} dx^i dx^j + e^{2\psi(r,t)} d\Omega^2, \\
F&=Q \sin \theta d\phi \wedge d\theta,
\eea
with $i,j=1,2$, $x^1=t$ ,$x^2=r$. Using some standard formulae\footnote{A nice collection can be found in \cite{Headrick}.} for warped product spaces and after a partial integration one obtains the action
\bea
\label{eq:RNredaction}
S_{\rm Einstein-Maxwell} &\sim \frac{4\pi}{\ell_P^2} \int dt dr \sqrt{-h} \left[ e^{2\psi}(R_h+2(\partial \psi)^2)+2-\frac{1}{2}e^{-2\psi}Q{^2 \ell_P^2}\right] \\
&=\frac{4\pi}{ \ell_P^2}  \int dt dr \sqrt{-h} \left[ \Phi^2 R_h+2(\partial \Phi)^2+2-\frac{1}{2}\Phi^{-2}Q{^2 \ell_P^2}\right],
\eea 
where in the last line we have defined $\Phi=e^{\psi}$. 

This action is a special instance of a class of dilaton-gravity models (studied extensively in \cite{Almheiri:2014cka})
\beq
\label{eq:almhpolch}
I=\frac{1}{16\pi G_N} \int d^2 x \sqrt{-h} \left[\Phi^2 R_h+\lambda (\partial \Phi)^2-U(\Phi^2/d^2) \right],
\eeq
with an arbitrary scalar potential $U$ and dimensionless coefficient $\lambda$ for the kinetic term. Here, $\Phi^2$ is called a dilaton field, basically because it multiples the Ricci scalar. Doing a similar dimensional reduction as the one presented here for the RN black hole, but in gravity theories with more matter content in many cases can lead to an action of this form. These theories typically admit black hole solutions with multiple charges which are also solutions to this action, and in case they are extremal, have an AdS$_2$ factor in their near horizon geometry. In \eqref{eq:almhpolch}, we are assuming a four dimensional parent theory, so that $G_N$ is of dimension (length)$^2$, cancelling the dimensions coming from $d^2x$. Notice that the dilaton $\Phi^2$ plays the role of the radius squared in \eqref{eq:warpedprod} so it is of dimension (length)$^2$ which is required for the first two terms to be correctly dimensionless. In the potential term $U$, we have introduced a length parameter $d$ to make the argument dimensionless.

Now we are in position to discuss the general backreaction problem. Let us couple the model \eqref{eq:almhpolch} to some matter fields by considering the action
\beq
S=I+S_{\rm matter}.
\eeq
We can write the equation of motion for the metric component $h^{++}$ in conformal gauge\footnote{In two dimensions, every metric can be put into conformal gauge $ds^2 = -e^{2\omega(u^+,u^-)} du^+ du^-$ by a coordinate transformation.} as
\beq
-e^{2\omega} \partial_+ (e^{-2\omega} \partial_+ \Phi^2) = T^{\rm matter}_{++},
\eeq
where we have defined $T^{\rm matter}_{++}$ so that it also incorporates the contribution of the kinetic term for the dilaton, proportional to $\lambda$. We can consider this equation in an asymptotically AdS$_2$ metric and integrate it along a null line $u^-=0$ from one boundary to the other
\beq
\label{eq:backderive}
\int_0^\pi du^+ e^{-2\omega} T^{\rm matter}_{++} = \left[e^{-2\omega} \partial_+ \Phi^2 \right]|_{u^+\rightarrow 0}-\left[e^{-2\omega} \partial_+ \Phi^2 \right]|_{u^+\rightarrow \pi}.
\eeq
Now classically, for any excitation we must have $T^{\rm matter}_{++}>0$.\footnote{In a quantum field theory one can have states with $T^{\rm matter}_{++}<0$ locally, so this argument does not apply. There is presumably a quantum version of the argument using something called the averaged null energy condition.} The l.h.s. is then an integral of a positive function and therefore is positive (and zero iff $T_{++}\equiv 0$). On the other hand, from \eqref{eq:ad2coords} we see that for an asymptotically AdS$_2$ space, near the boundaries and on the null line $u^-=0$ we must have that the conformal factor behaves as
\bea
e^{2\omega} \sim \frac{1}{\sin^2(u^+)} & \sim \frac{1}{(u^+)^2} && \text{for} \;\; u^+\rightarrow 0, \\
 & \sim \frac{1}{(u^+-\pi)^2} && \text{for} \;\; u^+\rightarrow \pi.
\eea
It follows that to obtain a nonzero result on the r.h.s. of \eqref{eq:backderive} the dilaton must diverge linearly near at least one of the boundaries
\bea
\label{eq:dilatonblow}
\Phi^2|_{u^+\rightarrow 0} & \sim \frac{1}{u^+}  + \text{const},\\
\Phi^2|_{u^+\rightarrow \pi} & \sim \frac{1}{u^+-\pi}  + \text{const}.
\eea
Presence of a nonzero matter stress tensor therefore basically destroys the assumed asymptotic region. Now, you might say that in usual AdS/CFT, bulk fields can diverge near the boundary as well. The point is that the divergences are associated to sources on the boundary, and therefore deformations of the boundary theory. The bulk fields in that case have both normalizable and non-normalizable modes, corresponding to the fact that there is room for nontrivial Lorentzian dynamics even when all sources are set to zero. There is no room for such dynamics in the case of AdS$_2$.

\subsection{Holographic interpretation}
\label{sec:holo}

So what does this all mean for AdS$_2$/CFT$_1$? One argues for conventional AdS/CFT via examining the low energy excitations of a stack of $D$-branes from both an open and a closed string perspective and observing that free supergravity in flat space decouples from both descriptions in the low energy limit. AdS/CFT arises from identifying the remaining systems with each other.\footnote{An old but gold review of the basics of AdS/CFT is \cite{Aharony:1999ti}.} We can play the same game with extermal black holes, the difference is that while a planar extended object has a continuous spectrum, a zero dimensional object has a gap (the black hole gap of \eqref{eq:gap}). Therefore, in the low energy limit, we are left with the ground states of the object. This is what the CFT$_1$ describes. Recall that scale invariance requires the energy momentum tensor to be traceless. In one dimension, this actually implies a vanishing Hamiltonian: CFT$_1$ is just a theory of a constraint. The standard AdS/CFT dictionary can be used in this case to relate the extremal entropy (with higher derivative and quantum corrections) to the number of states in this theory. We refer to the original works \cite{Sen:2008vm,Sen:2005wa} on this without giving further details here.

Can we get beyond describing only the ground states? The existence of the black hole gap tells us that to do this, we must not go all the way with our low energy limit, but ``zoom out" a little to see some excitations above the gap. Of course this means that there is no complete decoupling from the asymptotic, flat part of the spacetime. We can ask in general for AdS/CFT what happens if we back off a little from the decoupling limit. To intuitively understand this, it is useful to think in terms of the holographic renormalization group \cite{Susskind:1998dq,deBoer:1999tgo}, where the radial direction in AdS is interpreted as an energy scale in the CFT, with the boundary being the UV and the deep interior of AdS being the IR. In this picture, empty AdS corresponds to an RG flow which stays at the CFT fixed point forever. We can also consider finite energy excitations of the CFT. These look like the vacuum from a very UV point of view, and they differ only as we flow into the IR. Therefore, they correspond to geometries which approach AdS at the boundary, but are different in the interior. Finally, we may consider turning on some irrelevant\footnote{We use the standard Wilsonian terminology: operators with $\Delta <d$ are relevant, with $\Delta=d$ are marginal, and with $\Delta>d$ are irrelevant. These names describe whether the associated coupling grows (relevant) or decreases (irrelevant) as we flow towards the IR.} deformations to the CFT
\beq
\label{eq:deform}
S_{CFT} \rightarrow S_{CFT} + \int J O,
\eeq
where $O$ are operators with dimension $\Delta >d$. In this case, the RG trajectory is deflected from the CFT fixed point as we track it back to the UV. This is what happens for example when we do not take a complete decoupling limit for our $D$-branes: flowing towards the boundary (UV), we are deflected from the AdS geometry (CFT) and continue to flow out in some specific irrelevant direction which might correspond to flat space or some other geometry. In particular, when our $D$-branes themselves also sit in a big AdS' space, we eventually hit another fixed point corresponding to CFT' and we may interpret the original branes as finite energy states in this theory.

Of course, we know how to deal with deformations of the form \eqref{eq:deform} in AdS/CFT. We just turn on the non-normalizable modes of the dual bulk fields. The fact that these modes blow up near the boundary (precisely when $\Delta>d$) is just the dual statement that we are deflected from the fixed point as we flow to the UV. We solve this by introducing a radial (IR) cutoff in AdS, which corresponds to a UV cutoff in the field theory. Any state bellow the cutoff is on the ``IR-part" of the RG flow and thus is described by gravity in AdS with the ``blow-up" boundary conditions.

Given this far from precise, but intuitive understanding, it is clear what we have to do for AdS$_2$. The models \eqref{eq:almhpolch} describe the nonextremal solutions too, and the blowing up of the dilaton \eqref{eq:dilatonblow} in AdS$_2$ just signals that any excitation wants to deflect the RG-flow from the $CFT_1$ fixed point and couple it back to the asymptotic region. Different choices of parameters and potential in \eqref{eq:almhpolch} lead to different UV-parts for this deflected flow. However, there is clearly a part of this flow which stays close to the fixed point in the IR. This corresponds to the dual statement, that the geometry stays close to AdS$_2$ in the deep interior. As we will see, if we cut off the RG-flow sufficiently close to the fixed point, the dynamics is largely universal. For this purpose, we need to study cutoff AdS$_2$ spaces with blowing up boundary conditions on the dilaton to which we turn to next.

\subsection{Jackiw-Teitelboim theory}
\label{sec:JT}

We will consider the family of actions \eqref{eq:almhpolch} as the UV completion for the cutoff AdS$_2$ space and argue that inside the cutoff surface, the dynamics is largely universal, given some conditions on the cutoff are met. This universality then also turns into the universality of the low energy dynamics for the class of near extremal black holes that can be described by these models. We begin by quoting \eqref{eq:almhpolch} here again
\beq
I=\frac{1}{16\pi G_N} \int d^2 x \sqrt{-h} \left[\Phi^2 R_h+\lambda (\partial \Phi)^2-U(\Phi^2/d^2) \right].
\eeq
When the dilaton is constant, this action has AdS$_2$ solutions. Let this constant be $\Phi^2=\phi_0$. Extremality of the action requires this to satisfy
\beq
\label{eq:nulleom}
\frac{2}{\ell_{\rm AdS}^2}+\frac{1}{d^2} U'(\phi_0/d^2)=0,
\eeq
which estabilishes a relation between $\ell_{\rm AdS}$, $d$ and $\phi_0$. Note that $d$ is an external length scale coming from the UV theory. For example, it is $d=\ell_P Q$ for the Reisner-Nordstr\"om example of \eqref{eq:RNredaction}. Now we wish to look for solutions when the dilaton is a small deformation of this constant value\footnote{Not having the square on the right hand side is not a typo!}
\beq
\Phi^2 = \phi_0+\phi.
\eeq
By the discussion in sec. \ref{sec:backreact}, we expect the deformation to blow up as $\phi \sim 1/z$ as we approach the boundary $z\rightarrow 0$ in Poincar\'e coordinates. We want to cut off the space at $z=\epsilon$ such that
\beq
\frac{\phi(\epsilon)}{d^2}\equiv \eta \ll 1.
\eeq
In this case we can expand the potential in \eqref{eq:almhpolch} around $\Phi^2=\phi_0$
\bea
\label{eq:actionexpand}
I=\frac{1}{16\pi G_N} & \Big[ \int d^2x \sqrt{-h}(\phi_0 R_h-U(\phi_0/d^2))\\
&+\int d^2 x \sqrt{-h} \phi \left(R_h+\frac{2}{\ell_{\rm AdS}^2} \right)\\
&+\int d^2 \sqrt{-h} \frac{\lambda}{4} \frac{(\partial \phi)^2}{\phi_0+\phi}\Big] + O(\eta^2).
\eea
Here, we have used \eqref{eq:nulleom}. We now break down the three lines of this action.
\begin{itemize}
\item The action in the first line of \eqref{eq:actionexpand} is basically Einstein gravity in two dimensions. There is an IR divergent volume term $\int d^2 \sqrt{-h}U(\phi_0/d^2)$, which we can imagine removing with a local counter term. After adding the appropriate boundary term, required for manifolds with boundaries,\footnote{We will say a little more about this very soon.} this just gives the Euler-character of the manifold due to the Gauss-Bonnet theorem. This is a topological invariant. This action is the bulk description when we really stay at the $CFT_1$ fixed point. 

There is a way of inerpreting the divergences in this action. They are related to the nonzero energy of the ground state. After removing this contribution, the finite part can be related to the extremal entropy. See \cite{Sen:2008vm,Sen:2005wa} for the details of this.
\item The action in the second line of \eqref{eq:actionexpand} is called the Jackiw-Teitelboim theory \cite{Jackiw:1984je,Teitelboim:1983ux}, which is the universal dynamics we were after. This model will be the main focus of the next section.
\item The third line in \eqref{eq:actionexpand} is the full derivative term in \eqref{eq:almhpolch} which we have not expanded yet in $\eta$. To do this, we need to determine the order of the gradient of $\phi$ in $\eta$. We can do this, by recalling the discussion in \ref{sec:backreact}: the dilaton blows up as $1/z$ as we approach the boundary. First we need to restore dimensions: $\phi$ is of dimension (length)$^2$, $z$ is of dimension (length), so there must be a dimensionful coefficient between the two that we are missing. To do this, recall e.q. \eqref{eq:backderive} which we used to derive the blowing up of the dilaton. This comes from an Einstein's equation, so it must contain a $G_N=\ell_P^2$ multiplying the matter stress tensor, which itself must be proportional to the excitation energy $E$. So we must have $\phi \sim \ell_P^2 E/z$. This combination is dimensionless, and the missing dimensions can only come from the geometry, so they have to be made up by the AdS radius. We conclude
\beq
\phi \sim \frac{\ell_P^2 \ell_{\rm AdS}^2 E}{z}.
\eeq
Now we can estimate the gradient term in \eqref{eq:actionexpand}
\bea
\frac{(\partial \phi)^2}{\phi_0+\phi} &\sim \frac{g^{zz}(\frac{\ell_P^2 \ell_{\rm AdS}^2 E}{z^2})^2}{\phi_0+\phi}\\
&=\frac{\phi^2}{\ell_{\rm AdS}^2} \frac{1}{\phi_0 + \phi},
\eea
where we have used $g^{zz}=z^2/\ell_{\rm AdS}^2$. At this point, it is convenient to set $\phi_0=d^2$ which is a choice allowed by \eqref{eq:nulleom} and sets the AdS radius in terms of $d$ as $2\ell_{\rm AdS}^{-2} = -d^{-2} U'(1)$.\footnote{We clearly need $U'(1)<0$ to describe AdS, which is satisfied for example by the potential associated to the RN black hole \eqref{eq:RNredaction}.} This leads to 
\bea
\frac{(\partial \phi)^2}{\phi_0+\phi} &\sim \frac{|U'(1)|}{1+\frac{\phi}{d^2}}\left( \frac{\phi}{d^2}\right)^2\\
&=|U'(1)| \eta^2\big(1+O(\eta)\big).
\eea
This is indeed $O(\eta^2)$, so we can neglect it in \eqref{eq:actionexpand}.
\end{itemize}
The universal dynamics that we were after inside the cutoff surface is therefore governed by the action:
\beq
\label{eq:JTlorentz}
I=\frac{\phi_0}{16 \pi G_N} \int d^2 x \sqrt{-h} R_h +\frac{1}{16 \pi G_N} \int d^2 x \sqrt{-h} \phi \left(R_h+2 \right),
\eeq
where we have already dropped the volume term from the first piece and set $\ell_{\rm AdS}^{-2}=1$ to ease the notation when we further analyse this model. 

\section{Nearly AdS$_2$ spaces}
\label{sec:Nads}

In this section, we are going to discuss the model \eqref{eq:JTlorentz}. 
We will see that the configuration space of the model is simpler than it looks at first sight: it just consists of cutouts of different shapes from AdS$_2$. We will derive the action on this space and discuss the gravitational backreaction of matter fields. 

This section is based mainly on \cite{Maldacena:2016upp}. 

\subsection{Euclidean Jackiw-Teitelboim}

First, we are going to move over the Euclidean signature. This is a natural thing to do when we study holography, as ultimately we are interested in correlation functions of the boundary theory, which are naturally defined in Euclidean signature. Lorentzian correlators are then obtained via different analytic continuations.

\begin{figure}[h!]
\centering
\includegraphics[width=0.4\textwidth]{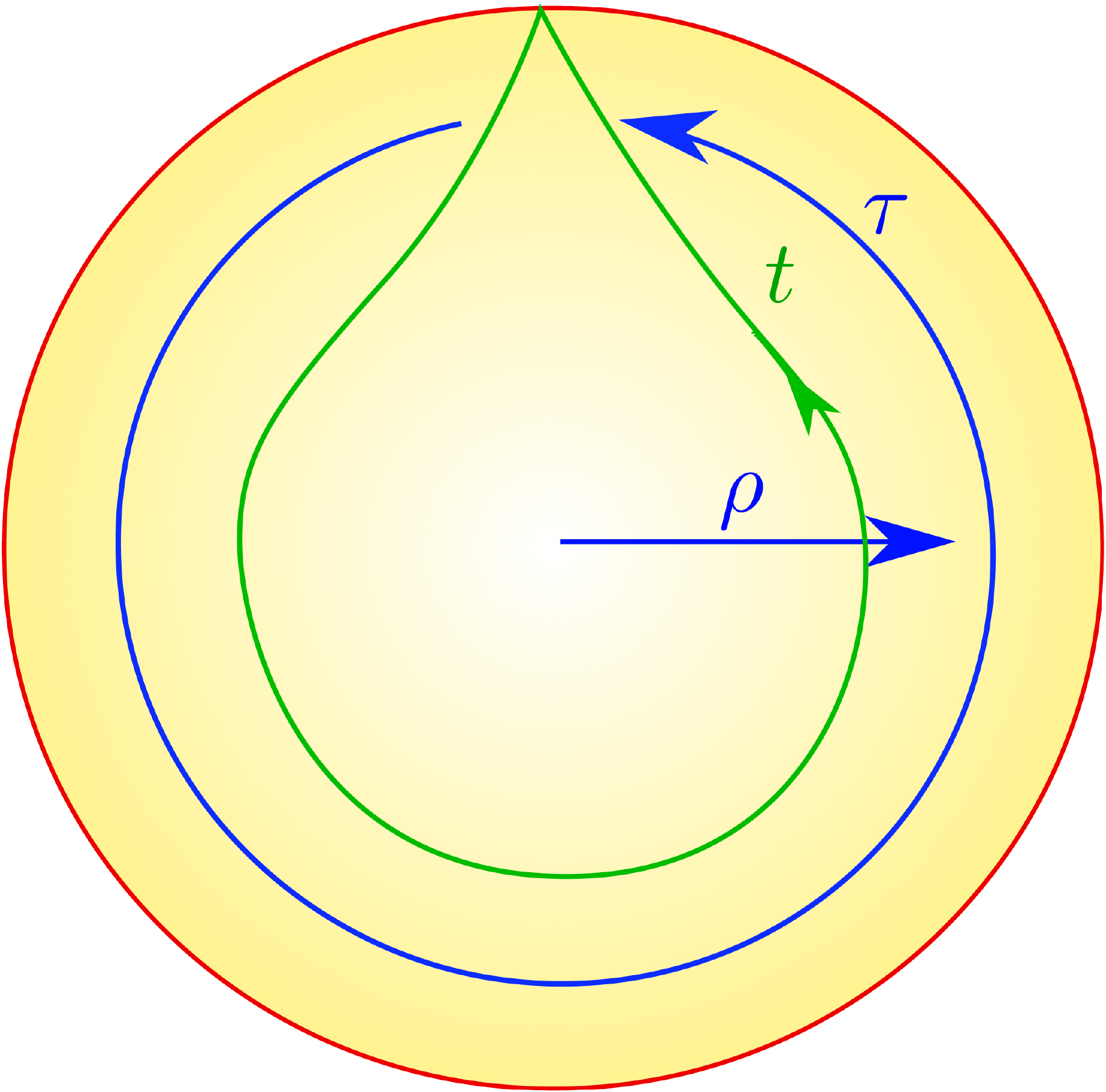} \hspace{1cm}\includegraphics[width=0.4\textwidth]{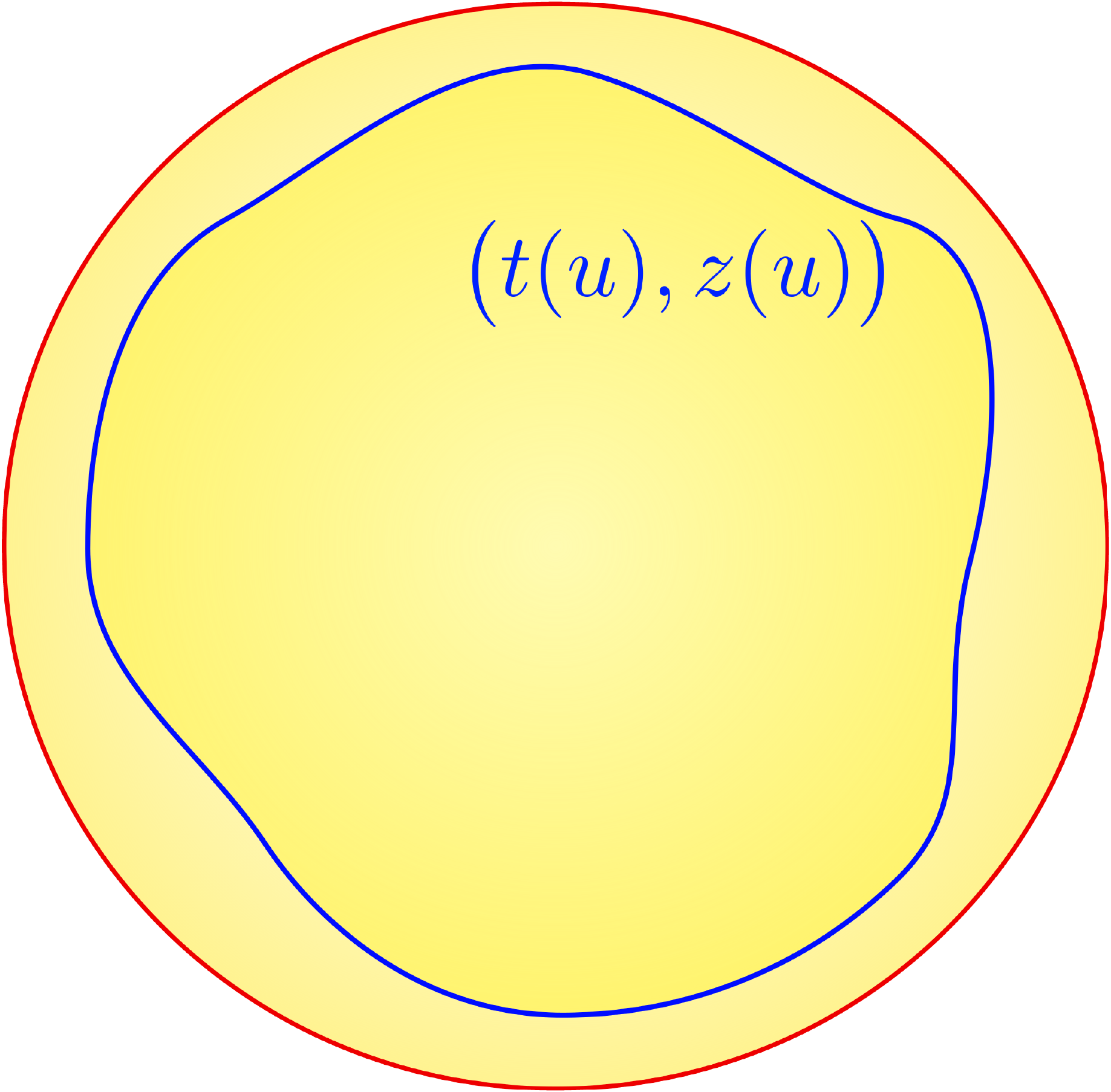}
\caption{Left: Coordinates on the hyperbolic disk. Right: A cutout from the hyperbolic disk.}
\label{fig:2}
\end{figure}

In Euclidean signature, AdS$_2$ is just the hyperbolic disk. Two sets of natural coordinates are obtained by Wick rotating the Poincar\'e coordinates in \eqref{eq:ad2coords}, or the Rindler coordinates of \eqref{eq:Rindler}, by $t_{\rm Lorentz}= -it_{\rm Euclidean}$, $\tau_{\rm Lorentz}= -i\tau_{\rm Euclidean}$ respectively
\bea
\label{eq:euclidmetrics}
ds^2 &= \frac{dt^2+dz^2}{z^2} && \text{Poincar\'e} \\
&=d\rho^2+\sinh^2 \rho d\tau^2 && \text{Rindler}.
\eea
Both of these coordinates cover the entire hyperbolic disk, as opposed to the Lorentzian case. The Poincar\'e time $t$ runs from $-\infty$ to $\infty$, while the Rindler time $\tau$ is $2\pi$ periodic and is a proper angular coordinate on the hyperbolic disk, see left of Fig. \ref{fig:2}.

Let us move on to the Euclidean version of the action \eqref{eq:JTlorentz}. This is straightforwardly obtained by Wick rotation. In addition, we are going to add the Gibbons-Hawking-York boundary terms, which are needed for the variational principle when we wish to put Dirichlet boundary conditions at the boundary of the manifold. As we have discussed in sec. \ref{sec:holo}, we are ultimately interested in cutoff versions of AdS$_2$, so we definitely need these. The action then reads as
\bea
\label{eq:JTeuclid}
I &=-\frac{\phi_0}{16 \pi G_N}\left[ \int_M d^2 x \sqrt{h} R_h +2\int_{\partial M} K\right] \\ &-\frac{1}{16 \pi G_N} \left[ \int_M d^2 x \sqrt{h} \phi \left(R_h+2 \right)+2\int_{\partial M} \phi_b K \right],
\eea
The boundary condition on $\phi$ is given by $\phi_b$. The boundary terms involve the trace of the extrinsic curvature (or second fundamental form), $K$. For codimension-one boundaries, the extrinsic curvature is given by
\beq
\label{eq:fundform}
K(T_1,T_2)=-h(T_1,\nabla_{T_2} n),
\eeq
where $n$ is the unit normal to the boundary surface and $T_{1,2}$ are tangent vectors to the surface (pushed forward to the tangent space of $M$ with the embedding of the surface) satisfying $h(T_{1,2},n)=0$.\footnote{Here, $h(X,Y)=h_{ab}X^a Y^b$, where $h_{ab}$ is the usual metric tensor on $M$.} In our case, the boundary is one dimensional, so there is a single tangent vector $T$. The trace of the bilinear form \eqref{eq:fundform} is then easily evaluated as
\beq
\label{eq:extrinsic}
K=-\frac{h(T,\nabla_{T} n)}{h(T,T)}.
\eeq

\subsection{Configuration space}

We are interested in cutoff versions of Euclidean AdS. There are many different ways to cut out a chunk of the hyperbolic disk, see right of Fig. \ref{fig:2}. The first line of the action \eqref{eq:JTeuclid} is the Einstein-Hilbert part and it is proportional to the Euler character of $M$, due to the Gauss-Bonnet theorem. Whenever $M$ is a simply connected chunk of the hyperbolic disk, this action is the same. This is an enormous amount of symmetry: all deformations of these chunks into each other are zero modes of the action. The role of the second piece in \eqref{eq:JTeuclid}, involving the dilaton $\phi$, will be to lift this degeneracy. 

We may describe the simply connected cutouts of Fig. \ref{fig:2} by fixing Poincar\'e coordinates on the hyperbolic disk and cutting it off at a boundary trajectory
\beq
\label{eq:param}
\big(t(u),z(u)\big).
\eeq
We want to think about the parameter $u$ of the curve as a time in the boundary theory. To make this uniform, we demand that the induced metric satisfies\footnote{Note that fixing the induced metric on the boundary is where we pick the Dirichlet boundary condition for the bulk metric, even though in this discussion we have gauge fixed this to be the Poincar\'e metric. We could imagine a coordinate change such that the cutoff surface is at a fixed coordinate position. In this description each cutout would appear to correspond to a different bulk metric, but all of which still satisfies \eqref{eq:induced}.} 
\beq
\label{eq:induced}
g|_{\rm bndy} = \frac{1}{\epsilon^2}.
\eeq
We take $\epsilon$ to be small. This is the parameter that plays the role of the UV cutoff in the boundary theory, as discussed in sec. \ref{sec:holo} and \ref{sec:JT}. Condition \eqref{eq:induced} implies that our parametrization \eqref{eq:param} satisfies 
\beq
\label{eq:z}
z=\epsilon \sqrt{(t')^2+(z')^2} = \epsilon t' + O(\epsilon^3),
\eeq
so it is determined by a single function $t(u)$. We may think about $t(u)$ as the dynamical variable in our gravitational model \eqref{eq:JTeuclid}. The fact that the Einstein-Hilbert part of the action is the same for all $t(u)$ can be interpreted as a symmetry under reparametrizations of the boundary time $u\mapsto f(u)$.\footnote{In the language of asymptotic symmetries, discussed by Max and C\'eline in their lectures \cite{Riegler:2017fqv}, these reparametrizations correspond to the asymptotic diffeomorphisms generated by
\beq
\zeta [\eta]= \eta(t)\partial_t + z \eta'(t)\partial_z.
\eeq
These generate the reparametrizations $t(u) \mapsto t(u)+\eta(t(u))$.
}
Notice that most of these reparametrizations of $t(u)$ map a given cutout of the hyperbolic disk to a different one. This might seem odd at first, but it is because of the condition \eqref{eq:induced} for the induced metric: by reparametrizing $u$ we are not just changing the parametrization of the boundary curve, but instead map it to a new one. This is the case for most reparametrizations however, not all of them. Translations and rotations of a fixed shape around in the hyperbolic disk does not change the chunk that we are actually cutting out. These correspond to the reparametrizations\footnote{To see this, use that the Poincar\'e metric, when written in terms of complex coordinates $w=t+i z$, is invariant under the M\"obius transformations $w\mapsto \frac{a w+b}{c w+d}$. The rule for $t$ follows by noting that $z$ is subleading in $\epsilon$ because of \eqref{eq:z}.}
\beq
\label{eq:sl2}
t(u) \mapsto \frac{a t(u)+b}{c t(u)+d}, \;\;\; ad-bc=1,
\eeq
which form an $SL(2,\mathbb{R})$ subgroup of all reparametrizations. Therefore, all the cutouts from the hyperbolic disk \textit{spontaneously break} the reparametrization symmetry down to $SL(2,\mathbb{R})$. We can think about the $t(u)$ as the Goldstone modes associated to this symmetry breaking. However, as $t(u)$ lives in one dimension, it is not just massless but actually has zero action.\footnote{The usual Goldstone theorem basically relies on the fact that the zero wavelength modes of fields that correspond to moving in the space of degenerate vacua do not appear in the action. Here, as there is no spatial direction, the only thing a field has is its zero wavelength mode.}

\subsection{Schwarzian theory}

The role of the second line in the action \eqref{eq:JTeuclid}, depending on a dilaton field $\phi$, is precisely to break explicitly this reparametrization symmetry and give a finite action for $t(u)$. In the context of the extremal black holes of sec. \ref{sec:motiv}, this translates to the very near horizon region enjoying the reparametrization symmetry, but as we move away a little bit towards the completion of the spacetime, the symmetry is broken explicitly. In the language of the holographic renormalization group of sec \ref{sec:holo}, the reparametrization symmetry is just the conformal symmetry of the fixed point CFT$_1$ and follows from the vanishing of the Hamiltonian. But once we move a little to the UV along some irrelevant direction, the symmetry is explicitly broken. The role of sec. \ref{sec:JT} was to argue that the holographic description of this irrelevant deformation is captured by \eqref{eq:JTeuclid} for a large class of UV completions.

Since the dilaton $\phi$ appears linearly in \eqref{eq:JTeuclid} and without derivatives, it is basically just a Lagrange multiplier. Integrating it out leads to the constraint $R=-2$, which just tells us that the metric is AdS$_2$. The action is therefore a boundary term
\bea
I&=-\frac{1}{8\pi G_N} \int_{\partial M} \phi_b K \\
&=-\frac{1}{8\pi G_N} \int \frac{du}{\epsilon^2} \phi_r(u) K.
\eea 
In the second line we have used \eqref{eq:induced} and set the boundary condition for the dilaton as
\beq
\phi_b = \frac{\phi_r(u)}{\epsilon},
\eeq
which is motivated by the linear blowing up of the dilaton near the boundary described in sec. \ref{sec:backreact}. The function $\phi_r(u)$ plays the role of a source for the operator dual to the dilaton, as in usual AdS/CFT. We can use \eqref{eq:extrinsic} to compute the extrinsic curvature with
\bea
T^a=\big(t',z'\big), && n^a=\frac{z}{\sqrt{{t'}^2+{z'}^2}}\big( -z',t' \big).
\eea
The result is
\bea
K&=\frac{t'({t'}^2+{z'}^2+z'z'')-z z' t''}{({t'}^2+{z'}^2)^{3/2}} \\ &= 1+S(t(u),u) \epsilon^2 + O(\epsilon^4).
\eea
In the second equality, we have used \eqref{eq:z} and defined the \textit{Schwarzian derivative}
\beq
\label{eq:schwartz}
S(t(u),u)=\frac{2 t' t'''-3{t''}^2}{2{t'}^2}.
\eeq
By generously neglecting the field independent divergent term, we arrive at the result that the Jackiw-Teitelboim theory reduces to a boundary theory with action\footnote{Further reading on this action and its gravitational context includes \cite{Jensen:2016pah,Engelsoy:2016xyb,Cvetic:2016eiv,Stanford:2017thb}.}
\beq
\label{eq:schaction}
I_{\rm Sch}=-\frac{1}{8\pi G_N} \int du \phi_r(u) S(t(u),u).
\eeq
We will assume the boundary value of the dilaton to be a constant $\phi_r(u)=\bar \phi_r$ in the following. The appearance of the Schwarzian derivative in the action is not very surprising. It is the lowest derivative local expression that is invariant under $SL(2,\mathbb{R})$ transformations \eqref{eq:sl2}.

So what are the solutions to \eqref{eq:schaction}? One can check that the equation of motion is
\beq
\frac{[S(t,u)]'}{t'}=0,
\eeq
so we are looking for nonconstant functions with constant Schwarzian. The $SL(2,\mathbb{R})$ transformations \eqref{eq:sl2} qualify, but we have already seen that these are not dynamical modes, but more of a redundancy in the description of the cutouts, a gauge symmetry. To find different maps with a constant Schwarzian, we can use the composition law\footnote{The quickest way to derive this is to think about the transformation rule of the stress tensor in 2d CFTs. One applies a conformal transformation $g$ and then another one $f$ and demands that the result is the same as applying $f\circ g$ at the beginning.}
\beq
S(f\circ g,t)={g'}^2 S(f,g)+S(g,t).
\eeq
Setting $t(u)=\tan \frac{\tau(u)}{2}$ results in the relation
\beq
\label{eq:rindlerschw}
S(t,u)=S(\tau,u)+\frac{1}{2}{\tau'}^2.
\eeq
We see that when $\tau$ is a linear function of $u$, the Schwarzian $S(t,u)$ is constant, so we have a solution. Notice that changing from $t$ to $\tau$ is nothing but the boundary limit of \eqref{eq:Rindler}, i.e. the change between Poincar\'e and Rindler coodinates. Since the Euclidean Rindler time $\tau$ is periodic with period $2\pi$, these coordinates are well fit to describe thermal solutions. These solutions are
\beq
\label{eq:thermalsol}
\tau(u)=\frac{2\pi}{\beta}u.
\eeq
Indeed, periodicity of $\tau$ requires the boundary time to be periodic with period $\beta$, so we can interpret $\beta$ as the inverse temperature.

\subsubsection{Thermal entropy}

We can obtain the thermal entropy of the boundary theory the following way. The holographic dictionary tells us that as $G_N \rightarrow 0$, the boundary partition function is given by
\beq
Z(\beta) = e^{-I_{\rm grav}},
\eeq
where $I_{\rm grav}$ is the on-shell Euclidean action in the bulk, with boundary conditions set by sources in the boundary theory. We can evaluate the Schwarzian part \eqref{eq:schaction} of the action for the thermal solutions \eqref{eq:thermalsol}:
\beq
\label{eq:schwfreeenerg}
I_{\rm Sch}=-2\pi^2 C \frac{1}{\beta}, \;\;\; C=\frac{\bar \phi_r}{8\pi G_N}.
\eeq
We have mentioned before, that the topological Einstein-Hilbert term in \eqref{eq:JTeuclid} can be interpreted as a ground state entropy, let the value of this be $-S_0$. The total on-shell action is then\footnote{You might remember that we have dropped a divergent term when we have derived the Schwarzian action. We have mentioned briefly in sec. \ref{sec:JT} that divergent contributions to the topological piece are related to the ground state energy of the $CFT_1$ fixed point. The term that we have dropped when deriving the Schwarzian action is just the leading correction to the ground state energy when we are deforming away from the fixed point. You can also check that all these divergent terms we have droppend along the way are proportional to $\beta$, so they do not contribute to the entropy.}
\beq
I_{\rm grav}=-S_0-2\pi^2 C \frac{1}{\beta}.
\eeq
Using this, we obtain the thermal entropy
\beq
S_{\rm th} = (1-\beta\partial_\beta)\log Z = S_0+4\pi^2 \frac{C}{\beta}.
\eeq
Based on what we have learned in the motivating section \ref{sec:motiv}, we may interpret this as the entropy of a near extremal black hole, which is correctly linear in temperature.

\subsubsection{Linearized theory}

Now that we have the nice action \eqref{eq:schaction} at hand we would like to put it to use and do some basic calculations. The simplest thing we can do is to study the linearized theory
\beq
\tau(u)=u+\varepsilon(u),
\eeq
with $\varepsilon$ being small. We are studying fluctuations around the $\beta=2\pi$ solution for simplicity. Quantizing the resulting quadratic theory corresponds to the tree level gravitational dynamics and gives the leading perturbative result in $G_N$.

\begin{figure}[h!]
\centering
\includegraphics[width=0.4\textwidth]{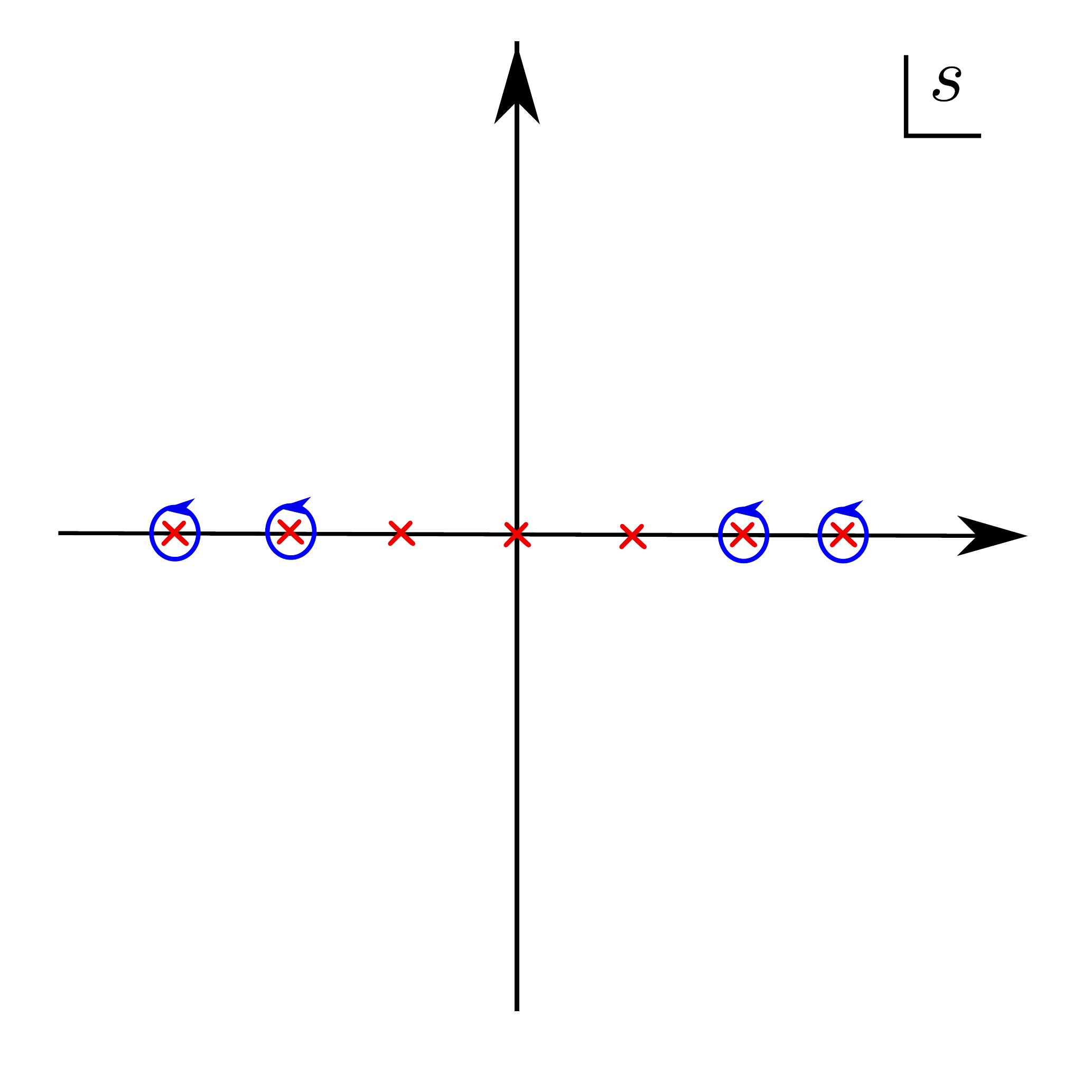} \hspace{1cm}\includegraphics[width=0.4\textwidth]{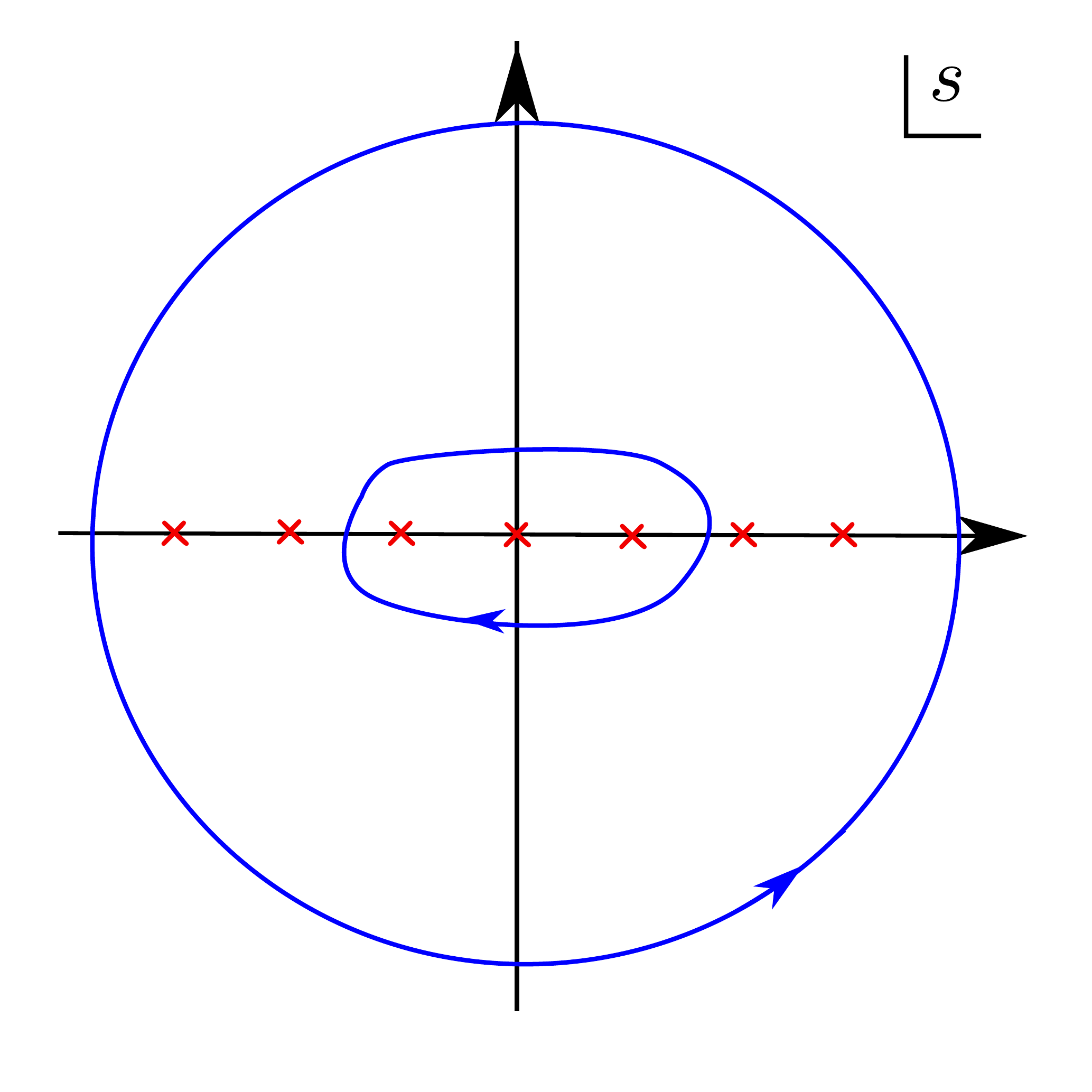}
\caption{Left: Original contour. Right: Deformed contour.}
\label{fig:3}
\end{figure}

To expand the action \eqref{eq:schaction}, we make use of \eqref{eq:rindlerschw}:
\beq
S(\tau,u)+\frac{1}{2}{\tau'}^2 = \frac{1}{2} + (\varepsilon'+\varepsilon'')+\big( \frac{1}{2}{\varepsilon'}^2-\frac{1}{2}{\varepsilon''}^2-(\varepsilon'' \varepsilon')'\big) + O(\varepsilon^3).
\eeq
Dropping total derivatives, the action then reads as
\beq
\label{eq:linearizedschw}
I_{\rm Sch}=\frac{C}{2} \int_0^{2\pi}du \big({\varepsilon''}^2-{\varepsilon'}^2 \big).
\eeq
When we quantize this linearized theory, the two point function is given by the inverse of the Gaussian kernel. To obtain this, we must diagonalize the kernel with a Fourier transformation
\beq
\varepsilon(u)=\sum_{n\in \mathbb{Z}} \varepsilon_n e^{i n u} \;\; \rightarrow \;\;I=\frac{C}{2}\sum_{n\in \mathbb{Z}}(n^4-n^2)\varepsilon_n \varepsilon_{-n}.
\eeq
At first sight, we might think that we have a problem as this kernel is not invertible: it is zero for $n=0, \pm 1$. However, we recall that $SL(2,\mathbb{R})$ transformations of a solution are just redundancies, therefore we should not integrate over them in the path integral. At the linearized level, these correspond to the variables $\varepsilon_0$, $\varepsilon_{\pm 1}$.
Dropping these modes, the propagator reads as
\beq
\label{eq:2ptfourier}
\langle \varepsilon(u) \varepsilon(0) \rangle = \frac{2}{C}\sum_{n \neq 0,\pm 1} \frac{e^{i n u}}{n^2(n^2-1)}.
\eeq
We can evaluate this sum by writing it as a contour integral
\beq
\langle \varepsilon(u) \varepsilon(0) \rangle = \frac{2}{C} \oint_{\mathcal{C}} \frac{ds}{e^{2\pi i s}-1} \frac{e^{i su}}{s^2(s^2-1)},
\eeq
where the contour $\mathcal{C}$ is the union of small circles, running counter clockwise around integer values of $s$, except $0, \pm 1$. We can deform this contour into the union of a big circle at infinity, running counter clockwise, and a clockwise contour encircling only the poles at $s=0,\pm 1$. See Fig. \ref{fig:3} for an illustration. Our aim is to drop the integral on the circle at infinity. Examining the behaviour of the integrand, we can easily see that we can do this when $0<u<2\pi$. This is enough, as we know from the form \eqref{eq:2ptfourier} that the propagator is symmetric under $u\rightarrow -u$. All we need to do then, is picking up the residues at $s=0,\pm 1$. The result is
\beq
\label{eq:epsilon2pt}
\langle \varepsilon(u) \varepsilon(0) \rangle = \frac{2\pi}{C} \left( -\frac{(u-\pi)^2}{2}+(u-\pi)\sin u+1+\frac{\pi^2}{6}+\frac{5}{2} \cos u\right).
\eeq
We can extend this to $u<0$ by substituting $u\rightarrow |u|$ everywhere.

\subsection{Coupling to matter}

The next step is to understand how the theory \eqref{eq:schaction} couples to matter. For this purpose, we add a free massive scalar $\chi$ in the bulk with action
\beq
\label{eq:matter}
I_{\rm matter}=\frac{1}{2}\int d^2 x \sqrt{h} \big(h^{ab}\partial_a \chi \partial_b \chi +m^2 \chi^2 \big),
\eeq
which is coupled to gravity in the usual way. The usual AdS/CFT dictionary tells us that the partition function of the dual theory acquires a dependence on the boundary value $\tilde \chi_r(t)$ of the field, defined via
\beq
\label{eq:matterbc}
\chi(z,t) = z^{1-\Delta} \tilde \chi_r(t) + \cdots, \text{   as   } z\rightarrow 0.
\eeq
We interpret $\tilde \chi_r(t)$ as a source for an operator with scaling dimension\footnote{We assume that $\Delta>1/2$, to avoid complications with possible alternate quantizations, as these are not the questions in focus here.}
\beq
\Delta=\frac{1}{2}\left( 1+\sqrt{1+4m^2} \right).
\eeq
The log of the boundary partition function will contain a term quadratic in $\tilde \chi_r$, coming from the on-shell evaluation of the action \eqref{eq:matter}\footnote{The quadratic piece in the sources generates the two point functions of the boundary theory, whose form is fixed by conformal invariance. This is enough to guess the form of $I_{\text{matter on-shell}}$ up to the proportionality constant. For the readers unfamiliar with the bulk derivation of this formula, we refer e.g. to \cite{Freedman:1998tz}.}
\beq
\label{eq:matteronshell}
I_{\text{matter on-shell}} = -D \int dt dt' \frac{\tilde \chi_r(t)\tilde \chi_r(t')}{|t-t'|^{2\Delta}}, \;\;\;\;\; D=\frac{(\Delta-\frac{1}{2})\Gamma(\Delta)}{\sqrt{\pi} \Gamma(\Delta-\frac{1}{2})}.
\eeq
In higher dimensions, there is an entirely analogous formula and here we basically have the same formula with $d=1$. But there is a big difference. The derivation of formula \eqref{eq:matteronshell} assumes that the background metric for the scalar is just the Poincar\'e metric \eqref{eq:euclidmetrics} of AdS. But we have seen that in two dimensions \textit{all} gravitational configurations can be described by this metric. These configurations are labelled by the boundary curve $t(u)$, and \eqref{eq:matteronshell} secretly depends on this. To see this, we rewrite the boundary condition \eqref{eq:matterbc} in terms of the boundary time $u$ using \eqref{eq:z} as
\bea
\chi(z,t) &=z(u)^{1-\Delta} \tilde \chi_r(t(u)) + \cdots \\
&= \epsilon^{1-\Delta} (t'(u))^{1-\Delta} \tilde \chi_r(t(u))+ \cdots \\
&=\epsilon^{1-\Delta}  \chi_r(u),
\eea
where the last equations defines $\chi_r(u)$, which we see transforms as a conformal primary of dimension $1-\Delta$ under reparametrizations. Using this, we can write \eqref{eq:matteronshell} as
\beq
\label{eq:matterbackreact}
I_{\text{matter on-shell}} = -D \int du du' \left[ \frac{t'(u)t'(u')}{\big( t(u)-t(u')\big)^2} \right]^{\Delta} \chi_r(u) \chi_r(u').
\eeq
This formula describes the complete gravitational coupling of $\chi$ to the metric degree of freedom $t(u)$. It implies that to leading order in $G_N$, the boundary partition function with source $\chi_r(u)$ turned on for a dimension $\Delta$ operator is given by
\beq
\label{eq:sourceddict}
Z[\chi_r(u)]=e^{-S_0-I_{\rm Sch}-I_{\text{matter on-shell}}}, 
\eeq
where $I_{\rm Sch}$ is the Schwarzian action of \eqref{eq:schaction} and the $t(u)$ that we need to use on the r.h.s. is obtained by extremizing the total action $I_{\rm Sch}+I_{\text{matter on-shell}}$. Notice that this $t(u)$ in general depends on $\chi_r(u)$ which makes the exponential non-quadratic in $\chi_r(u)$.

\subsection{Correlation functions}
\label{sec:4ptSchw}

In the case when $\Delta$ is not too large (the bulk scalar $\chi$ is not too heavy), we can basically neglect $I_{\text{matter on-shell}}$ in the determination of the saddle in $t(u)$. This is because the Schwarzian action \eqref{eq:schaction} comes with an extra $1/G_N$ factor. Since the coefficient in \eqref{eq:matteronshell} goes as $\Delta^{3/2}$ for large $\Delta$, the backreaction from $\chi_r(u)$ is suppressed as long as $\Delta$ grow slower than $G_N^{-2/3}$ as $G_N \rightarrow 0$. In this case, the dual field $V(u)$ to the source $\chi_r$ is effectively free, as all of its connected correlators vanish except for the two point function which is
\beq
\langle V(u)V(u') \rangle \sim \left[ \frac{t'(u)t'(u')}{\big( t(u)-t(u')\big)^2} \right]^{\Delta},
\eeq
with $t(u)$ being a saddle for the Schwarzian theory. 

There are two possible sources of $G_N$ corrections to this behaviour. First, the bulk scalar $\chi$ can have self interaction terms and it can also source other matter fields via bulk couplings. The backreaction from $\chi_r$ to the saddle $t(u)$ is also in this class.  Second, there are loop corrections coming from the fact that for finite $G_N$ in \eqref{eq:sourceddict}, we are supposed to integrate over the bulk fields $t(u)$ and $\chi(z,t)$ on the right hand side. The interesting part in the present context, which works in a much simpler way than in higher dimensions, is of course the gravitational loop corrections coming from $t(u)$. The way we can calculate the first loop correction, is simply via expanding around the thermal saddle 
\beq
t(u)=\tan \frac{u+\varepsilon(u)}{2}
\eeq
and using the propagator \eqref{eq:epsilon2pt} to contract quadratic appearances of $\varepsilon$. For this, we will need to make use of the expansion
\bea
\label{eq:2ptexpand}
\left[ \frac{t'(u_1)t'(u_2)}{\big( t(u_1)-t(u_2)\big)^2} \right]^{\Delta} &= \frac{1}{\left(2\sin \frac{u_{12}}{2}\right)^{2\Delta}} \Big[ 1+\mathcal{B}(u_1,u_2) + \mathcal{C}(u_1,u_2) + O(\varepsilon^3)\Big],
\eea
where we have introduced the shorthand $u_{ij}=u_i-u_j$ and denoted the linear and quadratic contributions respectively with
\bea
\mathcal{B}(u_1,u_2) &= \Delta \Big( \varepsilon'(u_1)+\varepsilon'(u_2)-\frac{\varepsilon(u_1)-\varepsilon(u_2)}{\tan \frac{u_{12}}{2}}\Big)\\
\mathcal{C}(u_1,u_2)&=\frac{\Delta}{\left( 2\sin \frac{u_{12}}{2}\right)^2}\big[ (1+\Delta+\Delta \cos u_{12})(\varepsilon(u_1)-\varepsilon(u_2))^2\\
&+2 \Delta \sin u_{12} (\varepsilon(u_2)-\varepsilon(u_1))(\varepsilon'(u_1)+\varepsilon'(u_2))\\
&-(\cos u_{12}-1)\big((\Delta-1)(\varepsilon'(u_1)^2+\varepsilon'(u_2)^2)+2\Delta \varepsilon'(u_1)\varepsilon'(u_2)\big)\big]
\eea
In order to extract the leading $G_N$ correction to the generating functional, we need to expand the exponential of \eqref{eq:matterbackreact} to quadratic order\footnote{Since the action comes with a prefactor $G_N^{-1}$, fluctuations around the saddle have typical size $\varepsilon \sim G_N^{1/2}$.} in $\varepsilon$ and take the expectation value in the linearized Schwarzian theory. This results, for the generator of connected correlators, in the expansion
\bea
\log \langle e^{-I_{\text{matter on-shell}}}\rangle &= D \int du_1 du_2\left( 1+\langle \mathcal{C}(u_1,u_2) \rangle \right) \frac{\chi_r(u_1)\chi_r(u_2)}{(2\sin \frac{u_{12}}{2})^{2\Delta}}\\
&+\frac{D^2}{2} \int du_1 du_2 du_3 du_4 \frac{\chi_r(u_1)\chi_r(u_2)\chi_r(u_3)\chi_r(u_4)}{{(2\sin \frac{u_{12}}{2})^{2\Delta}}{(2\sin \frac{u_{34}}{2})^{2\Delta}}}\langle \mathcal{B}(u_1,u_2)\mathcal{B}(u_3,u_4) \rangle \\
&+O(G_N^2).
\eea
Here, we have made use of the fact that the one point function of $\varepsilon$ vanish, so $\langle \mathcal{B}(u_1,u_2)\rangle=0$. 

\subsubsection{Two point function}

First, let us focus on the correction $\langle \mathcal{C}(u_1,u_2)\rangle$ to the two point function.  The expectation values are obtained by using the propagator \eqref{eq:epsilon2pt}. Assuming $u_1>u_2$ we obtain\footnote{Some useful relations to evaluate this quicker are \bea
\langle \varepsilon(u_1)\varepsilon(u_2)\rangle &= G(|u_{12}|), && \langle \varepsilon'(u_1)\varepsilon(u_2)\rangle &= \text{sgn} u_{12} G'(|u_{12}|), && \langle \varepsilon'(u_1)\varepsilon'(u_2)\rangle &=-G''(|u_{12}|).
\eea}
\bea
\langle \mathcal{C}(u_1,u_2) \rangle &= \frac{1}{2\pi C} \frac{\Delta}{\left( 2\sin \frac{u_{12}}{2}\right)^2}\Big[ 2+4\Delta+u_{12}(u_{12}-2\pi)(\Delta+1)\\
&+\big( \Delta u_{12} (u_{12}-2\pi)-4\Delta-2\big)\cos u_{12}
+2(\pi-u_{12})(2\Delta+1)\sin u_{12}\Big].
\eea

\subsubsection{Four point function}

Now we want to evalute $\langle \mathcal{B}(u_1,u_2)\mathcal{B}(u_3,u_4) \rangle$. Before doing this, let us introduce another operator $W$ with the same conformal weight $\Delta$ as $V$, which has vanishing two point function with $V$. The only purpose of this is that the connected four point function
\beq
F=\frac{\langle V(u_1)V(u_2)W(u_3)W(u_4)\rangle -\langle V(u_1)V(u_2) \rangle \langle W(u_3) W(u_4)\rangle}{\langle V(u_1)V(u_2) \rangle \langle W(u_3) W(u_4)\rangle}.
\eeq
is now directly given by 
\beq
F= \langle \mathcal{B}(u_1,u_2)\mathcal{B}(u_3,u_4) \rangle,
\eeq
and we do not need to worry about the cross channels which are present if all four operators are the same. Conceptually, there is no difference.

We can use \eqref{eq:epsilon2pt} again to evaluate the expectation value. However, now we see that the result actually depends on the operator ordering in a significant way. This is because $\langle \varepsilon(u_1)\varepsilon(u_2)\rangle$ only depends on $|u_{12}|$, and we have terms like $\langle \varepsilon'(u_1)\varepsilon(u_2)\rangle$, which are therefore proportional to $\text{sgn} u_{12}$. When we have the ordering $u_1>u_2>u_3>u_4$ between Euclidean times, we obtain a fairly simple expression
\beq
\label{eq:FTO}
F_{VVWW}= \frac{\Delta^2}{2\pi C}\left( \frac{u_{12}}{\tan \frac{u_{12}}{2}}-2\right)\left( \frac{u_{34}}{\tan \frac{u_{34}}{2}}-2\right).
\eeq
However, when we consider the other ordering $u_1>u_3>u_2>u_4$ we end up with a different result\footnote{Operators are always ordered in correlators so that larger Euclidean time is to the left. This is because $e^{-\tau H}$ is bounded only for positive $\tau$ so Euclidean time evolution is only possible in a single direction. Therefore this second Euclidean time ordering corresponds to the operator order $VWVW$.}
\beq
\label{eq:FOTO}
F_{VWVW} = F_{ VVWW}+\frac{\Delta^2}{2\pi C} \left[ 2\pi\frac{\sin \frac{u_{12}+u_{34}}{2}-\sin \frac{u_{23}+u_{14}}{2}}{\sin \frac{u_{12}}{2}\sin \frac{u_{34}}{2}} + \frac{2\pi u_{23}}{\tan \frac{u_{12}}{2}\tan \frac{u_{34}}{2}}\right].
\eeq
The main physical difference for this alternating ordering is the appearance of the cross distances $u_{14}$ and $u_{23}$, which are absent in \eqref{eq:FTO}. The significance of this subtle difference will be the topic of the next section.

\subsection{Relation to chaos}

To understand the significance of the cross distances in \eqref{eq:FOTO}, we need to venture off a little bit and talk about semiclassical chaos. This short discussion is mainly based on \cite{Maldacena:2015waa,Polchinski:2015cea}.

\begin{figure}[h!]
\centering
\includegraphics[width=0.3\textwidth]{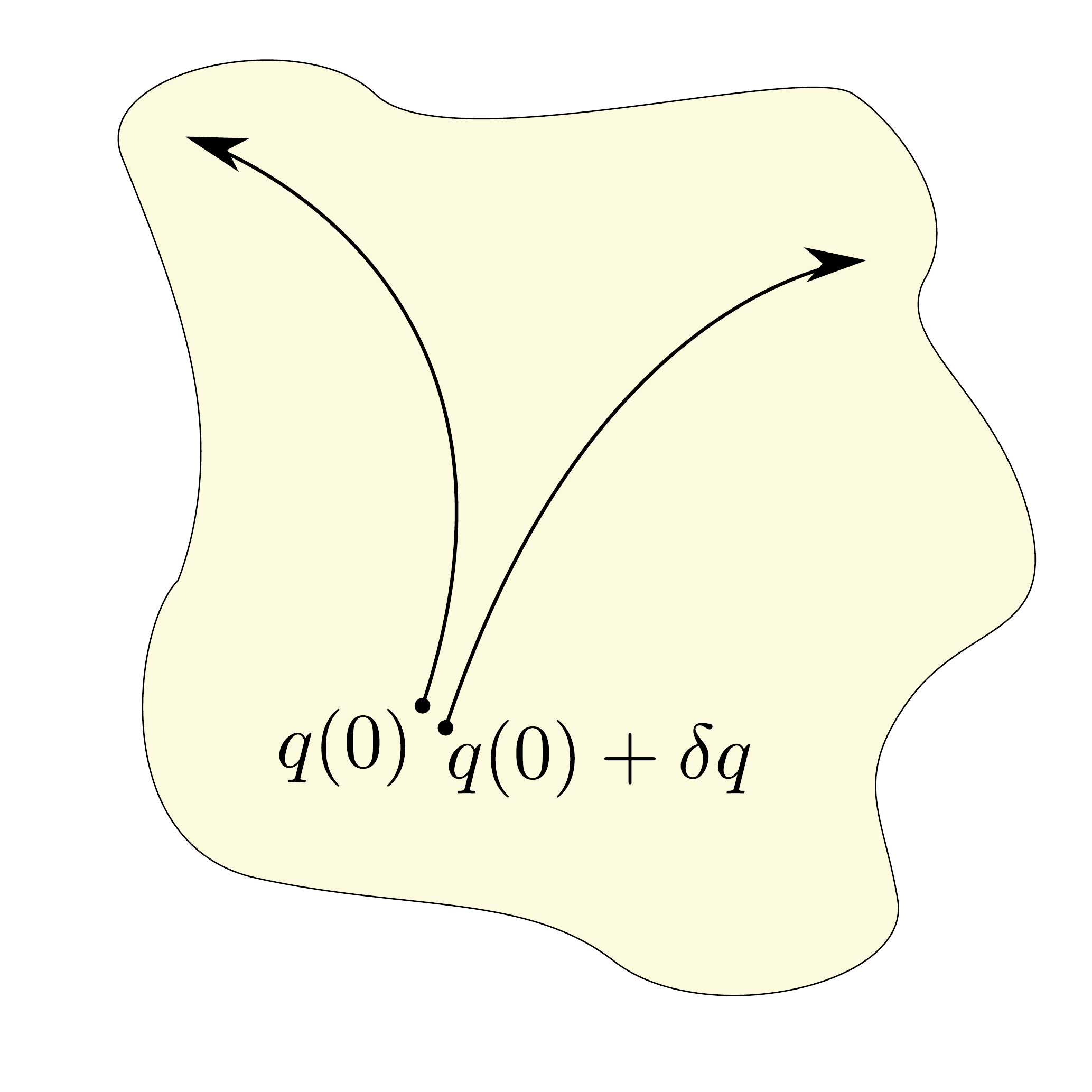} \hspace{1cm}\includegraphics[width=0.5\textwidth]{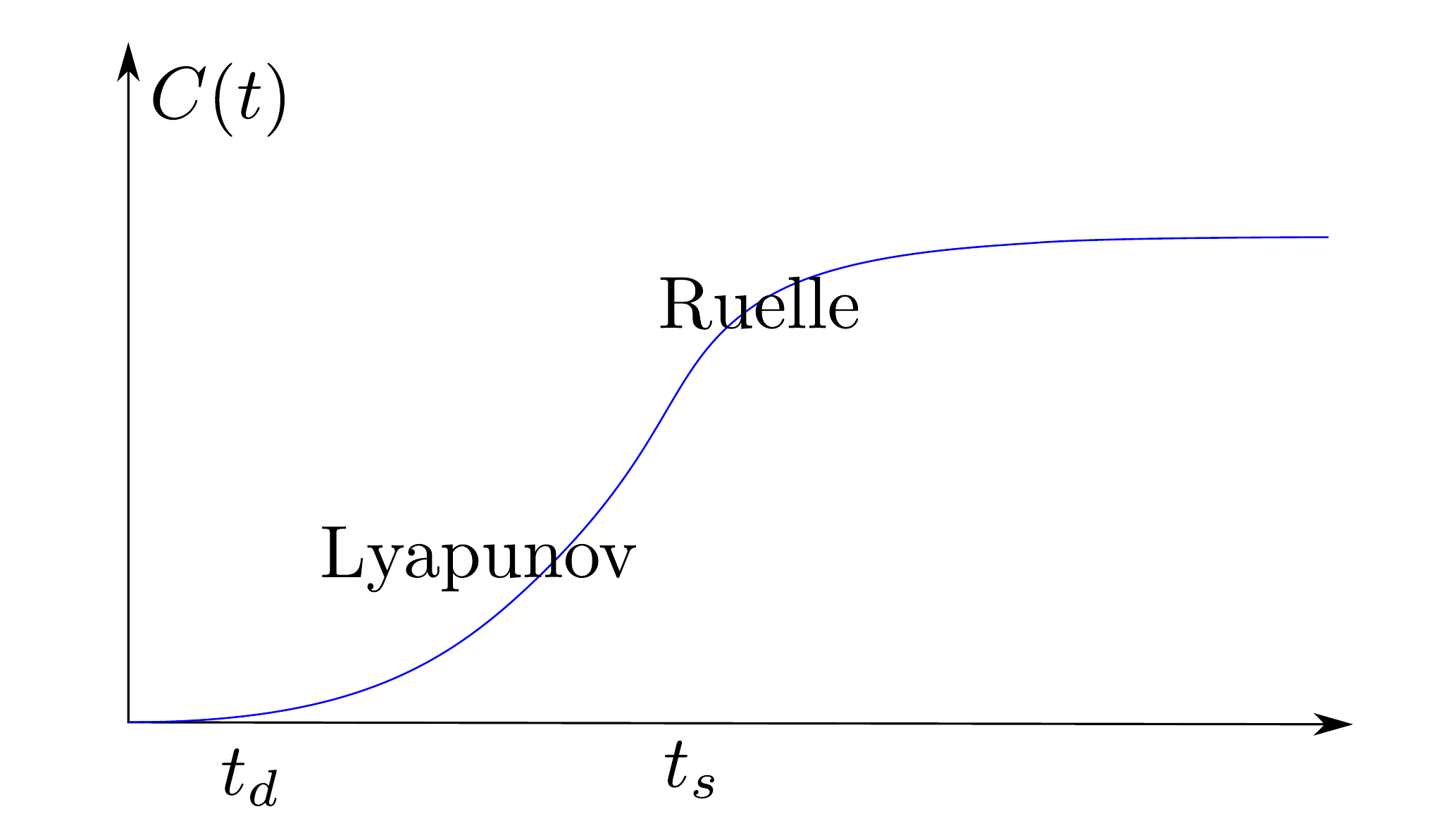}
\caption{Left: Divergence of trajectories in phase space. Right: Typical behaviour of \eqref{eq:commsq}.}
\label{fig:4}
\end{figure}

In a classical system, like a chaotic billiard, a simple diagnostic of chaos is the high dependence of trajectories on the initial conditions (see left Fig. \ref{fig:4}). This is called the butterfly effect, for reasons I am sure everyone is familiar with at least from popculture. In a chaotic system, nearby trajectories typically diverge exponentially fast in time, for some time period
\beq
\label{eq:semiclass}
\frac{\partial q(t)}{\partial q(0)}=\lbrace q(t),p(0)\rbrace \sim e^{\lambda_L t}.
\eeq
The exponent $\lambda_L$ is called the Lyapunov exponent and $\lbrace .,.\rbrace$ denotes the Poisson bracket. 

In a semiclassical quantum system with $\hbar \ll 1$, the Poisson bracket is well approximated by the commutator $\frac{1}{i\hbar} [q(t),p(0)]$. This gives a way to translate this diagnostic to quantum systems with some caveats. First, we need to square the commutator to avoid phase cancellations, and second we need to take an expectation value in some state. A general choice for this is the thermal state, which leads one to consider the quantity
\beq
\label{eq:commsq}
C(t)=-\langle [W(t),V(0)]^2\rangle_\beta,
\eeq
where $\langle .\rangle_\beta=\text{Tr} e^{-\beta H }/Z$ is the thermal expectation value. 
By expanding the square of the commutator, we find four terms, all of which are four point functions of $VVWW$. By using the KMS relation\footnote{KMS stands for Kubo-Martin-Schwinger and it expresses the Euclidean periodicity condition on thermal correlators, e.g. on the two point function $\text{Tr}\big(e^{-\beta H} A(t)B(0)\big)=\text{Tr}\big(e^{-\beta H}B(0)A(t+i\beta)\big)$.} to move operators cyclically, we can easily see that two of these are in Lorentzian time order, while two of them are out of time order correlators of the form
\beq
\label{eq:otoc}
\langle V(0)W(t)V(0)W(t) \rangle.
\eeq
The typical behaviour of $C(t)$ in a chaotic system is illustrated on the right of Fig. \ref{fig:4}. After a short collusion time $t_d\sim \beta$, there is usually a region of Lyapunov growth. This ends when the commutator obtains macroscopic values (it is initially $\sim \hbar$), which happens at the Ehrenfest time, $t_s \sim \frac{1}{\lambda_L} \log \frac{1}{\hbar}$. After this, $C(t)$ starts saturating exponentially to its late time average. This is called the Ruelle region.

It is also important to realize, that the Lyapunov region is not well defined in a theory where there is no large separation between the collusion and the scrambling time. The phase space argument that we gave shows that this large separation is pretty much guaranteed when we are looking at a theory with a classical limit,\footnote{By this we mean that there is some parameter $\chi$, such that for small $\chi$ the dynamics of some set of observables reduces to classical Hamiltonian dynamics on some phase space. This $\chi$ could be $\hbar$, or $N^{-2}$ for gauge theories with a holographic dual.} in which case we expect the growth of the Lyapunov part to be really related to the butterfly effect. Of course, many theories, like spin chain models, are not in this class.

With this in mind, one can study the behaviour of $C(t)$ in a theory holographically dual to Einstein gravity by studying certain shockwaves sent into an AdS-Schwartzschild black hole. In this case, the classical limit is governed by $G_N$, which takes over the role of $\hbar$. The original reference on this is \cite{Shenker:2013pqa}. Without going into any detail, it turns out that the growth of the Lyapunov part is a result of the exponential redshift near the horizon. On the other hand, the Ruelle region, i.e. the saturation to equilibrium, is described by the quasinormal ringdown of perturbations to the black hole.

Now after this detour, let us return to discussing the four point function \eqref{eq:FOTO}. We can reintroduce the temperature by rescaling $u_i \rightarrow \frac{2\pi}{\beta}u_i$ and we continue to Lorentzian by sending $u=i \hat u$ after this. We then parametrize
\bea
\hat u_1=a && \hat u_3=b+\hat u, && \hat u_2=0 && \hat u_4 =\hat u,
\eea
and insert all this into the four point function \eqref{eq:FOTO} with alternating ordering to obtain the out of time order correlator \eqref{eq:otoc}. We focus on the $\hat u \gg \beta$ regime. In this case we have
\bea
\hat u_{23}\sim \hat u_{14} \sim -\hat u, && \hat u_{12} \sim a, && \hat u_{34} \sim b.
\eea
Notice that only the cross-distances grow with $\hat u$, the ones that appear only in the out of time order case \eqref{eq:FOTO} and not in \eqref{eq:FTO}. At late times, we have
\beq
F_{VWVW}\sim \beta \frac{\Delta^2}{C}e^{\frac{2\pi}{\beta} \hat u}.
\eeq
This suggests that we have a Lyapunov exponent $\lambda_L=2\pi/\beta$. This is the value that one also gets for higher dimensional black holes. It is argued in \cite{Maldacena:2015waa} that this is the maximal Lyapunov exponent that a chaotic quantum system with a classical limit can have.

You might wonder where is the Ruelle region, the saturation to the late time value. The answer is that it is invisible in \eqref{eq:FOTO} because this is a perturbative result in $G_N$. Any higher powers in $G_Ne^{\hat u}$ are neglected. These terms become large precisely around the scrambling time $t\sim \log 1/G_N$. These are called secular terms, and they are typical in any kind of real time perturbation theory. To see the Ruelle region, one must evaluate the Schwarzian path integral nonperturbatively in $G_N$. This is hard, but the interested reader can find an illustrative (but non controled) approximation in \cite{Maldacena:2016upp}.

\section{SYK model}

The primary sources when writing this section were \cite{Maldacena:2016hyu,Polchinski:2016xgd} and some seminar talks by Douglas Stanford, but we will point out many additional references along the way. 

\subsection{The model}
\label{sec:sykdef}

The model that we are going to study in the remainder of these lectures is an ensemble of simple, finite dimensional quantum mechanical models.\footnote{The model was first proposed and studied by Kitaev \cite{Kitaev}, and it is based on a simplification of a condensed matter model introduced by Sachdev and Ye \cite{Sachdev:1992fk,Sachdev:2015efa}.} The members of the ensemble are specified by their Hamiltonians, which are just finite, Hermitian matrices. We are going to build up these Hamiltonians from gamma matrices, familiar from fermionic (half spin) representations of the Lorentz group. The difference here, will be that we need these representations for the orthogonal group (i.e. the Eulidean version), and for general dimension $N$, not just four. So let us give a lightning review how to build these representations.

We want to find representations of the Clifford algebra
\beq
\label{eq:cliff}
\lbrace \psi_i, \psi_j \rbrace = \delta_{ij}, \;\;\;\;i,j=1,...,N,
\eeq
where $\lbrace.,.\rbrace$ is the usual anticommutator. Let us restrict our attention to $N=2K$ even.\footnote{We will not need odd $N$, but in that case, the representation is built out from the $N-1$ case by adding an analogue of $\gamma_5$, the product of all gamma matrices.} We are going to look for Hermititan representations $\psi_i=\psi_i^\dagger$. In this case, we can build a representation of the algebra by introducing a new, complex basis
\bea
c_i=\frac{1}{\sqrt{2}}(\psi_{2i} - i \psi_{2i+1}), && c_i^\dagger=\frac{1}{\sqrt{2}}(\psi_{2i} + i \psi_{2i+1}), && i=1,...,K,
\eea
which satisfy
\bea
\lbrace c_i,c_j\rbrace=\lbrace c_i^\dagger,c_j^\dagger \rbrace=0, && \lbrace c_i, c_j^\dagger\rbrace=\delta_{ij}.
\eea
We recognize the familiar canonical anticommutation relations for fermionic modes, which are thus equivalent with the Clifford algebra relation \eqref{eq:cliff}. But we know how to build a representation of this: we pick a vacuum annihilated by all the modes $c_i|0\rangle=0$, and build the basis of the representation as
\beq
(c_1^\dagger)^{n_1}...(c_K^\dagger)^{n_K}|0\rangle, \;\;\;\; n_k=0,1.
\eeq
There are $2^K=2^{\frac{N}{2}}$ such states, corresponding to whether a given mode is occupied by a fermion or not. It is a general result, that this representation is the only irreducible representation of \eqref{eq:cliff} up to unitary equivalence. 

One can also give an explicit recursion relation for the representation matrices, which may come in handy for numerical realizations. This is given as
\bea
\label{eq:recursion}
\psi^{(K)}_i &=\psi^{(K-1)}_i \otimes \left( \begin{array}{cc} -1 & 0 \\ 0 & 1 \end{array}\right), && i=1,...,N-2,\\
\psi^{(K)}_{N-1}&= \frac{1}{\sqrt{2}} I_{2^{K-1}}  \otimes \left( \begin{array}{cc} 0 & 1 \\ 1 & 0 \end{array}\right), \\
\psi^{(K)}_{N}&=\frac{1}{\sqrt{2}}  I_{2^{K-1}}  \otimes \left( \begin{array}{cc} 0 & -i \\ i & 0 \end{array}\right),
\eea
where the superscript indicates the dimension $N=2K$, and $I_d$ is the $d\times d$ identity matrix. It is clear that $\psi^{(K)}_i$ are $2^K \times 2^K$ matrices if we start the recursion with $2\times 2$ matrices. These initial matrices are given by
\bea
\psi^{(1)}_1&=\frac{1}{\sqrt{2}}\left( \begin{array}{cc} 0 & -i \\ i & 0 \end{array}\right), && \psi^{(1)}_2&=\frac{1}{\sqrt{2}}\left( \begin{array}{cc} 0 & 1 \\ 1 & 0 \end{array}\right).
\eea
We are going to suppress the upper $K$ label in what follows. 

A member of the SYK ensemble has the Hamiltonian
\beq
\label{eq:sykham}
H=\sum_{ijkl=1}^N J_{ijkl} \psi_i \psi_j \psi_k \psi_l,
\eeq
which is therefore a matrix of size $2^{\frac{N}{2}}\times 2^{\frac{N}{2}}$. The numbers $J_{ijkl}$ are all-to-all couplings which are different for different members of the ensemble. Each of them are picked randomly and independently from a Gaussian ensemble with mean $\mu=0$ and variance $\sigma = \sqrt{3!} J/N^{3/2}$. Here, $J$ is a fixed number, a parameter of the ensemble, and the scaling of the variance with $N$ is essential for the model to have an interesting large $N$ behaviour, we will see this very explicitly as we proceed.

We will also study here and there a mild generalization of these models with $q$-body interactions
\beq
\label{eq:sykhamq}
H=i^{q/2} \sum_{1 \leq i_1 <...<i_q\leq N}J_{i_1...i_q} \psi_{i_1}...\psi_{i_q},
\eeq
where the variance of the zero mean Gaussian couplings is now $\sigma=\sqrt{(q-1)!}J/N^{\frac{q-1}{2}}$ and $q$ is an even integer. This model will have an additional analytic handle in the form of an $1/q$ expansion, which will be useful to illustrate some of the general features.

\subsubsection{Brief review}

Having the model properly defined, let us give a brief review of some of its most exciting properties. Reviewing how some of these are established will be a primary focus for the next sections.

\begin{figure}[h!]
\centering
\includegraphics[width=0.4\textwidth]{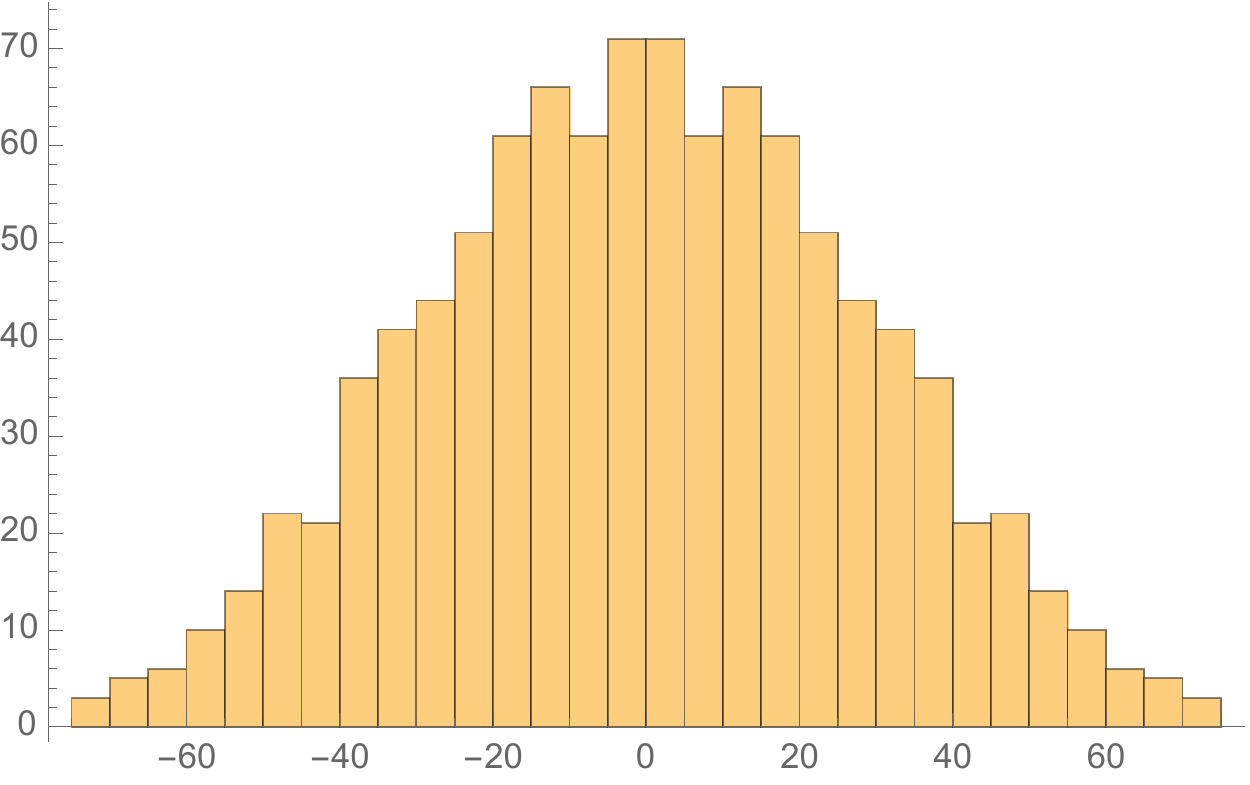} \hspace{1cm}\includegraphics[width=0.4\textwidth]{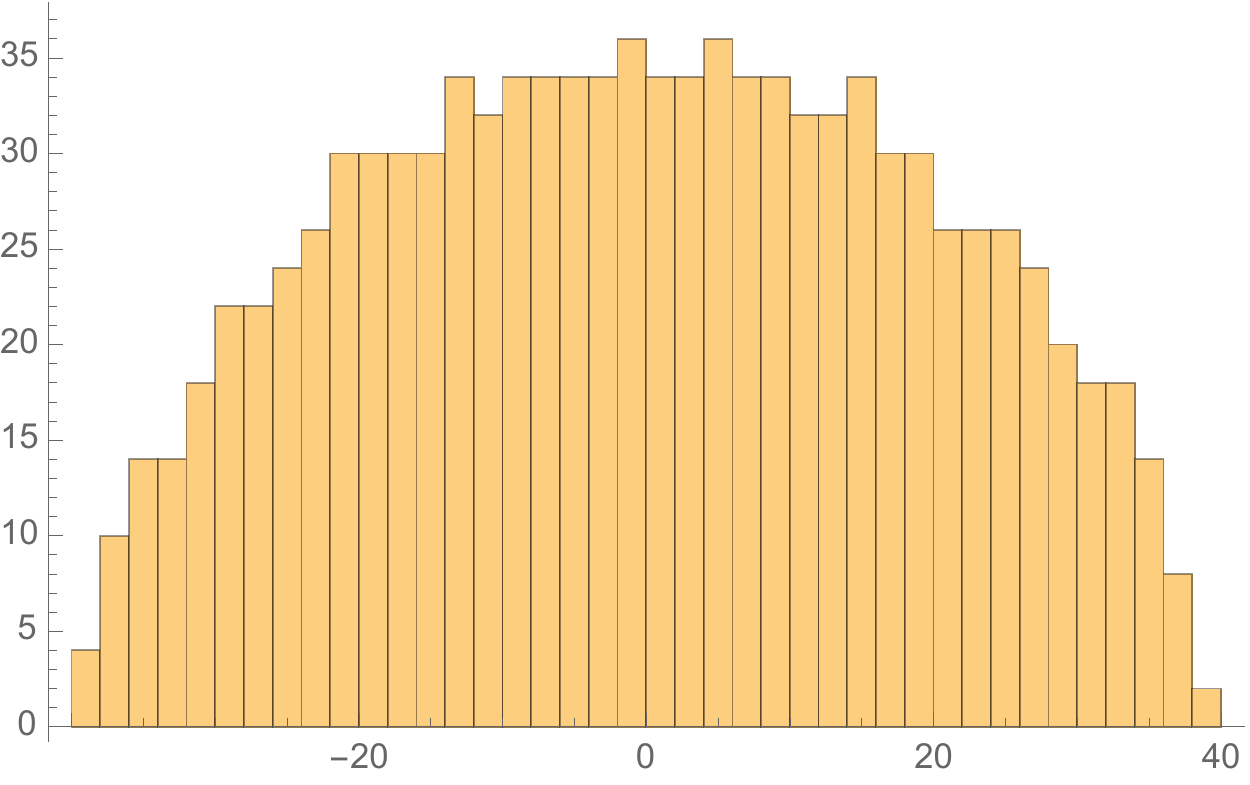}
\caption{Left: Histogram of the spectrum of the $q=2$ model. Right: Histogram of the spectrum of the $q=4$ model. Both plots are for $N=20$ and can be produced with the use of the recursion relations \eqref{eq:recursion} on a laptop while drinking a smaller cup of coffee.}
\label{fig:5}
\end{figure}

\begin{itemize}
\item The model classicalizes in the large $N$ limit and is therefore solvable by solving a set of classical equations of motion for some master fields $G$ and $\Sigma$ living in two dimensions.
\item In this large $N$ limit, there is an emergent time reparametrization symmetry $t\mapsto f(t)$ in the low energy sector of the model. This is spontaneously broken by the vacuum down to $SL(2,\mathbb{R})$. The reparametrization modes can be thought of as Goldstone modes, and they aquire a nonzero action because the symmetry is actually explicitly broken when we move away from the IR. Notice that this symmetry breaking pattern is identical to the one that we encountered for the Jackiw-Teitelboim theory in section \ref{sec:Nads}. There, we already gave a holographic argument that this pattern is generic in the IR for a large class of models. We will see that SYK is likely in this universality class, even though it is a UV completion that will turn out to be not dual to any of the holographic UV completions that we have considered in \eqref{eq:almhpolch}.
\item The spectrum of the model also has a very interesting behaviour. We plot the density of states coming from \eqref{eq:sykhamq} for a single choice of couplings in Fig \ref{fig:5} for $q=2$ and $q=4$ and with $N=20$. The $q=2$ model is quadratic, so it is a free theory (the Hamiltonian is just a mass term). The spectrum is correspondingly similar to what is usual in integrable theories, with a long tail at low energies. On the other hand, the spectrum ends rather abruptly for the $q=4$ model, not unlike the edges of the semicircle law for Gaussian random matrices.\footnote{A spectrum sharing features with Gaussian random matrices is typical for quantum chaotic systems. In the context of SYK see for example \cite{Garcia-Garcia:2016mno,Cotler:2016fpe}. } This means that there are a lot of states just above the ground state. Even more interestingly, as we increase $N$, the edge of the spectrum gets steeper and steeper. In fact, we have
\beq
\label{eq:SYKentropylimit}
\lim_{\beta \rightarrow \infty} \lim_{N\rightarrow \infty} S(\beta) \sim N,
\eeq
where $S(\beta)$ is the thermal entropy at inverse temperature $\beta$. This formula is surprising, because it shows that the two limits do not commute: for any finite $N$ the model has very few ground states,\footnote{For a generic choice of couplings $J_{ijkl}$ which is not invariant under any subgroup of $O(N)$ acting on its indices, the ground state is at most twofold degenerate. The degeneracy can come about because of the existence of an antiunitary map (the time reversal $T$ in the context of spacetime $\gamma$ matrices) that commutes with the Hamiltonian. When $T^2=-1$, this leads to Kramers doubling (orthogonality of $|0\rangle$ and $T|0\rangle$). When $T^2=1$, there is still a matrix for even $N$ (the analogue of $\gamma_5$) that commutes with the Hamiltonian and satisfies $T\gamma_5T^{-1} =\pm \gamma_5^*$, with the sign depending on $N$. Having the minus sign leads to a degeneracy because in this case $T$ must exchange the eigenspaces of $\gamma_5$ (the spinor chiralities). These signs are controlled by $N \mod 8$ (the Bott periodicity of Clifford algebras), see e.g. appendix B of \cite{Polchinski:1998rr} for the case of Lorentzian signature.} so the zero temperature entropy is small. This weird property is actually very appealing. If we want to regard the SYK model as a toy model for black holes, \eqref{eq:SYKentropylimit} gives the extremal entropy of the black hole. This way, the SYK model explicitly illustrates how a quantum black hole can appear to have a macroscopic ground state degeneracy in the classical limit, without violating the third law of thermodynamics.\footnote{Of course, the aformentioned Gaussian random matrices with the semicircle law have the same feature, so this property alone would not make the SYK model that interesting.}
\end{itemize}

\subsection{Large $N$ diagrammatics}

We now begin analysing the model \eqref{eq:sykhamq}. The first thing we do, is that we examine conventional perturbation theory in the coupling $J$. It will turn out, that we can resum this for large $N$. You might ask why do we look at a $J=0$ limit at the first place? The fermions $\psi_i$ have power counting dimension $[\psi]=0$, which makes the interaction term in \eqref{eq:sykhamq} a relevant coupling. Therefore, we expect the theory to be asymptotically free at very large energies.

We will work exclusively in Euclidean signature. The Euclidean time ordered two point function of the fermions will be denoted as
\beq
G_{ij}(\tau) = \langle T \psi_i(\tau)\psi_j(0)\rangle \equiv \Theta(\tau)\langle  \psi_i(\tau)\psi_j(0)\rangle - \Theta(-\tau)\langle  \psi_j(0)\psi_i(\tau)\rangle,
\eeq
where $\Theta$ is the Heaviside theta function, and
\beq
\psi_i(\tau)=e^{\tau H} \psi_i e^{-\tau H}.
\eeq
A central quantity will be the normalized trace of the above two point function, we will denotes this with
\beq
\label{eq:syk2pt}
G(\tau)=\frac{1}{N} \sum_{i=1}^N G_{ii}(\tau).
\eeq
For zero coupling, we have $H=0$ so $\psi_i(\tau) \equiv \psi_i$. We can then use the Clifford algebra relation \eqref{eq:cliff} to evaluate the two point function
\bea
\label{eq:freeprop}
G^{\rm free}_{ij}(\tau) = \frac{1}{2} \delta_{ij}\text{sgn}\tau, && G^{\rm free}(\tau)=\frac{1}{N} \sum_i G^{\rm free}_{ii}=\frac{1}{2} \text{sgn}\tau,
\eea
where $\text{sgn}\tau=\Theta(\tau)-\Theta(-\tau)$ is just the sign function. It is also useful to give it in Fourier space\footnote{Notice that this is just the inverse of the derivative kernel $\partial_\tau$ in the kinetic term of the action of a fermion, as it should be.}
\beq
G^{\rm free}_{ij}(\omega) = \int_{-\infty}^{\infty} d\tau e^{i\omega \tau} G^{\rm free}_{ij}(\tau) = - \frac{\delta_{ij}}{i\omega},
\eeq
 Now perturbation theory proceeds as follows.

\begin{itemize}
\item For each realization of the model, there is a four leg vertex
\begin{center}
\includegraphics[width=0.1\textwidth]{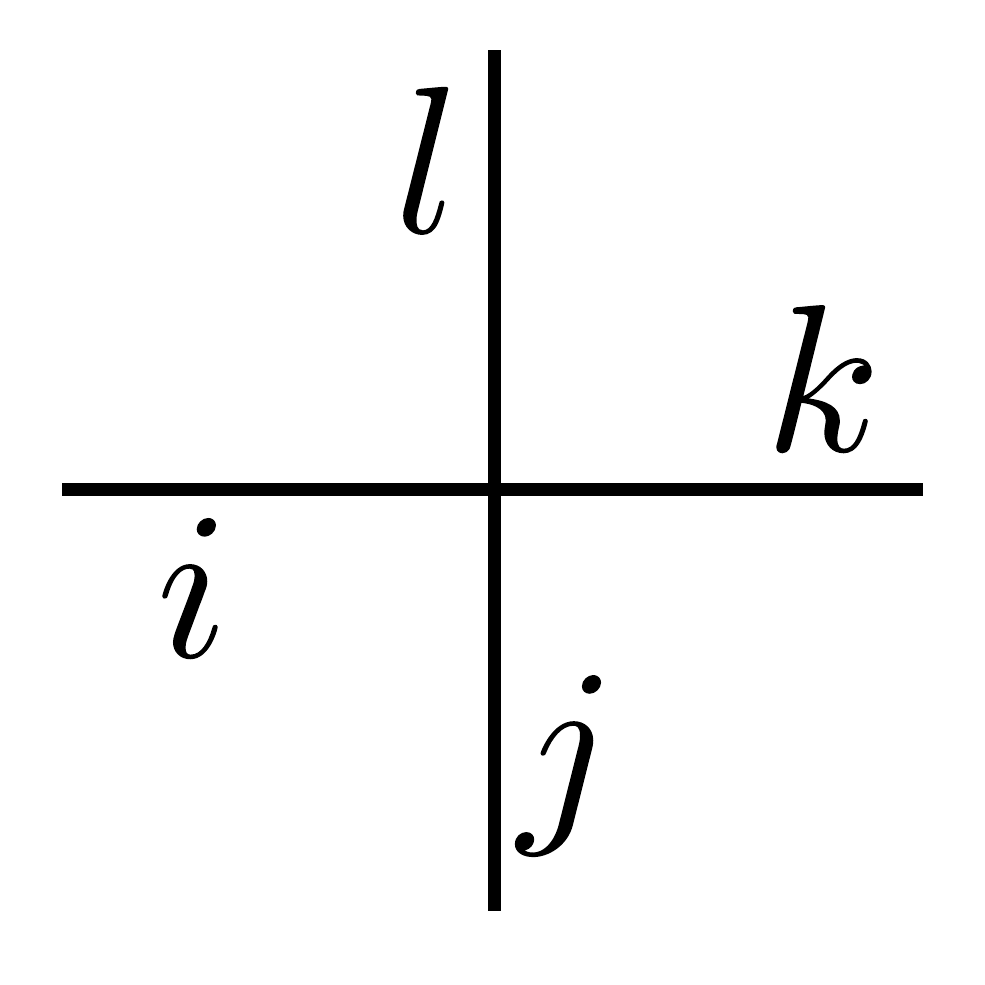}
\end{center}
proportional to $J_{ijkl}$. 
\item We calculate Feynman diagrams in each realization of the model. Then we average the diagram over disorder. Each diagram will contain a number of $J_{ijkl}$ and since the distribution is Gaussian, we evaluate these expectation values using Wick's theorem and the two point expectation value
\beq
\label{eq:disorder2pt}
\langle J_{i_1j_1k_1l_1}J_{i_2j_2k_2l_2} \rangle_J = 3! \frac{J^2}{N^3} \delta_{i_1i_2}\delta_{j_1j_2}\delta_{k_1k_2}\delta_{l_1l_2}
\eeq
\item The first contribution to the two point function is the tadpole
\begin{center}
\includegraphics[width=0.35\textwidth]{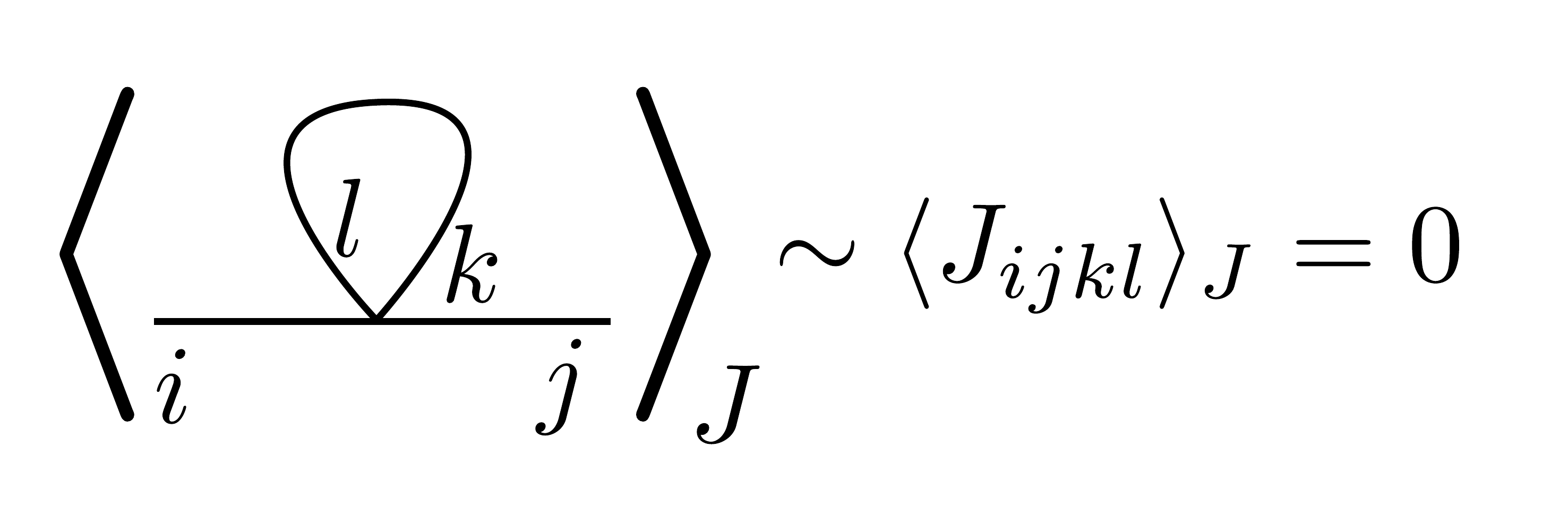}.
\end{center}
However, this is zero, since it is linear in $J_{ijkl}$.
\item The next diagram is the melon (or settling sun)
\begin{center}
\includegraphics[width=0.9\textwidth]{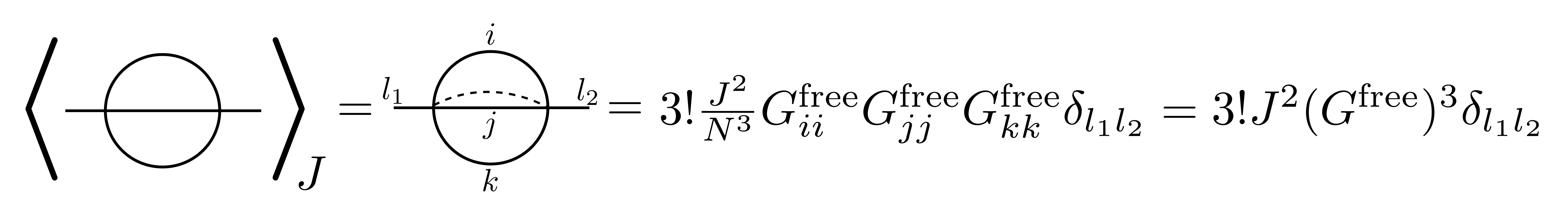},
\end{center}
where the dashed line denotes the disorder pairing with \eqref{eq:disorder2pt}, and doubled indices are summed over. Notice that the result does not scale with $N$.
\item There are multiple order $J^4$ diagrams, with multiple pairings. For illustration, let us check one diagram with two different types of pairings. When the pairing happens inside the melon we get
\begin{center}
\includegraphics[width=0.9\textwidth]{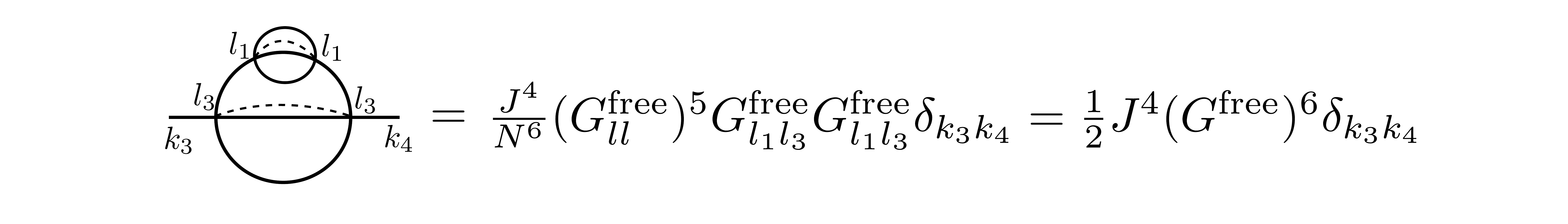},
\end{center}
where we have used that $G_{l_1l_3}^{\rm free}G_{l_1l_3}^{\rm free}=\frac{1}{2}N G^{\rm free}$, because of \eqref{eq:freeprop}. This again does not scale with $N$. The other pairing is
\begin{center}
\includegraphics[width=0.9\textwidth]{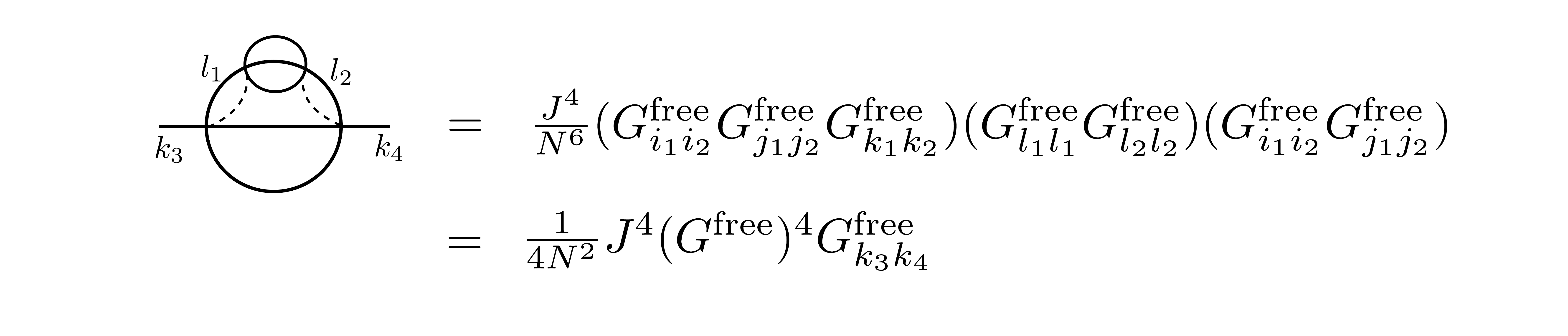}.
\end{center}
We see that this is suppressed as $N^{-2}$ compared to the previous nonzero diagrams.
\end{itemize}

This last lesson is a general one. The claim is that the only diagrams that are not suppressed by some power of $N^{-1}$ are the ones when the disorder average pairs vertices inside a single melon. Therefore, the two point function has an iterated structure
\begin{center}
\includegraphics[width=0.9\textwidth]{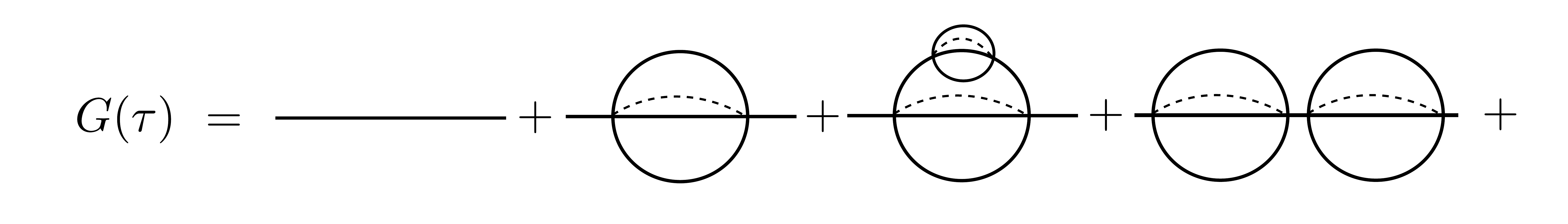}.
\end{center}
We can summarize this in the following closed set of consistency equations
\begin{center}
\includegraphics[width=0.9\textwidth]{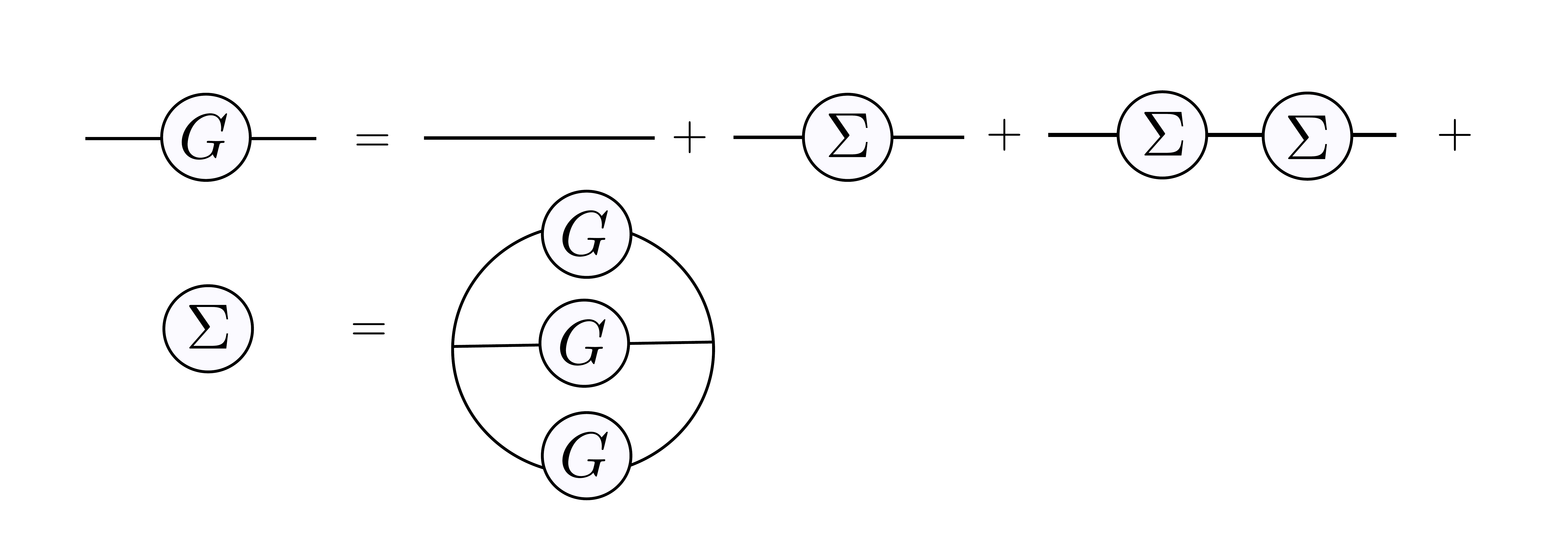},
\end{center}
where the object $\Sigma$ is called the self energy and is defined to contain all the iterated melon diagrams.

We can write these pictorial equations down easily by introducing a matrix multiplication notation for the bilinear kernels:
\beq
(AB)(\tau,\tau')=\int d\tau'' A(\tau,\tau'')B(\tau'',\tau').
\eeq
Using this, the first equation reads as
\bea
G &= G^{\rm free} + G^{\rm free} \Sigma G^{\rm free} + G^{\rm free} \Sigma G^{\rm free} \Sigma G^{\rm free} + \cdots \\
&=G^{\rm free} \big[ 1+\Sigma G^{\rm free} + \Sigma G^{\rm free} \Sigma G^{\rm free} + \cdots \big] \\
&=G^{\rm free} \big[1-\Sigma G^{\rm free} \big]^{-1}\\
&=\big[ (G^{\rm free})^{-1}-\Sigma \big]^{-1},
\eea
where we have resummed a geometric series in the third line. These type of resummed diagrammatic equations are called Schwinger-Dyson equations. Now the inverse of the free propagator is just the kernel in the bilinear kinetic term in the action
\beq
(G^{\rm free})^{-1}(\tau,\tau')=\delta(\tau-\tau')\partial_{\tau'},
\eeq
so we will usually write the shorthand
\beq
\label{eq:schwd1}
G=\big[ \partial_\tau -\Sigma \big]^{-1}.
\eeq
The second equation is very easy to write, it is just
\beq
\label{eq:schwd2}
\Sigma(\tau,\tau')=J^2 \big[ G(\tau,\tau')\big]^3.
\eeq

In summary, to leading order in $1/N$, the two point function is given by solving a set of integral equations. For general $q$, only the second equation changes, and we have
\bea
\label{eq:schwdq}
G&=\big[ \partial_\tau -\Sigma \big]^{-1},\\
\Sigma(\tau,\tau')&=J^2 \big[ G(\tau,\tau')\big]^{q-1}.
\eea

\subsection{Master fields}

We have mentioned previously, that the SYK model becomes classical in the large $N$ limit. This is not obvious from the analysis of the previous section, so here we review an alternative derivation of the equations \eqref{eq:schwdq} directly from the path integral. For simplicity, let us restrict again to $q=4$. The partition function associated to the Hamiltonian \eqref{eq:sykham} is written as a path integral\footnote{Here, $D\psi_i$ denotes a standard Berezin integral for each $i=1,...,N$ and each Euclidean time, we refer to Wikipedia for a crash course, if needed.}
\beq
\label{eq:partfunc1}
Z(J_{ijkl})=\int D\psi_i \exp \Big(-\int d\tau\big[\frac{1}{2}\sum_i \psi_i \partial_\tau \psi_i + \sum_{1\leq i<j<k<l \leq N} J_{ijkl}\psi_i\psi_j\psi_k\psi_l\big]\Big).
\eeq
Now we want to realize the average over the disorder ensemble for $J_{ijkl}$. There are two physically different ways of doing this:
\begin{itemize}
\item One can average directly the partition function: $\langle Z \rangle_J$. This is called annealed disorder, and in this case $J$ is similar to a microscopic degree of freedom.
\item Instead, one can average the free energy: $\langle \log Z \rangle_J$. This is more complicated to do, but bears more physical relevance in condensed matter theory: this is the right thing to do in many instances of disorder, e.g. when one describes the effect of lattice errors in a crystal. Technically, one deals with this by the replica trick: we use the formula $\log Z = \lim_{n\rightarrow 0} \partial_n Z^n$ and obtain $Z^n$ by introducing $n$ copies of each of our fields in the path integral and do an analytic continuation to nonintegral $n$. 
\end{itemize}
For simplicity, we will go with the annealed disorder here, but it is important to point out that the two approaches give the same result to leading order in $1/N$ \cite{Sachdev:2015efa}.

Averaging the partition function means that we need to do the Gaussian expectation values with variances that we have specified in sec. \ref{sec:sykdef} where we have defined the model
\beq
\label{eq:disorderaverage}
\langle Z \rangle_J \sim \int dJ_{ijkl} \exp \left( -\sum_{1\leq i<j<k<l \leq N} \frac{J^2_{ijkl}}{2\frac{3!J^2}{N^3}}\right)Z(J_{ijkl}).
\eeq
Since in \eqref{eq:partfunc1} the coupling appears linearly in the exponent, we can perform these integrals easily,\footnote{The master formula is $\int dx e^{-a x^2+bx}=\sqrt{\pi/a}e^{\frac{b^2}{4a}}$, as usual.} leading to
\bea
\langle Z \rangle_J &\sim \int D\psi_i \exp \Big(-\int d\tau \frac{1}{2}\sum_i \psi_i \partial_\tau \psi_i \\
&+ \sum_{1\leq i<j<k<l \leq N}  \frac{3 J^2}{N^3} \int \int d\tau d\tau' (\psi_i\psi_j\psi_k\psi_l)(\tau)(\psi_i\psi_j\psi_k\psi_l)(\tau') \Big).
\eea
Now we can write the sums in an alternative way: $\sum_{1\leq i<j<k<l \leq N}=\frac{1}{4!}\sum_{i\neq j \neq k \neq l}$. We can use this to decouple the sums as
\beq
 \sum_{1\leq i<j<k<l \leq N}(\psi_i\psi_j\psi_k\psi_l)(\tau)(\psi_i\psi_j\psi_k\psi_l)(\tau') = \frac{1}{4!} \Big[ \sum_i \psi_i(\tau)\psi_i(\tau')\Big]^4 .
\eeq
Here, we have used that in the path integral the Grassmannian variables satisfy $\psi_i(\tau)^2=0$. The next trick is to insert 1 into the path integral in a clever way:
\bea
1&=\int DG \delta\big(NG(\tau,\tau')-\sum_i \psi_i(\tau)\psi_i(\tau')\big) \\ &\sim \int DG D\Sigma \exp \Big(-\frac{N}{2}\int \int d\tau d\tau' \Sigma(G-\frac{1}{N}\sum_i \psi_i\psi_i)\Big).
\eea  
The field $G$ is just a new name for the fermion bilinear, while $\Sigma$ plays the role of a Lagrange multiplier enforcing the delta constraint. This way, we can write
\bea
\langle Z \rangle_J &\sim  \int D\psi_i DGD\Sigma \exp \Big( -\int \frac{1}{2}\sum_i \psi_i \partial_\tau \psi_i -\frac{1}{2} \int \int N \Sigma (G-\frac{1}{N} \sum_i \psi_i \psi_i) \\
&+ \frac{J^2 N}{2\cdot 4} \int \int G^4\Big).
\eea
The goal of this gymnastic was to obtain an exponential that is bilinear in the fermion variables $\psi_i$. Therefore, we can do the Gaussian Berezin integral\footnote{We remind you that the Gaussian Berezin integral gives a positive power of the determinant: $\int d\psi e^{-\frac{1}{2}\psi A \psi}=\sqrt{\det A}$, opposed to normal integration. Here, $A$ is antisymmetric.} with the result
\bea
\label{eq:masterfieldpathint}
\langle Z \rangle_J &\sim   \int DGD\Sigma \big[ \det(\partial_\tau-\Sigma) \big]^{\frac{N}{2}}\exp \Big( -\frac{N}{2} \int \int (\Sigma G-\frac{1}{4}J^2 G^4) \Big)\\
&=\int DGD\Sigma e^{-N I[G,\Sigma]},
\eea
with
\beq
\label{eq:masteraction}
I[G,\Sigma]=-\frac{1}{2} \log \det (\partial_\tau-\Sigma)  + \frac{1}{2} \int \int (\Sigma G-\frac{1}{4}J^2 G^4).
\eeq
The purpose of all this suffering was the factor of $N$ in front of the action in the second line of \eqref{eq:masterfieldpathint}. Notice that the precise scaling of the variance in the Gaussian average of \eqref{eq:disorderaverage} was crucial to get this. It is now clear that $N$ plays the role of $\hbar^{-1}$ and the large $N$ limit is a classical limit. For the generalized model \eqref{eq:sykhamq} with $q$-body interactions, this action generalizes to
\beq
\label{eq:masteractionq}
I[G,\Sigma]=-\frac{1}{2} \log \det (\partial_\tau-\Sigma)  + \frac{1}{2} \int \int (\Sigma G-\frac{1}{q}J^2 G^q).
\eeq
The classical equations of motion can be derived by looking for extrema of this action. It is easy to show that these are identical to the Schwinger-Dyson equations \eqref{eq:schwdq}.\footnote{The identity $\log \det A=\text{Tr}\log A$ is useful here.}

\subsection{Conformal limit}
\label{sec:conflim}

Now that we have derived equations \eqref{eq:schwdq} in two different ways, it is probably time to think about what can we say about its solutions. In fact, it is cheap to solve \eqref{eq:schwdq} numerically,\footnote{See \cite{Maldacena:2016hyu} for some more details.} so in this sense the SYK model is solvable at large $N$. 

Let us discuss the IR properties of this solution. The coupling $J$ has dimension energy, so low energy means small frequencies compared to $J$. We can write \eqref{eq:schwd1} in Fourier space
\beq
\frac{1}{G(\omega)} = -i\omega-\Sigma(\omega),
\eeq
and notice from \eqref{eq:schwd2} that $\Sigma$ is up by $J^2$ compared to $G$.\footnote{$G(\tau)$ is dimensionless in naive power counting so its Fourier transform has dimension of time. Similarly, $\Sigma(t)$ has dimension (energy)$^2$ and $\Sigma(\omega)$ has dimension of energy.} This means that for frequencies $\omega \ll J$, we should be able to drop the $-i\omega$ term. Doing so leads to a new set of IR equations
\bea
\label{eq:schwdconf}
\int d\tau'' G(\tau,\tau'')\Sigma(\tau'',\tau') &=-\delta(\tau-\tau') \\
\Sigma(\tau,\tau') &=J^2 G(\tau,\tau')^{q-1}.
\eea 
A crucial observation is that these equations have an extra symmetry. They are invariant under reparametrizations $\tau \mapsto \phi(\tau)$ provided we transform the fields as
\bea
\label{eq:sykreparam}
G(\tau,\tau') &\mapsto \big[\phi'(\tau)\phi'(\tau') \big]^\Delta G(\phi(\tau),\phi(\tau')), \\
\Sigma(\tau,\tau') &\mapsto \big[\phi'(\tau)\phi'(\tau') \big]^{\Delta(q-1)} \Sigma(\phi(\tau),\phi(\tau')).
\eea
Both of these guys therefore transform as conformal two point functions, and the conformal dimension is given by $\Delta=1/q$. It is trivial to see this for the second equation in \eqref{eq:schwdconf}, and very easy to show for the first equation, by changing integration variable $\tilde \phi=\phi(\tau'')$:
\bea
\int d\tau'' \big[\phi'(\tau)\phi'(\tau'') \big]^{\frac{1}{q}} &G(\phi(\tau),\phi(\tau'')) \big[\phi'(\tau'')\phi'(\tau') \big]^{1-\frac{1}{q}} \Sigma(\phi(\tau''),\phi(\tau')) \\ &=\int d\tilde \phi G(\phi(\tau),\tilde \phi) \Sigma(\tilde \phi,\phi(\tau')) \phi'(\tau') \left[ \frac{\phi'(\tau)}{\phi'(\tau')}\right]^{\frac{1}{q}} \\
&=-\phi'(\tau') \delta(\phi(\tau)-\phi(\tau')) \\
&=-\delta(\tau-\tau').
\eea
This reparametrization symmetry is emergent in the IR, and valid when we ask questions on scales $|\tau-\tau'| \gg J^{-1}$. However, it is explicitly broken by the presence of the derivative term $\partial_\tau$ in \eqref{eq:schwdq} compared to \eqref{eq:schwdconf}.

We still need to find at least one solution to \eqref{eq:schwdconf}. Since both $G$ and $\Sigma$ transform as conformal two point functions, it makes sense to look for a solution which has the form of a conformal two point function on the line:\footnote{Note that in the vacuum or the thermal state we must have translation invariance: $G(\tau,\tau')=G(\tau-\tau')$. However, general reparametizations \eqref{eq:sykreparam} might break this.}
\beq
\label{eq:confsol1}
G_c(\tau)=\frac{b}{|\tau |^{2\Delta}} \text{sgn}\tau.
\eeq
We can write our ansatz in a compact way as
\bea
G_c(\tau)=b d_\Delta(\tau), && \Sigma_c(\tau)=J^2 b^{q-1}d_{\Delta(q-1)}(\tau), && d_\Delta(\tau)= \frac{\text{sgn}\tau}{|\tau |^{2\Delta}}.
\eea
This automatically solves the second equation in \eqref{eq:schwdconf}. We want to insert this ansatz into the Fourier transform of the first equation, for which we need the Fourier transform of $d_\Delta$. This is given as
\bea
d_\Delta(\omega) &=\int_{-\infty}^\infty d\tau e^{i\omega \tau} \frac{\text{sgn}\tau}{|\tau |^{2\Delta}}\\
&=2i\text{Im} \Big[ \int_0^\infty d\tau e^{i\omega \tau} \tau^{-2\Delta}\Big].
\eea
For $\omega>0$ we can rotate the contour to the positive part of the imaginary axis and get an integral defining the gamma function
\bea
d_\Delta(\omega)&=2i\text{Im} \Big[ \left(\frac{i}{\omega} \right)^{1-2\Delta}\Gamma(1-2\Delta)\Big] , \\
&=2i \cos(\pi \Delta) \Gamma(1-2\Delta) \frac{1}{\omega^{1-2\Delta}}.
\eea
It is clear from the definition that $d_\Delta(-\omega)=-d_\Delta(\omega)$. We can use this to extend the formula to $\omega<0$. Inserting this ansatz into the equation $G(\omega)=-1/\Sigma(\omega)$ yields
\bea
\label{eq:b}
\Delta &=\frac{1}{q}\\
b^q&= \frac{1}{\pi J^2} \left( \frac{1}{2}-\frac{1}{q} \right) \tan \frac{\pi}{q}.
\eea
Several comments are in order:
\begin{itemize}
\item The solution \eqref{eq:confsol1} also tells us what form the exact solution of \eqref{eq:schwdq} on the line must take when $\tau \gg J^{-1}$. We also know that $G$ must agree with the free two point function $\frac{1}{2}\text{sgn}\tau$ in the UV, when $\tau \ll J^{-1}$. The only regime where we do not know it analytically is $\tau \sim J^{-1}$.
\item Reparametrizations of \eqref{eq:confsol1} of the form \eqref{eq:sykreparam} are also solutions. Notice that when $\phi$ is a M\"obius transform like \eqref{eq:sl2}, the solution \eqref{eq:confsol1} does not change. Therefore, \eqref{eq:confsol1} spontaneously breaks the reparametrization invariance down to $SL(2,\mathbb{R})$.
\item The two point function on the circle $\tau \sim \tau + \beta$ (or finite temperature two point function) can be obtained from \eqref{eq:confsol1} by applying the reparametrization $\phi(\tau)=\tan \frac{\pi \tau}{\beta}$. The result is
\beq
\label{eq:confsolfiniteT}
G_c(\tau)=b \left[ \frac{\pi}{\beta \sin \frac{\pi \tau}{\beta}} \right]^{2\Delta} \text{sgn}\tau.
\eeq
\end{itemize}

\subsection{Large $q$ and low temperature entropy}

You might recall that we made the claim earlier in our quick summary of properties of the SYK model that 
\beq
\label{eq:sykzerotempent}
\lim_{\beta \rightarrow \infty} \lim_{N\rightarrow \infty} S(\beta) \sim N,
\eeq
and that the two limits do not commute. This is an unusual property, but one that we might demand from a theory describing the microscopics of near extremal black holes. One purpose of this section is to explain this statement a little bit more. We will need to evaluate the thermal free energy of the large $N$ SYK model for this, which is easier to do for the generalized model \eqref{eq:sykhamq} in a large $q$ expansion. This expansion makes life easier in tackling many other questions that we do not discuss here, so it is useful to review it anyway.

The logic behind the statement that the limits in \eqref{eq:sykzerotempent} do not commute is that for finite $N$ and $J_{i_1...i_q}$ we do not expect a large degeneracy in a single realization of the model. This is because a generic $J_{i_1...i_q}$ does not allow\footnote{For some values of $N$, there is a twofold degeneracy coming from properties of the Clifford algebra. This is of course not a large degeneracy.} for global symmetries in \eqref{eq:sykhamq}. Given this, showing that \eqref{eq:sykzerotempent} holds also shows that the limits do not commute. If we are only interested in showing this for large $q$, the strategy is to show that the large $q$ limit is a free limit and the dependence on the coupling $J$ enters only at level $1/q$. Since the $J=0$ model has zero Hamiltonian, the total Hilbert space (which has dimension $2^{\frac{N}{2}}$) is degenerate and the ground state entropy is therefore $\frac{N}{2}\log 2$, in line with the claim \eqref{eq:sykzerotempent}. We will see that this receives corrections in the $1/q$ expansion. 

There are some objections that one can raise against the reasoning above. First, by doing an $1/q$ expansion we are formally having a $q\rightarrow \infty$ limit too in \eqref{eq:sykzerotempent}, in which case the ground state entropy is large anyway, so we might be afraid that only the $q$ and $\beta$ limits do not commute. To settle this, we will present a formula for the large $N$ ground state entropy, which is exact in $q$, towards the end of this section. 
The second objection is that the large $N$ action \eqref{eq:masteraction} is only valid when one averages over the disorder, in which case it is problematic to talk about degeneracy of energy levels. One can however show that there are some special fixed choices of the couplings $J_{i_1...i_q}$ which allow for analytic (diagrammatic) derivation of the melonic Schwinger-Dyson equations \eqref{eq:schwdq}, while still not allowing for large global symmetries. These models are called melonic tensor models in the literature and we will comment a little more on them towards the end of the lectures.

So let us review the solution of \eqref{eq:schwdq} in the $1/q$ expansion. We start by parametrizing the solution as
\beq
\label{eq:largeqsol1}
G(\tau)=\frac{1}{2}\text{sgn}(\tau)e^{\frac{g(\tau)}{q-1}} = \frac{1}{2}\text{sgn}(\tau) \left( 1 + \frac{g(\tau)}{q} + \cdots \right).
\eeq
The second equation in \eqref{eq:schwdq} then fixes
\beq
\label{eq:largeqsigma}
\Sigma(\tau) = J^2 \frac{1}{2^{q-1}} \text{sgn}(\tau)e^{g(\tau)},
\eeq
so we only need to solve the first equation in \eqref{eq:schwdq} for the function $g(\tau)$. This is done by going over to Fourier space
\bea
\frac{1}{G(\omega)} & = \frac{1}{(-i\omega)^{-1}+ \frac{1}{2q}\big[  \text{sgn} \times g \big] (\omega) } \\
&=-i\omega \left( 1 + \frac{i \omega}{2q}\big[ \text{sgn} \times g \big] (\omega) + \cdots \right),
\eea
and comparing this with the equation to solve, $G(\omega)^{-1}=-i\omega-\Sigma(\omega)$ leading to
\beq
\Sigma(\omega)=-\frac{\omega^2}{2q}\big[ \text{sgn} \times g \big] (\omega).
\eeq
Writing this in real time and comparing to \eqref{eq:largeqsigma} gives the differential equation for $g$
\beq
J^2 \frac{1}{2^{q-1}} \text{sgn}(\tau)e^{g(\tau)}=\partial_\tau^2 \left( \frac{1}{2q} \text{sgn}(\tau) g(\tau) \right).
\eeq
The general solution is
\beq
e^{g(\tau)} = \frac{c^2}{\mathcal{J}^2} \frac{1}{\big[ \sin\big(c(|\tau|+\tau_0)\big)\big]^2}, \;\;\;\; \mathcal{J}= \frac{\sqrt{q}J}{2^{\frac{q-1}{2}}},
\eeq
where $c$ and $\tau_0$ are integration constants. For finite temperature solutions, we need to enforce the boundary conditions\footnote{This is because $G(0^+)=\frac{1}{2}$ is enforced by the Clifford algebra.} $g(0)=g(\beta)=0$. This leads to
\beq
\label{eq:largeqsol2}
e^{g(\tau)} = \left( \frac{\cos \frac{\pi v}{2}}{\cos \big[\pi v (\frac{1}{2}-\frac{|\tau|}{\beta} )\big]}\right)^2, \;\;\;\; \beta \mathcal{J}=\frac{\pi v}{\cos \frac{\pi v}{2}}.
\eeq
The parameter $v$ controls the dimensionless coupling $\beta \mathcal{J}$, which runs between 0 and $\infty$ as $v$ runs from 0 to 1.

\begin{figure}[h!]
\centering
\includegraphics[width=0.6\textwidth]{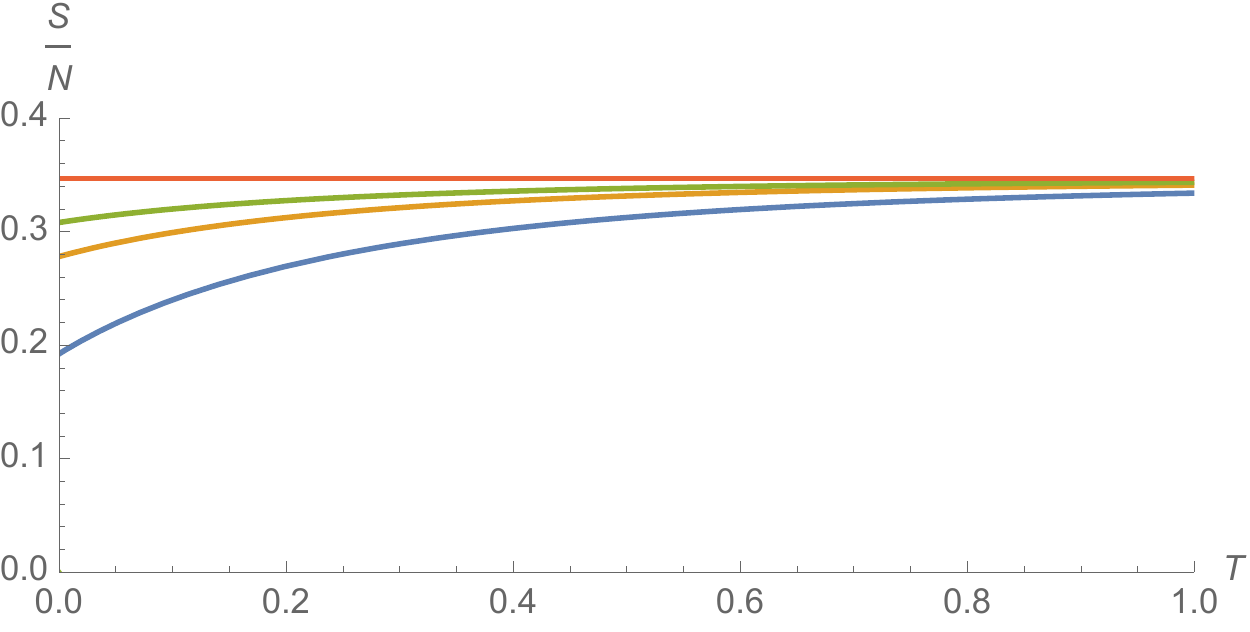}
\caption{Temperature dependence of the thermal entropy of the SYK model, obtained in the large $q$ expansion, for $q=4,6,8$ and $\infty$. Larger $q$ has larger ground state entropy.}
\label{fig:6}
\end{figure}

Now let us use this solution to evaluate the free energy. At large $N$, the partition function is dominated by the saddle point
\beq
e^{-\beta F} = Z\sim e^{-N I[G_*,\Sigma_*]},
\eeq
where $F$ is the free energy and the action $I$ is given by \eqref{eq:masteraction}. The star is to indicate the saddle point configuration. The large $q$ saddle is given explicitly by \eqref{eq:largeqsol1}, \eqref{eq:largeqsol2}. However, it is difficult to evaluate \eqref{eq:masteractionq} directly because of the determinant term. We can get around this difficulty by using a trick. We take the derivative
\beq
\label{eq:freeenergyderivative}
J\partial_J \left( -\frac{\beta F}{N}\right) = \frac{J^2\beta}{q} \int_0^\beta d\tau G_*(\tau)^q.
\eeq
The power of the trick here is that $\partial_J G_*$ and $\partial_J \Sigma_*$ do not enter this expression because $G_*$ and $\Sigma_*$ satisfy the equations of motion \eqref{eq:schwdq} coming from $I$. It also simply follows from the equations of motion that
\beq
-J^2\int_0^\beta d\tau G_*(\tau)^q = \lim_{\tau \rightarrow 0^+} \partial_\tau G_*(\tau).
\eeq
Note that this last expression is basically the expectation value of the energy. One way to see this is to note that $Z$ depends only on the dimensionless combination $\beta J$, so acting on $\log Z$ we have $J\partial_J=\beta\partial_\beta$. Another, more direct way to see this is to notice that the definition \eqref{eq:syk2pt} of the two point function implies that
\beq
 \lim_{\tau \rightarrow 0^+} \partial_\tau G_*(\tau) =\frac{1}{N} \sum_i \langle [H,\psi_i(0)]\psi_i(0) \rangle.
\eeq
One can then use the Clifford algebra relation \eqref{eq:cliff} to obtain the commutator $[H,\psi_i] \sim N \sum_{j_1...j_{q-1}}J_{i j_1...j_{q-1}}\psi_{j_1}...\psi_{j_{q-1}}$, which directly shows that $ \lim_{\tau \rightarrow 0^+} \partial_\tau G_*(\tau) \sim \langle H \rangle$. Using the large $q$ solutions \eqref{eq:largeqsol1}, \eqref{eq:largeqsol2}, we can directly evaluate
\beq
 \lim_{\tau \rightarrow 0^+} \partial_\tau G_*(\tau) = \frac{1}{2(q-1)} g'(0) = -\frac{\pi v}{\beta q} \tan \frac{\pi v}{2} + O(1/q^2).
\eeq
By writing
\beq
\label{eq:derivatives}
J\partial_J =\beta \mathcal{J} \partial_{\beta \mathcal{J}} = \frac{v}{1+\frac{\pi v}{2}\tan \frac{\pi v}{2}} \partial_v,
\eeq
we can cast \eqref{eq:freeenergyderivative} into the form of a differential equation for the free energy
\beq
\frac{v}{1+\frac{\pi v}{2}\tan \frac{\pi v}{2}} \partial_v \left( -\frac{\beta F}{N}\right)  = \frac{\pi v}{q^2} \tan \frac{\pi v}{2},
\eeq
which integrates to
\beq
\label{eq:largeqfreeenerg}
-\frac{\beta F}{N}= \frac{1}{2} \log 2 + \frac{\pi v}{q^2} \left( \tan \frac{\pi v}{2}-\frac{\pi v}{4}\right),
\eeq
with constant of integration being fixed by the $v\rightarrow 0$ value. This is the zero coupling limit, where we know that $Z=\text{Tr}(1)=2^{\frac{N}{2}}$. We can obtain the thermal entropy with the aid of \eqref{eq:derivatives}
\bea
\frac{S_{\rm therm}}{N} &= (1-\beta\partial_\beta)\left( -\frac{\beta F}{N} \right) \\
&= \frac{1}{2} \log 2 - \left(\frac{\pi v}{2 q} \right)^2,
\eea
which we plot on Fig. \ref{fig:6} as a function of the dimensionless temperature $T=(\beta \mathcal{J})^{-1}$. The claim is that the exact in $q$ large $N$ entropy has a qualitatively similar behaviour, in particular, the zero temperature limit is not zero.

It is instructive to compare \eqref{eq:largeqfreeenerg} in a strong coupling (or small temperature) expansion to the result for the Schwarzian theory \eqref{eq:schwfreeenerg}. We expand in $(\beta \mathcal{J})^{-1}$ (around $v=1$)
\beq
\label{eq:largeqstrongcouple}
-\frac{\beta F}{N}= \frac{1}{2} \log 2 + \frac{1}{q^2} \left[ \beta \mathcal{J}-\frac{\pi^2}{4}+\frac{\pi^2}{2\beta \mathcal{J}} + \cdots\right].
\eeq
Terms proportional to $\beta \mathcal{J}$ correspond to the ground state energy, which we neglected in the gravitational setup of sec. \ref{sec:Nads} because of divergences. The piece $-\frac{\pi^2}{4q^2}$ is the correction to the ground state entropy, while the last term $\frac{\pi^2}{2q^2\beta \mathcal{J}}$ is responsible for the linear low temperature behaviour of the entropy, which is similar to \eqref{eq:schwfreeenerg}.

To close up this section, let us review a formula for the ground state entropy which is exact in $q$. The trick is the same as previously: we take a parametric derivative of the action \eqref{eq:masteractionq} and use that we do not need to deal with the derivatives of the saddle point solutions. The difference is that now we take this parameter to be $q$. This way we obtain
\beq
\partial_q \big( -\frac{\beta F}{N} \big) = -\frac{J^2\beta }{2 q} \int_0^\beta d\tau \Big[\log G_*(\tau)-\frac{1}{q} \Big]G_*(\tau)^q.
\eeq
Here, we have used that the saddle must be translational invariant to do one of the integrals. We are interested in this expression in the zero temperature $\beta \rightarrow \infty$ limit. In this case, the use of the conformal solution \eqref{eq:confsolfiniteT} is justified, since this should be valid for $ \tau/\beta \gg (\beta J)^{-1}$. In this sense, the zero temperature entropy is an infinite coupling quantity. Introducing $\theta = \pi \tau/\beta$ we obtain
\beq
\partial_q \big( -\frac{\beta F}{N} \big) = -\frac{J^2 \pi b^q }{2 q} \int_0^\pi d\theta \Big[\log \big( b[\pi \beta^{-1}\sin^{-1}\theta ]^{2/q}\big)-\frac{1}{q} \Big]\frac{1}{\sin^2\theta}.
\eeq
This integral is naturally divergent, because we have used the conformal solution. Let us cut it off some $\delta$ away from the UV sensitive points $0$ and $\pi$. We expect that the conformal solution is applicable if $\delta \sim (\beta J)^{-1}$. This results in
\bea
\partial_q \big( -\frac{\beta F}{N} \big) &= -\frac{J^2 \pi b^q }{2 q} \int_\delta^{\pi-\delta} d\theta \Big[\log \big( b[\pi \beta^{-1}\sin^{-1}\theta ]^{2/q}\big)-\frac{1}{q} \Big]\frac{1}{\sin^2\theta}\\
&=-\frac{J^2 \pi b^q }{2 q} \Big[\frac{2(-3+q\log b -2 \log \delta + 2\log \frac{\pi}{\beta})}{q\delta} + \frac{2\pi}{q} + O(\delta) \Big].
\eea
The first, divergent piece corresponds to the ground state energy part $E_0$ in the free energy $\beta F=\beta E_0-S_0$. This is because we had to pick $\delta \sim (\beta J)^{-1}$ so this piece is indeed proportional to $\beta$. Notice that the combination $q\log b -2 \log \delta + 2\log \frac{\pi}{\beta}$ is $\beta J$ independent so there are really no $\log \beta$ type terms. This is because $b^{q/2} \sim J^{-1}$ as can be seen from \eqref{eq:b}. We see that the ground state energy $E_0$ is sensitive to the UV part of the solution (which is manifested here by the dependence on $\delta$), but the ground state entropy $S_0$ is not and is given by the $\delta$ independent piece
\beq
\partial_q \big( \frac{S_0}{N} \big) = -\frac{J^2 \pi^2 b^q}{q^2},
\eeq
where $S_0=\lim_{\beta \rightarrow \infty} (\beta \partial_\beta-1)\beta F$ is the ground state entropy.
Using \eqref{eq:b} to express $b$ with $q$ we can integrate this to
\beq
\frac{S_0}{N} = \frac{1}{2}\log 2 - \int_0^{1/q} dx \pi \left(\frac{1}{2}-x\right)\tan \pi x,
\eeq
where the constant of integration is fixed from the entropy of the $q=\infty$ case which we have seen in \eqref{eq:largeqfreeenerg} to be $N\log 2/2$.

We can run two basic consistency checks on this formula. First, notice that for $q=2$ it gives zero. This is expected, as the quadratic model should not have the large ground state entropy, instead the long tails in the density of states that we have seen on the left of Fig. \ref{fig:5}. Second, it is easy to expand this formula in $1/q$ and find agreement with the $\beta$ independent part of the result \eqref{eq:largeqstrongcouple}.

\subsection{Schwarzian theory}

Let us return to the question of IR dynamics in the large $N$ model \eqref{eq:masteraction}. We have seen that there is a conformal limit $\tau \gg J^{-1}$ in which the solution is given by \eqref{eq:confsol1} (or \eqref{eq:confsolfiniteT} for finite temperature). We have also seen that in this case there is an emergent reparametrization symmetry, meaning that the transformation \eqref{eq:sykreparam} maps solutions to solutions. Therefore, the action \eqref{eq:masteraction} with the derivative term neglected, has a saddle manifold, parametrized by a reparametrization $\phi(\tau)$, instead of a saddle point. Of course, including the derivative term lifts this degeneracy and ensures that there is a single saddle. 

For dimensionless time $\theta=\frac{2\pi}{\beta}\tau$ the conformal limit requires $\theta \gg (\beta J)^{-1}$ which is basically exact in the infinite coupling limit $\beta J\rightarrow \infty$. Therefore we expect the action of these reparametrizations to be suppressed with the inverse coupling at least as $(\beta J)^{-1}$. This gives rise to an interesting situation in the path integral: while the action \eqref{eq:masteraction} comes with a large pre-factor of $N$, there is some small subset of path integral directions for which this factor is lowered by $(\beta J)^{-1}$, which is something that we can make parametrically small. 

There is a similar situation arising in holographic theories. For large $N$, there is a classical bulk theory since the action is proportional to $G_N^{-1}$. But some fluctuation of the classical saddles enjoy an enhancement compared to others: these are the massless stringy modes in the bulk and the enhancement is controlled by the string scale $\ell_{\rm string}/\ell_{\rm AdS}$. This parameter is related to a marginal coupling in the CFT. In SYK, we have a similar parametric separation of the reparametrization modes $\phi(\tau)$ from the rest, however as we will explain later, the spectrum of SYK is very different from that of a weakly coupled string theory in AdS$_2$. Still, it is natural to ask how can we separate the dynamics of the reparametrization modes $\phi(\tau)$ in a strong coupling expansion. Since the symmetry breaking pattern is identical to the one discussed in sec. \ref{sec:Nads}, it is natural to suspect that this dynamics is governed by the Schwarzian action \eqref{eq:schaction}, which is the unique lowest order in derivatives action that is $SL(2,\mathbb{R})$ invariant.

Since the terms in the large $N$ action \eqref{eq:masteraction} which are outside of the determinant are manifestly reparametrization invariant, we can consider an effective action for the reparametrizations given by
\beq
\label{eq:reparamaction}
I_{\rm eff}[\phi] = -\frac{1}{2}\log \det(\partial_\tau-\Sigma^\phi_*)+\frac{1}{2}\log \det(\partial_\tau-\Sigma_*).
\eeq
Here $\Sigma_*$ is the true saddle and $\Sigma^\phi_*$ is its reparametrization
\beq
\Sigma^\phi_*(\tau,\tau')=\big[ \phi'(\tau) \phi'(\tau') \big]^{1-\frac{1}{q}} \Sigma(\phi(\tau),\phi(\tau')).
\eeq
We have choose the $\phi$ independent terms in this action so that it is zero for $\phi(\tau)=\tau$. Writing this in terms of dimensionless time $\theta=\frac{2\pi}{\beta}\tau$ and using that $\Sigma \sim J^2$ it is easy to see that a formal expansion in the derivative $\partial_\tau$ is indeed identical to a $(\beta J)^{-1}$ expansion. So naively, all we need to do is to pick up the leading term in the expansion in $\partial_\tau$. 

However, this naive thinking actually does not work for a very deep reason. While \eqref{eq:reparamaction} must be finite because it comes from an exact rewriting of an average of perfectly fine quantum mechanical models, once we expand in $(\beta J)^{-1}$ each term becomes UV divergent. This is easy to see already at the leading order, where we naively have
\bea
\label{eq:schdiv}
I_{\rm eff}[\phi] &= \frac{1}{2} \left( \text{Tr}\big[ \partial_\tau (\Sigma_*^\phi)^{-1}\big]-\text{Tr}\big[ \partial_\tau (\Sigma_*)^{-1}\big]\right) + \cdots\\
&=\frac{1}{2} \left( \text{Tr}\big[ \partial_\tau G_*^\phi\big]-\text{Tr}\big[ \partial_\tau G_*\big]\right) + \cdots,
\eea
where in the second line we have used that $(\Sigma^\phi_*)^{-1}=G_*^\phi + O(\partial_\tau)$ which is a consequence of \eqref{eq:schwdq}.
The trace requires us to evaluate $\big[\partial_\tau  G_*^\phi(\tau,\tau')\big]|_{\tau'=\tau}$. However, we know that the true saddle in the UV must approach the zero coupling result
\beq
G_*(\tau,0) \approx \frac{1}{2} \text{sgn} \tau,
\eeq
so its derivative is proportional to a Dirac delta. The above expansion would instruct us to evaluate $\delta(0)$ which is clearly nonsensical and is a sign of a UV divergence.\footnote{A systematic way of regulating these divergences is presented in \cite{Gurau:2017qna}.} This situation is entirely analogous to what happens on the big stage in string theory. The theory is UV finite, but once an $\alpha'$ expansion is considered, each order is governed by a finite number of quantum fields which require regularization. The infinite tower of fields conspire in a way such that their joint contribution is finite.

At the time of writing, there were a number of attempts in the literature to obtain the Schwarzian action from \eqref{eq:reparamaction}, however most of them should be treated with some caution.\footnote{References \cite{Jevicki:2016bwu,Jevicki:2016ito} evaluate the leading correction in derivatives for $q=2$, find the Schwarzian action, and then treat different $q$ in an expansion around $q=2$. The problem with this approach is that we know that the Schwarzian action cannot dominate the physics in any limit of the $q=2$ model, since it leads to a maximal Lyapunov exponent, while the $q=2$ model is integrable. We will soon see what likely goes wrong here. There is also a different approach discussed in \cite{Jevicki:2016ito}, which relies on a particular regularization of \eqref{eq:schdiv}, it is however not clear here how much the result depends on the choice of regularization scheme. Also, there is still no distinction between the integrable $q=2$ case and the rest. We will do something here that is similar in spirit to this. In contrast to this, another reference \cite{Bagrets:2016cdf} observes that for a naive regularization, the $q>2$ cases have vanishing action for the leading order contribution in $\partial_\tau$, and derives the Schwarzian action from the next-to-leading order contribution, with coefficient $\log(J/\Delta)/J$, where $\Delta$ is the naive cutoff. However, we know that the free energy at finite temperature is only a function of $\beta J$, so the log term leads to different thermodynamical behaviour then what is observed from \eqref{eq:masteraction}, not to mention that the coefficient does not agree with the result of \cite{Maldacena:2016hyu} in the linearized case.} What we know for sure, is that for small fluctuations, at the quadratic level, the reparametrization modes have an action identical with what comes from the linearization \eqref{eq:linearizedschw} of the Schwarzian theory \cite{Maldacena:2016hyu}, and that the strong coupling expansion of the free energy \eqref{eq:largeqstrongcouple} is compatible with the nonlinear Schwarzian theory (even without the $1/q$ expansion).

We are not going to answer here whether the nonlinear Schwarzian action is a valid approximation for the reparametrization mode dynamics of the SYK model. Instead of that, we will solve a different problem: we discuss the $(\beta J)^{-1}$ expansion in a modification of the SYK model. Doing this serves two purposes. It will make it clear what can go wrong with a naive regularization in the derivative expansion in SYK, and at the same time illustrates in a simple way how the Schwarzian action can emerge. So let us consider a model with action \eqref{eq:masteraction}, with the derivative $\partial_\tau$ replaced by
\beq
\partial_\tau^\epsilon f(\tau) \equiv -\int d\tau' \big[\partial_{\tau'}\delta_\epsilon(\tau-\tau') \big]f(\tau'),
\eeq
where $\delta_\epsilon(\tau)$ is some Dirac delta approximating function, converging to the Dirac delta in distributional sense as $\epsilon \rightarrow 0$. The parameter $\epsilon$ is our cutoff. We choose it to have dimensions of time, and we want to set it as $\epsilon = a_0/J$, where $a_0$ is some $O(1)$ number. The purpose of this is that this way this modified model also has the conformal solutions and the reparametrization symmetry when $\tau \gg \epsilon$. Note that we could obtain this model from a fermionic theory with nonlocal kinetic kernel $\partial_\tau^\epsilon$.

One can imagine many choices for $\delta_\epsilon$. The most obvious one would be a Gaussian smearing $\delta_\epsilon(\tau)=(\sqrt{2\pi} \epsilon)^{-1} e^{-\tau^2/(2\epsilon^2)}$. However, this kernel is not invertible on $L^2(\mathbb{R})$ and invertibility is crucial for the free propagator to exist. This is because the inverse of its Fourier transform is a Gaussian with wrong sign. One could imagine a Lorentzian kernel $\delta_\epsilon=\epsilon \pi^{-1}(\tau^2+\epsilon^2)^{-1}$. We can invert the derivative of this, with result $[\partial \delta_\epsilon]^{-1}=\pi^{-1} \arctan \tau/\epsilon$. This kernel is fine, but it is dying off for large $\tau$ very slowly. For our purposes something that dies off faster than any polynomial would be more ideal. There is fortunately such a choice:
\bea
\label{eq:regdelta}
\delta_\epsilon(\tau) &= \frac{e^{-\frac{|\tau|}{\epsilon}}}{2\epsilon}, \\
[\partial \delta_\epsilon(\tau)]^{-1} &= \frac{1}{2}\text{sgn}\tau-\epsilon^2 \pi \delta '(\tau)
\eea
To evaluate the expansion of the modified action, we of course need the modified saddle point. We cannot determine this exactly, similarly as we cannot determine it for the SYK model. But we know some things. If we parametrize the saddle as
\beq
G_*^\epsilon(\tau) = b \frac{1}{|\tau|^{\frac{2}{q}}} \text{sgn}_\epsilon (\tau),
\eeq
then we know how the function $\text{sgn}_\epsilon (\tau)$ behaves both in the IR and the UV:
\bea
\label{eq:epsilonasimpt}
\text{sgn}_\epsilon (\tau) & = \text{sgn}(\tau) \;\;\;\; \text{ when } |\tau | \gg \epsilon, \\
\text{sgn}_\epsilon (\tau) & = |\tau|^{\frac{2}{q}} \frac{1}{b} [\partial \delta_\epsilon]^{-1}(\tau) \;\;\;\; \text{ when } |\tau | \ll \epsilon.
\eea
The function connects these two behaviours smoothly when $|\tau| \approx \epsilon$.
We will define the reparametrization of this as
\beq
G_*^{\phi,\epsilon}(\tau_1,\tau_2) = \left[ \frac{\phi'(\tau_1)\phi'(\tau_2)}{(\phi(\tau_1)-\phi(\tau_2))^2}\right]^{\frac{1}{q}}\text{sgn}_\epsilon (\tau_1-\tau_2),
\eeq
which is identical with \eqref{eq:reparamaction} for $|\tau_1-\tau_2| \gg \epsilon$, but it is different from it in the UV. This is again a choice that we make, but notice that this is an ambiguity present already in the SYK model, since \eqref{eq:reparamaction} is only a symmetry for $|\tau_1-\tau_2|  \gg J^{-1}$. We now want to evaluate 
\bea
I^\epsilon_{\rm eff}[\phi] &= \frac{1}{2} \left( \text{Tr}\big[ \partial^\epsilon_\tau (\Sigma_*^{\phi,\epsilon})^{-1}\big]-\text{Tr}\big[ \partial^\epsilon_\tau (\Sigma_*^\epsilon)^{-1}\big]\right) + \cdots\\
&=\frac{1}{2} \int d\tau_1 d\tau_2 [-\partial_{\tau_2} \delta_\epsilon(\tau_1-\tau_2)] \big( G_*^{\phi,\epsilon}(\tau_1,\tau_2) -G_*^{\epsilon}(\tau_1,\tau_2) \big) + \cdots.
\eea
For small $\epsilon$ (or large $J$), the integral is localized in the region $|\tau_1-\tau_2| \lesssim \epsilon$. In this case we can use the expansion
\beq
\left[ \frac{\phi'(\tau_1)\phi'(\tau_2)}{(\phi(\tau_1)-\phi(\tau_2))^2}\right]^{\frac{1}{q}}-\frac{1}{|\tau_1-\tau_2|^{\frac{2}{q}} }= |\tau_1-\tau_2|^{2-\frac{2}{q}} \Big[ \frac{1}{12 q} S(\phi,\tau_1) + O(\tau_1-\tau_2) \Big],
\eeq
where $ S(\phi,\tau_1)$ is the Schwarzian derivative \eqref{eq:schwartz}. This way, we may write
\bea
I^\epsilon_{\rm eff}[\phi] & \approx -\frac{b}{24 q} \int d\tau_1 S(\phi,\tau_1) \int_{-\infty}^\infty da [-\partial_a \delta_\epsilon(a)] |a|^{2-\frac{2}{q}} \text{sgn}_\epsilon (a)\\
&=-\frac{b}{24 q} \int d\tau_1 S(\phi,\tau_1) \left[ \int_{-\infty}^\infty da \frac{\text{sgn}a}{2\epsilon^2} e^{-\frac{|a|}{\epsilon}} |a|^{2-\frac{2}{q}} \text{sgn}_\epsilon (a) \right],
\eea
where we have changed to integration variable $a=\tau_1-\tau_2$ in the first line and used the choice \eqref{eq:regdelta} in the second line. We already see that the Schwarzian action very naturally appears in such a derivative expansion. 

To extract the coefficient, we would need to evaluate the $a$ integral. However, because of the exponential, it is dominated by the region $|a| \lesssim \epsilon$, which is precisely the region where we do not have access to the form of $\text{sgn}_\epsilon(\tau)$. In this sense, this integral is highly dependent on the physics at the cutoff scale $\epsilon$. It is instructive to evaluate the integral for both the extreme UV and IR cases of \eqref{eq:epsilonasimpt}. We obtain
\bea
I^\epsilon_{\rm eff}[\phi] &\approx -\frac{\epsilon}{24 q } \int d\tau_1 S(\phi,\tau_1) , && \text{when }\text{sgn}_\epsilon (a) \rightarrow \frac{|a|^{\frac{2}{q}}}{2b}\big[\text{sgn}a-2\epsilon^2 \pi \delta '(a)\big],\\
I^\epsilon_{\rm eff}[\phi] &\approx -\frac{b}{24 q } \Gamma(3-2/q) \epsilon^{1-\frac{2}{q}} \int d\tau_1 S(\phi,\tau_1) , && \text{when }\text{sgn}_\epsilon (a) \rightarrow \text{sgn}a.
\eea
Now let us recall that the we have set the cutoff as $\epsilon=a_0/J$ with $a_0$ being a dimensionless $O(1)$ number. The first case is therefore proportional to $J^{-1}$. In the second case, we need to recall from \eqref{eq:b} that the coefficient $b$ is of order $J^{-\frac{2}{q}}$. Using this, we end up with a coefficient in the second case which is again proportional to $J^{-1}$. The real $\text{sgn}_\epsilon(\tau)$ should smoothly connect these two extreme cases, so it is reasonable to think that the coefficient is proportional to $1/J$ in that case too. Therefore, we may safely assume that 
\beq
I^\epsilon_{\rm eff}[\phi] \sim \frac{1}{J}\int d\tau_1 S(\phi,\tau_1),
\eeq
to leading order in derivatives. The $1/J$ suppression of the action is what we were fighting for, this guarantees that these modes are the easiest to excite in the regime $1\ll \beta J \ll N$, moreover these modes have thermodynamics that is consistent with the large coupling expansion of SYK.\footnote{The subleading corrections in $(\beta J)^{-1}$ and their gravitational interpretation is studied in \cite{Kitaev:2017awl}.}

Let us summarize what we can learn from the above derivation and how does it connect to the physics of reparametrizations in the SYK model.
\begin{itemize}
\item First, we should recover the SYK model if we take $a_0 \rightarrow 0$. In this case it looks like the leading derivative contribution that we have calculated goes to zero, unless $q=2$. The finite contribution in the $q=2$ case has the same origin as the one observed in \cite{Jevicki:2016bwu,Jevicki:2016ito}. For the $q>2$ cases in SYK, one either needs to use a different regulator \cite{Jevicki:2016ito} or go to the next order in the derivative expansion \cite{Bagrets:2016cdf} to get a nonzero result.
\item It is very important that having to choose a dimensionful cutoff $\epsilon \sim J^{-1}$ turns the $1/J$ expansion into a non-systematic one. Indeed, we explicitly needed this to argue that the leading-in-derivative action is proportional to $J^{-1}$ for all $q$. Appearance of a divergence in $\epsilon$ for higher order terms can lead to unexpected improvements in the $J$ scaling in which case we cannot neglect some higher order terms in $\partial_\tau$ as they may still contribute as $J^{-1}$. In fact, we know that this must happen for $q=2$, since the Schwarzian action leads to a maximal Lyapunov exponent, while the $q=2$ theory is integrable. This is also an issue for SYK.
\end{itemize}

\subsection{Four point function}

The determination and analysis of the four point function is the main focus of references \cite{Maldacena:2016hyu,Polchinski:2016xgd}, where all the gory technical details on this topic can be found. Accordingly, we are going to be rather sketchy in this section, and focus on the overall logic and the physical understanding. This is opposed to the very basics of the SYK model that we have covered so far, where we tried to be a bit more detailed and pedagogical than the literature.

So why on earth are we so interested in the four point function? In the conformal limit, the two point function of the fermion fields $\psi_i(\tau)$ is given by \eqref{eq:confsol1}. This suggests that we can interpret $\psi_i(\tau)$ as a conformal primary of dimension $1/q$ in this limit. We also expect that the theory has other operators that are conformal primaries, and since we are dealing with an interacting model, they should have nontrivial anomalous scaling dimensions. In fact, in dimensions $d\geq 2$, a conformal field theory is entirely determined by the set of primary scaling dimensions and three point function (or operator product expansion, OPE in short) coefficients, and the exact spectrum is given by the set of scaling dimensions because of the state operator correspondence.\footnote{For those in a need for a crash course on CFTs, we refer to \cite{Ginsparg:1988ui}.} While there are some caveats in applying these results to $d=1$, still, obtaining the scaling dimensions and the three point function coefficients gets us pretty close to solving the theory. The four point function knows about both of these data, as it can be decomposed into conformal blocks\footnote{This decomposition is morally the same as just inserting a complete set of states between the two pairs of operators. Parts of such an expansion can be summed up into conformal blocks, because some of the states are related to each other by the action of conformal generators.}
\beq
\label{eq:confblock}
\langle \psi(\tau_1) \psi(\tau_2) \psi(\tau_3) \psi(\tau_4) \rangle = \langle \psi(\tau_1) \psi(\tau_2)\rangle \langle \psi(\tau_3) \psi(\tau_4) \rangle \sum_{h} (C_{\psi \psi}^h)^2 z^h {}_2 F_1(h,h,2h,z),
\eeq
where $h$ runs over the set of conformal primaries, $C_{\psi \psi}^h$ is the set of OPE coefficients, the variable
\beq
z=\frac{\tau_{12}\tau_{24}}{\tau_{13}\tau_{24}}, \;\;\;\; \tau_{ab}\equiv \tau_a-\tau_b,
\eeq
is the conformally invariant cross-ratio of the four insertion points, and $z^h {}_2 F_1(h,h,2h,z)$ is the conformal block summing the contribution of the  $SL(2,\mathbb{R})$ descendants of the primary $h$. If one succeeds in writing the four point function in this form, the scaling dimensions $h$ appearing in the sum, and the OPE coefficients $C_{\psi \psi}^h$ can be read off.

Let us now give a sketch of how this can be done in the SYK model. We are going to focus on the large $N$ limit, where the model classicalizes and has action \eqref{eq:masteraction}. The four point function of the fermions is just the two point function of the bilinear $G$, which at the classical level is of course the product of the on-shell value $G_*$
\bea
\frac{1}{N^2} \sum_{i,j} \langle T \psi_i(\tau_1)\psi_i(\tau_2)\psi_j(\tau_3)\psi_j(\tau_4) \rangle &= \langle G(\tau_{12})G(\tau_{34}) \rangle \\
&\sim G_*(\tau_{12})G_*(\tau_{34}).
\eea
In the conformal limit and the language of the conformal block decomposition \eqref{eq:confblock}, this is just the contribution of the identity operator $h=0$. We have learned therefore that for all the other operators, $C_{\psi \psi}^h$ must be suppressed by $1/\sqrt{N}$. The leading quantum correction can be extracted by studying small fluctuations around the saddle in the action \eqref{eq:masteraction}. It is useful to parametrize these corrections as
\bea
\frac{1}{N^2} \sum_{i,j} \langle T \psi_i(\tau_1)\psi_i(\tau_2)\psi_j(\tau_3)\psi_j(\tau_4) \rangle &= G_*(\tau_{12})G_*(\tau_{34})\\ &\times \Big[ 1 + \frac{1}{N}\mathcal{F}(\tau_1,\tau_2,\tau_3,\tau_4) + \cdots \Big].
\eea
We are after the function $\mathcal{F}$.
We write the fluctuation around the saddle as
\bea
G&=G_*+|G_*|^{\frac{2-q}{2}} g \\
\Sigma &=\Sigma_*+|G_*|^{\frac{q-2}{2}} \sigma.
\eea
so that 
\beq
\frac{1}{N}\mathcal{F}(\tau_1,\tau_2,\tau_3,\tau_4) =G_*(\tau_{12})^{-\frac{2}{q}}G_*(\tau_{34})^{-\frac{2}{q}} \langle g(\tau_1 ,\tau_2) g(\tau_3,\tau_4) \rangle.
\eeq
We can obtain this two point function by expanding the action \eqref{eq:masteraction} to quadratic order and doing the path integral for $\sigma$, which is now a Gaussian integral. This gives a  quadratic action for $g$ of the form
\beq
\label{eq:smallgaction}
I[g] = \frac{J^2(q-1)}{4} \int d\tau_1...d\tau_4 g(\tau_1,\tau_2) Q(\tau_1,\tau_2;\tau_3,\tau_4) g(\tau_3,\tau_4),
\eeq
with.
\bea
\label{eq:ladderkernel}
Q=\tilde K^{-1}-1, && \;\; \tilde K(\tau_1,\tau_2;\tau_3,\tau_4) = -J^2(q-1) |G_*(\tau_{12})|^{\frac{q-2}{2}} G_*(\tau_{13}) G_*(\tau_{24}) |G_*(\tau_{34})|^{\frac{q-2}{2}}.
\eea
Here, we are again using matrix notation for the kernels with ``row indices" being the first two variables and ``column indices" being the second two variables. The two point function of $g$ is then just the inverse of the kernel $Q$, so that we have
\beq
\label{eq:leadingcorr}
\mathcal{F}(\tau_1,\tau_2,\tau_3,\tau_4) = G_*(\tau_{12})^{-\frac{2}{q}}G_*(\tau_{34})^{-\frac{2}{q}} Q^{-1}(\tau_1,\tau_2;\tau_3,\tau_4).
\eeq
The task is therefore to understand the inverse of the kernel $Q$. 

\begin{figure}[h!]
\centering
\includegraphics[width=0.6\textwidth]{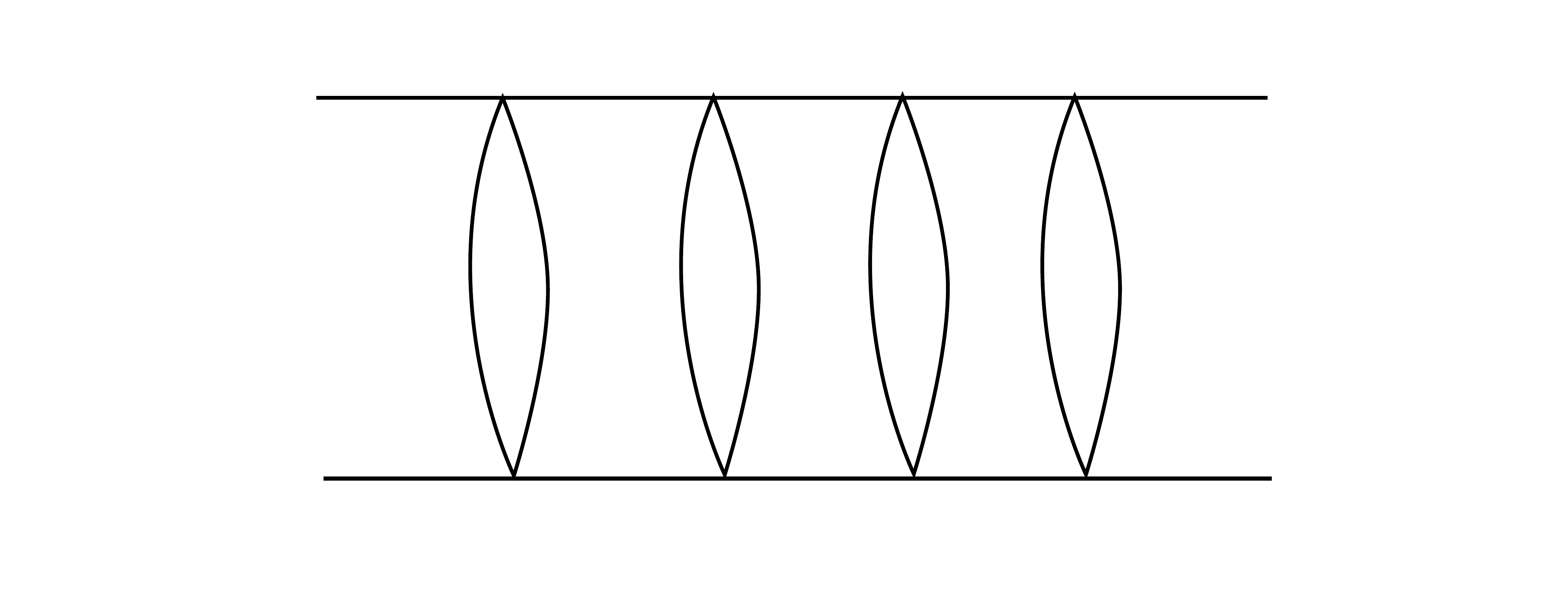} 
\caption{A four point ladder diagram.}
\label{fig:ladder}
\end{figure}

Before doing this, we want to make a quick point on the diagrammatic understanding of the four point function. We can write
\beq
Q^{-1}=(\tilde K^{-1}-1)^{-1} \equiv (1-\tilde K)^{-1} \tilde K = \sum_n \tilde K^n \tilde K.
\eeq
In this expansion, each term can be interpreted as a four point function ``ladder" diagram with $n$ rungs, such as the one on Fig. \ref{fig:ladder}. Just like melons for the two point function, these ladders give the leading in $1/N$ diagrams contributing to the four point function.

Now let us return to the question of understanding $Q^{-1}$ in the conformal limit. The strategy is to diagonalize it and write it as a spectral decomposition, by finding a suitable complete set of eigenfunctions to the kernel $\tilde K$ of \eqref{eq:ladderkernel}. Given such eigenfunctions $\Psi_\lambda (\tau_1,\tau_2)$ satisfying
\beq
\label{eq:Ktildeeigen}
\tilde K \Psi_\lambda = k_c(\lambda)\Psi_\lambda,
\eeq
and some completeness relation\footnote{Here by $1$ we mean the identity distribution on the space of antisymmetric functions of two variables.}
\beq
\sum_\lambda \frac{1}{\langle \Psi_\lambda | \Psi_\lambda \rangle}|\Psi_\lambda \rangle \langle \Psi_\lambda |=1,
\eeq
we can write an expansion of the form
\beq
\label{eq:formalQinv1}
Q^{-1} = \sum_\lambda \frac{k_c(\lambda)}{1-k_c(\lambda)}\frac{1}{\langle \Psi_\lambda | \Psi_\lambda \rangle}|\Psi_\lambda \rangle \langle \Psi_\lambda | .
\eeq
Of course, these formulas are very schematic at this level. We have not specified on what values of the formal label $\lambda$ we sum over, or what the inner product $\langle .|.\rangle$ is. A detailed derivation of these important details can be found in \cite{Maldacena:2016hyu}. Here, we just summarize the result. First, notice that it is conformal invariance again that makes it possible to find the eigenfunctions of $\tilde K$. More specifically, the $SL(2,\mathbb{R})$ generators
\beq
L_p^{(12)} =|\tau_{12}|^{-1}\big[ \tau_1^p \partial_{\tau_1}+\tau_2^p \partial_{\tau_2}\big]|\tau_{12}|, \;\;\; [L_p^{(12)},L_q^{(12)}]=(q-p)L_{p+q-1}^{(12)}, \;\;\; p=0,1,2,
\eeq
satisfy $L_p^{(12)} \tilde K = \tilde K L_p^{(34)}$. Therefore, $\tilde K$ commutes with the Casimir 
\beq
\label{eq:casimir}
C^{(12)}={L_1^{(12)}}^2-\frac{1}{2}(L_0^{(12)}L_2^{(12)}+L_2^{(12)}L_0^{(12)})
\eeq
 and eigenfunctions of $C^{(12)}$ are also eigenfunctions of $\tilde K$.  But $C^{(12)}$ is a second order differential operator which is much easier to diagonalize. More specifically, given
 \beq
C^{(12)}\Psi_\lambda = \lambda(\lambda-1)\Psi_\lambda,
 \eeq
 equation \eqref{eq:Ktildeeigen} also holds. This equation has multiple solutions for any $\lambda \in \mathbb{C}$, the general solution is a linear combination of conformally invariant three point functions
\beq
\label{eq:casimirsol}
\Psi_\lambda(\tau_1,\tau_2) = \int d\tau_0 g_\lambda(\tau_0) f^{\tau_0}_\lambda (\tau_1,\tau_2), \;\;\;\; f^{\tau_0}_\lambda (\tau_1,\tau_2) = \frac{\text{sgn}(\tau_{12}) }{|\tau_{01}|^\lambda |\tau_{02}|^\lambda |\tau_{12}|^{1-\lambda}}.
\eeq 
Any one of these solutions allows one to determine $k_c(\lambda)$ by directly evaluating \eqref{eq:Ktildeeigen}. For example\footnote{See \cite{Maldacena:2016hyu} for the general formula in terms gamma functions.}
 \bea
 \label{eq:kcex}
 k_c(\lambda) &= -\frac{3}{2} \frac{\tan \frac{\pi(\lambda-\frac{1}{2})}{2}}{\lambda-\frac{1}{2}}, && q&=4,\\
  k_c(\lambda) &=\frac{2}{\lambda(\lambda-1)}, && q&=\infty,\\
   k_c(\lambda) &= -1 , && q&=2.
 \eea
This was actually the easy part.  Finding a subset of the solutions \eqref{eq:casimirsol} which form a complete basis of antisymmetric eigenfunctions for a suitable choice of inner product is the nontrivial task. Without further details, it turns out that the set of required $\lambda$ is\footnote{This is compatible with restricting the eigenvalue of the Casimir, $\lambda(\lambda-1)$ to be real.}
\bea
\lambda =\frac{1}{2} + i s, \;\;\; s\in \mathbb{R}, \;\;\;\text{and}\;\;\; \lambda=2n,\;\;\; n\in \mathbb{Z}^+.
\eea
This allows one to write down a formula for $Q^{-1}$ of the form
\bea
\label{eq:Qinv2}
Q^{-1}(\tau_1,\tau_2;\tau_3,\tau_4) &= \int_{-\infty}^\infty ds \alpha_{\lambda_s}(\tau_1,...,\tau_4) \frac{k_c(\lambda_s)}{1-k_c(\lambda_s)} + \sum_{n=1}^\infty \beta_{\lambda_n}(\tau_1,...,\tau_4)  \frac{k_c(\lambda_n)}{1-k_c(\lambda_n)} , \\
\lambda_s &=\frac{1}{2} + i s, \;\;\;\;
\lambda_n =2n.
\eea
The functions $\alpha$ and $\beta$ incorporate the dependence on the eigenfunctions and the measure factors coming from the inner product in \eqref{eq:formalQinv1}.\footnote{As before, they can be found in \cite{Maldacena:2016hyu}.} An important obstacle is that the $n=1$ term in the discrete part diverges because $k_c(2)=1$. This can be explicitly checked for the examples \eqref{eq:kcex}. This is not unexpected, this is the eigenvalue associated to the reparametrization modes, which we have seen to be zero modes of the action in the conformal limit in sec. \ref{sec:conflim}. We know therefore that these eigenvalues must be shifted from this value by $(\beta J)^{-1}$ corrections. In fact, their contribution can be shown to be that of the linearized Schwarzian theory \eqref{eq:linearizedschw}. This means that we have actually already calculated their leading in $(\beta J)^{-1}$ contribution to the four point function in section \ref{sec:4ptSchw}.\footnote{Specifically, their contrition is given by \eqref{eq:FTO} or \eqref{eq:FOTO}, with $C\sim (\beta J)^{-1}$.} In terms of some putative bulk dual for SYK, one can think about this as the contribution of gravity to the four point function, while all the rest in \eqref{eq:Qinv2} is coming from matter fields.

\begin{figure}[h!]
\centering
\includegraphics[width=0.7\textwidth]{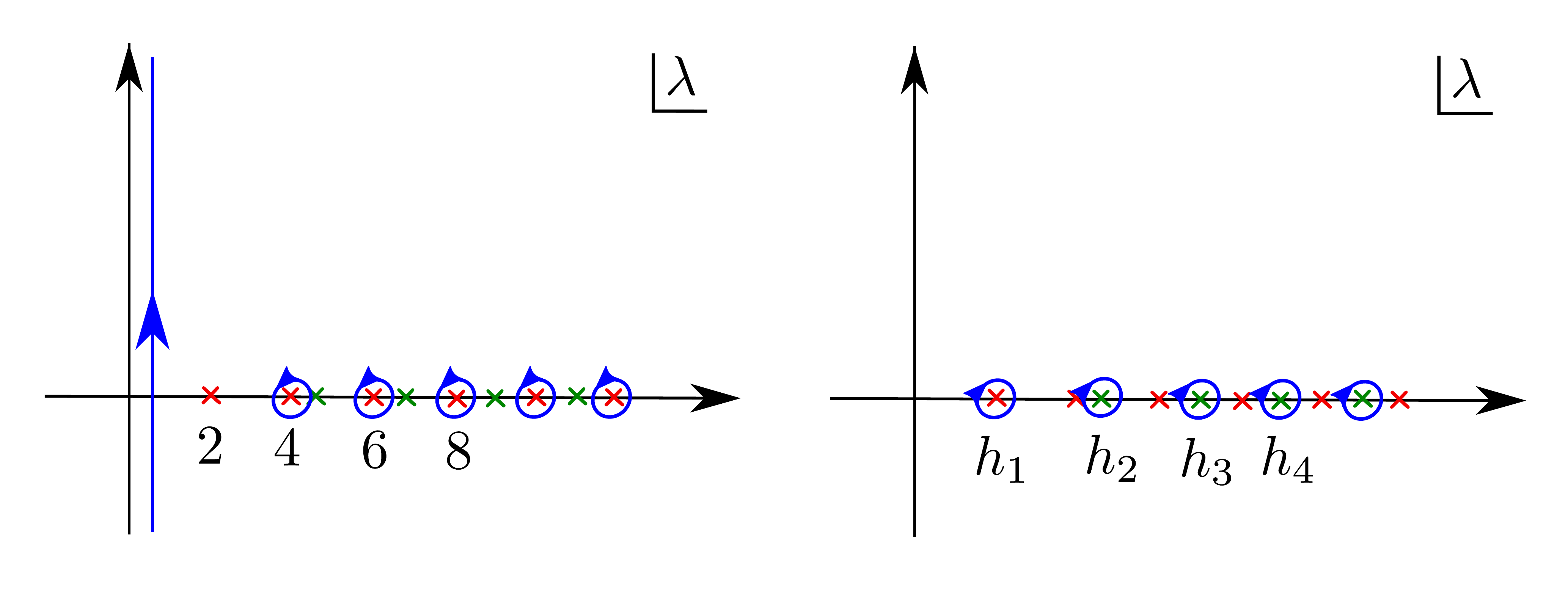} 
\caption{Contours for the four point function.}
\label{fig:4ptfunc}
\end{figure}

So how can we extract from \eqref{eq:Qinv2} the scaling dimensions and the OPE coefficients? This equation is somewhat reminiscent to \eqref{eq:confblock} as it is a decomposition indexed by eigenvalues of the conformal Casimir, but it is also different because the would-be scaling dimensions are not real and no conformal blocks appear. There is one further step to cast \eqref{eq:Qinv2} into the required form. The claim is that $\beta_\lambda$ is such that \eqref{eq:Qinv2} (with the $n=1$ term neglected) can be rewritten in the form
\beq
\label{eq:Qinv3}
Q^{-1}(\tau_1,\tau_2;\tau_3,\tau_4) = \int_{-\infty}^\infty ds \alpha_{\lambda_s}(\tau_1,...,\tau_4) \frac{k_c(\lambda_s)}{1-k_c(\lambda_s)} + \sum_{n=2}^\infty \text{Res}\Big[\alpha_{\lambda}(\tau_1,...,\tau_4)  \frac{k_c(\lambda)}{1-k_c(\lambda)}\big]|_{\lambda=2n}.  \\
\eeq
This expression can be interpreted as a single contour integral over a contour which is the union of a line and small distinct circles around the poles of $\alpha_{\lambda}$, see the left of Fig. \ref{fig:4ptfunc}. The claim is that $\alpha_{\lambda}$ has all its poles at the values $\lambda=2n$, $n\in \mathbb{Z}$. We can then deform the integration contour as in the right of Fig. \ref{fig:4ptfunc} so that we annihilate the vertical line with the circle contours, at the price of picking up the poles at the positive solutions of the equation
\beq
\label{eq:spectrum}
k_c(h_m)=1.
\eeq
One can show that this contour deformation yields an expression for the leading $1/N$ correction to the four point funciton, \eqref{eq:leadingcorr}, of the form
\beq
\mathcal{F}(\tau_1,...,\tau_4) = \sum_{m=1}^\infty c_{h_m}^2 z^{h_m} {}_2 F_1(h_m,h_m,2h_m,z),
\eeq
where the coefficients $c_{h_m}^2$ are certain known analytic functions of $h_m$ (and $q$) and are related to the OPE coefficients in \eqref{eq:confblock} as $C^{h_m}_{\psi\psi} = c_{h_m}/\sqrt{N}$. We see that \eqref{eq:spectrum} indeed gives the set of scaling dimensions for the conformal primaries appearing in the $\psi \times \psi$ OPE. There is no explicit expression for $h_m$, but e.g. for $q=4$ and $m\gg 1$ it behaves as
\beq
\label{eq:integerspectrum}
h_m \approx 2\Delta + 1+2m + \frac{3}{2\pi m},
\eeq
which shows that these operators consist of two $\psi$s and $2m+1$ derivatives, plus an anomalous piece due to the interactions. These operators have been explicitly worked out in \cite{Gross:2017hcz} and have the form
\beq
\label{eq:SYKops}
\mathcal{O}_m = \frac{1}{\sqrt{N}} \sum_{i=1}^N \sum_{k=0}^{2m+1} d_{mk} \partial_\tau^k \psi_i \partial^{2m+1-k}\psi_i,
\eeq
with $d_{mk}$ some coefficients that are not needed here explicitly.

There is a simple alternative way to see that \eqref{eq:spectrum} gives the right set of propagating modes for the SYK model, directly from the quadratic action \eqref{eq:smallgaction}. In a free theory with quadratic action, the Euclidean two point function is the inverse of the kernel of the quadratic action. This kernel is generally invertible in Euclidean signature, simply because it is usually some elliptic differential operator $D$, for which $Df=0$ has a unique solution for a given set of boundary conditions on $f$. The situation is different in Lorentzian signature, where the equation $Df=0$ has propagating wave solutions, which renders the kernel non-invertible. Correspondingly, the Euclidean two point function, when analytically continued to complex times, has singularities on the real time axis. For the Lorentzian version of the large $N$ quadratic SYK action \eqref{eq:smallgaction}, propagating solutions are precisely the eigenfunctions $\tilde K f=f$, since $Q=\tilde K^{-1}-1$.

\subsection{Bulk dual?}

A central question from the point of view of AdS$_2$ holography is of course that to what extent can we describe the physics of SYK as a gravitational theory in two dimensions. We have already seen that there is a similar pattern of symmetry breaking going on in the strong coupling limit, as for the Jackiw-Teitelboim theory discussed in sec. \ref{sec:Nads}. There are some reparametrization modes that dominate the low energy limit when $1\ll \beta J \ll N$ and to some extent, the physics of these modes is described by the Schwarzian action, which we have seen to be tightly related to the Jackiw-Teitelboim theory. We would then expect the set of operators in \eqref{eq:SYKops} to be dual to some matter fields coupling to this gravitational theory. Moreover, the fundamental degrees of freedom in the large $N$ limit appear to be the bilocal fields $G(\tau_1,\tau_2)$ and $\Sigma(\tau_1,\tau_2)$, which are already functions of two coordinates, so it is tempting to think about them as fields living in two dimensions. However, their action \eqref{eq:masteraction} is clearly nonlocal. Even the quadratic action \eqref{eq:smallgaction} for the small fluctuations $g$ is nonlocal. 

Let us give a simple illustration of how local fields can arise from \eqref{eq:smallgaction}. Consider a field configuration of the form
\beq
g_m(\tau_1,\tau_2)=\int d\lambda \zeta_m(\lambda)\int d\tau_0 g_\lambda(\tau_0) f^{\tau_0}_\lambda (\tau_1,\tau_2)
\eeq
with $f^{\tau_0}_\lambda (\tau_1,\tau_2)$ as in \eqref{eq:casimirsol}. We imagine the function $\zeta_m(\lambda)$ to be highly peaked at $\lambda=h_m$, a solution of \eqref{eq:spectrum}, but having some small width. To linear order in $\lambda-h_m$ we can write
\beq
\frac{1}{k_c(\lambda)} \approx 1- \frac{\lambda(\lambda-1)-h_m(h_m-1)}{2 h_m-1} k_c'(h_m),
\eeq
from which it follows that acting on such $g_m$ we have
\beq
\label{eq:localkernelapprox}
(\tilde K^{-1}-1)g_m \approx \frac{k_c'(h_m)}{2h_m-1}\big(C^{(12)}-M_m^2 \big) g_m ,
\eeq
where $C^{(12)}$ is the Casimir of \eqref{eq:casimir} and $M_m^2=h_m(h_m-1)$. Realizing that Lorentzian AdS$_2$ is the group manifold $SL(2,\mathbb{R})$, we see that $C^{(12)}$ is related to the Laplacian on it. More specifically,
\beq
\label{eq:kinematic}
|\tau_{12}|C^{(12)}|\tau_{12}|^{-1} = z^2(-\partial_t^2+\partial_z^2), \;\;\;\; t=\frac{\tau_1+\tau_2}{2}, \;\;\; z=\frac{\tau_1-\tau_2}{2}.
\eeq
Combining this with \eqref{eq:localkernelapprox} leads to a quadratic action for a redefined field $\phi_m=|\tau_{12}|g_m$ which is approximately that of a massive scalar of mass $M_m$ propagating on AdS$_2$
\bea
\int g_m Q g_m &\approx  \int dt dz \frac{1}{z^2} \phi_m \big[z^2(-\partial_t^2+\partial_z^2)-M_m^2\big] \phi_m\\ 
&\equiv \int d^2 x \sqrt{g_{\rm AdS_2}} \phi_m(\Box_{\rm AdS_2}-M_m^2)\phi_m,
\eea
 with the mass being related to the conformal weight via the $d=1$ version of the usual AdS/CFT relation. We will soon explain why this line of reasoning is a bit too naive, but first let us discuss what this means for the bulk dual of SYK. We have an infinite number of massive bulk fields, each with $O(1)$ mass. There is no parameter that we can use to make some of these fields very heavy. This is \textit{not} what we expect from weakly coupled local gravitational physics, where there are a finite number of light fields, and any other infinite tower of fields (e.g. KK modes or massive string modes) have parametrically large mass. In this sense, SYK is more analogous to a string theory with $\ell_{\rm string} \sim \ell_{\rm AdS}$. However, it has much less content than such a string theory, since while the spectrum is roughly integer spaced (see \eqref{eq:integerspectrum}), the number of states does not follow a Hagedorn growth.
 
 Now let us return to discuss why interpreting the space \eqref{eq:kinematic} as the holographic spacetime is too naive. The problem is that the resulting Laplacian is Lorentzian, even if we start out with the Euclidean boundary field theory, suggesting that we are not getting the holographic AdS$_2$ space right. This Laplacian acts on the space of pair of points of the boundary circle. This type of space has higher dimensional generalizations, which are called kinematic spaces in \cite{Czech:2016xec}. In the present case, it is equivalent with the space of boundary anchored geodesics of the Poincar\'e disc. This latter is the Euclidean version of the expected bulk. Fields living on kinematic space are argued in \cite{Czech:2016xec} to be dual to geodesic operators in AdS/CFT, local operators integrated along geodesics. So the bilocal $G$ is more likely related to some geodesic operators instead of local fields. The local operators can be obtained by so called inverse ``X-ray transforms". The precise correspondence is that so called OPE-blocks, contributions of a given conformal family to the OPE, are equal to geodesic operators of massive free fields in AdS. This suggests that field operators $\phi_m$ on Euclidean AdS$_2$ should have a definition of the form
 \beq
 C(\tau_{12},\partial_2) \mathcal{O}_m(\tau_2) \sim \int_{\gamma_{12}} \phi_m,
 \eeq
 where $\mathcal{O}_m$ are the primary operators \eqref{eq:SYKops}, $ C(\tau_{12},\partial_2)$ is a certain differential operator, fixed by conformal symmetry, and $\gamma_{12}$ is the geodesic of the Poincar\'e disc anchored at $\tau_1$ and $\tau_2$.\footnote{The details of a related idea have been worked out in \cite{Das:2017wae}.} Of course, this type of reasoning leads to the same mass spectrum $M_m$ and the same overall conclusions.

\subsection{Outlook}

There is a vast number of follow-up works on the SYK model and its generalizations. The goal of these lectures was merely to present the basics of the original model and the gravitational context that makes the model interesting for holography. To close the lectures, we say here a few words about a small, subjective list of follow-up works connecting to the SYK model.

\begin{itemize}
\item {\bf Higher dimensional generalizations:} In $d$ dimensions, a fermion with canonical kinetic term has power counting dimension $(d-1)/2$. This means that $q$ fermion interaction terms are dimensionless and therefore relevant in $d=1$, but give irrelevant interactions\footnote{More precisely, the $d=2$, $q=4$ case is marginal.} in $d\geq 2$ and $q\geq 4$. This makes it difficult to get SYK-like physics which is strongly coupled in the IR. There are two solutions so far to this problem
\begin{enumerate}
\item One can consider bosons instead of fermions. A boson has dimension $d-2$ so $\phi^q$ is relevant in $d=2$. This approach is taken by \cite{Murugan:2017eto}, where they also consider models based on superfields in two dimensions. These models differ from SYK-like physics as they do not have maximal Lyapunov exponent and they flow to a proper conformal fixed point in the IR. 
\item Another possibility is to modify the kinetic term for the fermions. Working with a two-derivative quartic kinetic term, Lorentz invariance can be preserved and the fermions will be dimensionless in two dimension. This is the approach of \cite{Turiaci:2017zwd}. Their model has large $N$ melonic dominance and flows to some CFT in the IR.
\end{enumerate}
\item {\bf Supersymmetry:} There are nice supersymmetric generalization of the one dimensional SYK model, constructed in \cite{Fu:2016vas}. The basic idea is to construct a supercharge, odd in the fermions, such as $Q=i C_{ijk}\psi_i \psi_j \psi_k$, and consider a Hamiltonian $H=Q^2$. Such a Hamiltonian has the same form as the SYK Hamiltonian with $J_{ijkl}$ being some bilinear combination of the $C_{ijk}$. Choosing $C_{ijk}$ randomly from a Gaussian ensemble leads to SYK-like physics, with unbroken SUSY in the ground state for large $N$.
\item {\bf Tensor models:} A valid objection against the SYK model is that it is not really a quantum mechanical model because one needs to average over the couplings $J_{ijkl}$. This objection is avoided by the fact that a single realization of the couplings $J_{ijkl}$, which is ``sufficiently random", leads to the same physics as what we have discussed for the case when we average over the couplings. There is actually some analytical handle on this statement. It is possible to write down models without disorder, which are built from Majorana fermions with additional index structure, and these additional indices are contracted in a certain way. For example, there is a so called colored tensor model, proposed in \cite{Witten:2016iux}, and an uncolored tensor model proposed in \cite{Klebanov:2016xxf}. These models contain no disorder average, but it is possible to show that they satisfy the same large $N$ Schwinger-Dyson equations as the SYK model. By grouping tensor indices on the fermions into a single index, one can realize both these models as an SYK Hamiltonian with $J_{ijkl}$s choosen to be ones and zeros at certain places. So one can think about them as choices of couplings which are somewhat regular (preserve some larger global symmetry) but still random enough to produce the relevant physics. Even the fine structure of the spectrum of these models (modulo the degeneracies coming from the global symmetries associated to the special choice of $J_{ijkl}$s) is similar to the case of generic random choice of couplings \cite{Krishnan:2016bvg,Krishnan:2017ztz}.
\item {\bf Bulk dual:} While we have seen that the SYK model is neither dual to a weakly coupled local gravitational theory nor a string theory, the large $N$ solvability of the model makes it in principle possible to systematically derive a bulk Lagrangian theory in an $1/N$ expansion with an infinite number of weakly interacting fields. This approach is pioneered in \cite{Gross:2017hcz,Gross:2017aos}.
\item {\bf Towards top-down models:} Finding a limit of some stringy model of a black hole that is solvable at large $N$ in some similar way as the SYK model would be amazing. Some steps are taken in this direction in \cite{Ferrari:2017ryl,Azeyanagi:2017mre}. Here, it is shown that certain matrix models admit a double expansion in $N$ (size of the matrices) and $D$ (number of the matrices, interpreted as the dimensionality of some target space), such that they are solvable and dominated by melonic diagrams. Similar matrix models appear in string theory in many places, albeit with $D$ being a fixed $O(1)$ number. Still, one might hope that some of the results can be extrapolated to such smaller values of $D$.
\end{itemize}

\subsection*{Acknowledgment}

First, I would like to thank the organizers for inviting me to give these lectures, it was a lot of fun to prepare, especially as being someone who were not directly involved in these developments. 
I also thank the participants of the school for many interesting questions and discussions. I am especially grateful for Tim De Jonckheere for reading these notes and providing valuable feedback. I thank Matt DeCross, Chen-Te Ma, Gr\'egoire Mathys, Andrew Rolph, Craig Thone and Edward Witten for reading carefully previous versions of these notes and sending valuable corrections.

My work was supported in part by a grant from the Simons Foundation (\#385592, Vijay Balasubramanian) through the It From Qubit Simons Collaboration, by the Belgian Federal Science Policy Office through the Interuniversity Attraction Pole P7/37, by FWO-Vlaanderen through projects G020714N and G044016N, and by Vrije Universiteit Brussel through the Strategic Research Program ``High-Energy Physics''.

\bibliographystyle{utphys}
\bibliography{syk}

\providecommand{\href}[2]{#2}\begingroup\raggedright\begin{thebibliography}{10}

\bibitem{Aharony:1999ti}
O.~Aharony, S.~S. Gubser, J.~M. Maldacena, H.~Ooguri, and Y.~Oz, ``{Large N
  field theories, string theory and gravity},''
  \href{http://dx.doi.org/10.1016/S0370-1573(99)00083-6}{{\em Phys. Rept.}
  {\bfseries 323} (2000) 183--386},
\href{http://arxiv.org/abs/hep-th/9905111}{{\ttfamily arXiv:hep-th/9905111
  [hep-th]}}.

\bibitem{Maldacena:1998uz}
J.~M. Maldacena, J.~Michelson, and A.~Strominger, ``{Anti-de Sitter
  fragmentation},'' \href{http://dx.doi.org/10.1088/1126-6708/1999/02/011}{{\em
  JHEP} {\bfseries 02} (1999) 011},
\href{http://arxiv.org/abs/hep-th/9812073}{{\ttfamily arXiv:hep-th/9812073
  [hep-th]}}.

\bibitem{Almheiri:2014cka}
A.~Almheiri and J.~Polchinski, ``{Models of AdS$_{2}$ backreaction and
  holography},'' \href{http://dx.doi.org/10.1007/JHEP11(2015)014}{{\em JHEP}
  {\bfseries 11} (2015) 014},
\href{http://arxiv.org/abs/1402.6334}{{\ttfamily arXiv:1402.6334 [hep-th]}}.

\bibitem{Townsend:1997ku}
P.~K. Townsend, ``{Black holes: Lecture notes},''
\href{http://arxiv.org/abs/gr-qc/9707012}{{\ttfamily arXiv:gr-qc/9707012
  [gr-qc]}}.

\bibitem{Maldacena:1997ih}
J.~M. Maldacena and A.~Strominger, ``{Universal low-energy dynamics for
  rotating black holes},''
  \href{http://dx.doi.org/10.1103/PhysRevD.56.4975}{{\em Phys. Rev.} {\bfseries
  D56} (1997) 4975--4983},
\href{http://arxiv.org/abs/hep-th/9702015}{{\ttfamily arXiv:hep-th/9702015
  [hep-th]}}.

\bibitem{Maldacena:1996ds}
J.~M. Maldacena and L.~Susskind, ``{D-branes and fat black holes},''
  \href{http://dx.doi.org/10.1016/0550-3213(96)00323-9}{{\em Nucl. Phys.}
  {\bfseries B475} (1996) 679--690},
\href{http://arxiv.org/abs/hep-th/9604042}{{\ttfamily arXiv:hep-th/9604042
  [hep-th]}}.

\bibitem{Headrick}
\url{{http://people.brandeis.edu/~headrick/HeadrickCompendium.pdf}}.

\bibitem{Sen:2008vm}
A.~Sen, ``{Quantum Entropy Function from AdS(2)/CFT(1) Correspondence},''
  \href{http://dx.doi.org/10.1142/S0217751X09045893}{{\em Int. J. Mod. Phys.}
  {\bfseries A24} (2009) 4225--4244},
\href{http://arxiv.org/abs/0809.3304}{{\ttfamily arXiv:0809.3304 [hep-th]}}.

\bibitem{Sen:2005wa}
A.~Sen, ``{Black hole entropy function and the attractor mechanism in higher
  derivative gravity},''
  \href{http://dx.doi.org/10.1088/1126-6708/2005/09/038}{{\em JHEP} {\bfseries
  09} (2005) 038},
\href{http://arxiv.org/abs/hep-th/0506177}{{\ttfamily arXiv:hep-th/0506177
  [hep-th]}}.

\bibitem{Susskind:1998dq}
L.~Susskind and E.~Witten, ``{The Holographic bound in anti-de Sitter space},''
\href{http://arxiv.org/abs/hep-th/9805114}{{\ttfamily arXiv:hep-th/9805114
  [hep-th]}}.

\bibitem{deBoer:1999tgo}
J.~de~Boer, E.~P. Verlinde, and H.~L. Verlinde, ``{On the holographic
  renormalization group},''
  \href{http://dx.doi.org/10.1088/1126-6708/2000/08/003}{{\em JHEP} {\bfseries
  08} (2000) 003},
\href{http://arxiv.org/abs/hep-th/9912012}{{\ttfamily arXiv:hep-th/9912012
  [hep-th]}}.

\bibitem{Jackiw:1984je}
R.~Jackiw, ``{Lower Dimensional Gravity},''
\href{http://dx.doi.org/10.1016/0550-3213(85)90448-1}{{\em Nucl. Phys.}
  {\bfseries B252} (1985) 343--356}.

\bibitem{Teitelboim:1983ux}
C.~Teitelboim, ``{Gravitation and Hamiltonian Structure in Two Space-Time
  Dimensions},''
\href{http://dx.doi.org/10.1016/0370-2693(83)90012-6}{{\em Phys. Lett.}
  {\bfseries 126B} (1983) 41--45}.

\bibitem{Maldacena:2016upp}
J.~Maldacena, D.~Stanford, and Z.~Yang, ``{Conformal symmetry and its breaking
  in two dimensional Nearly Anti-de-Sitter space},''
  \href{http://dx.doi.org/10.1093/ptep/ptw124}{{\em PTEP} {\bfseries 2016}
  no.~12, (2016) 12C104},
\href{http://arxiv.org/abs/1606.01857}{{\ttfamily arXiv:1606.01857 [hep-th]}}.

\bibitem{Riegler:2017fqv}
M.~Riegler and C.~Zwikel, ``{Canonical Charges in Flatland},''
\href{http://arxiv.org/abs/1709.09871}{{\ttfamily arXiv:1709.09871 [hep-th]}}.

\bibitem{Jensen:2016pah}
K.~Jensen, ``{Chaos in AdS$_2$ Holography},''
  \href{http://dx.doi.org/10.1103/PhysRevLett.117.111601}{{\em Phys. Rev.
  Lett.} {\bfseries 117} no.~11, (2016) 111601},
\href{http://arxiv.org/abs/1605.06098}{{\ttfamily arXiv:1605.06098 [hep-th]}}.

\bibitem{Engelsoy:2016xyb}
J.~Engelsoy, T.~G. Mertens, and H.~Verlinde, ``{An investigation of AdS$_{2}$
  backreaction and holography},''
  \href{http://dx.doi.org/10.1007/JHEP07(2016)139}{{\em JHEP} {\bfseries 07}
  (2016) 139},
\href{http://arxiv.org/abs/1606.03438}{{\ttfamily arXiv:1606.03438 [hep-th]}}.

\bibitem{Cvetic:2016eiv}
M.~Cveti\v{c} and I.~Papadimitriou, ``{AdS$_{2}$ holographic dictionary},''
  \href{http://dx.doi.org/10.1007/JHEP12(2016)008,
  10.1007/JHEP01(2017)120}{{\em JHEP} {\bfseries 12} (2016) 008},
  \href{http://arxiv.org/abs/1608.07018}{{\ttfamily arXiv:1608.07018
  [hep-th]}}.
[Erratum: JHEP01,120(2017)].

\bibitem{Stanford:2017thb}
D.~Stanford and E.~Witten, ``{Fermionic Localization of the Schwarzian
  Theory},'' \href{http://dx.doi.org/10.1007/JHEP10(2017)008}{{\em JHEP}
  {\bfseries 10} (2017) 008},
\href{http://arxiv.org/abs/1703.04612}{{\ttfamily arXiv:1703.04612 [hep-th]}}.

\bibitem{Freedman:1998tz}
D.~Z. Freedman, S.~D. Mathur, A.~Matusis, and L.~Rastelli, ``{Correlation
  functions in the CFT(d) / AdS(d+1) correspondence},''
  \href{http://dx.doi.org/10.1016/S0550-3213(99)00053-X}{{\em Nucl. Phys.}
  {\bfseries B546} (1999) 96--118},
\href{http://arxiv.org/abs/hep-th/9804058}{{\ttfamily arXiv:hep-th/9804058
  [hep-th]}}.

\bibitem{Maldacena:2015waa}
J.~Maldacena, S.~H. Shenker, and D.~Stanford, ``{A bound on chaos},''
  \href{http://dx.doi.org/10.1007/JHEP08(2016)106}{{\em JHEP} {\bfseries 08}
  (2016) 106},
\href{http://arxiv.org/abs/1503.01409}{{\ttfamily arXiv:1503.01409 [hep-th]}}.

\bibitem{Polchinski:2015cea}
J.~Polchinski, ``{Chaos in the black hole S-matrix},''
\href{http://arxiv.org/abs/1505.08108}{{\ttfamily arXiv:1505.08108 [hep-th]}}.

\bibitem{Shenker:2013pqa}
S.~H. Shenker and D.~Stanford, ``{Black holes and the butterfly effect},''
  \href{http://dx.doi.org/10.1007/JHEP03(2014)067}{{\em JHEP} {\bfseries 03}
  (2014) 067},
\href{http://arxiv.org/abs/1306.0622}{{\ttfamily arXiv:1306.0622 [hep-th]}}.

\bibitem{Maldacena:2016hyu}
J.~Maldacena and D.~Stanford, ``{Remarks on the Sachdev-Ye-Kitaev model},''
  \href{http://dx.doi.org/10.1103/PhysRevD.94.106002}{{\em Phys. Rev.}
  {\bfseries D94} no.~10, (2016) 106002},
\href{http://arxiv.org/abs/1604.07818}{{\ttfamily arXiv:1604.07818 [hep-th]}}.

\bibitem{Polchinski:2016xgd}
J.~Polchinski and V.~Rosenhaus, ``{The Spectrum in the Sachdev-Ye-Kitaev
  Model},'' \href{http://dx.doi.org/10.1007/JHEP04(2016)001}{{\em JHEP}
  {\bfseries 04} (2016) 001},
\href{http://arxiv.org/abs/1601.06768}{{\ttfamily arXiv:1601.06768 [hep-th]}}.

\bibitem{Kitaev}
A.~Kitaev, ``{A simple model of quantum holography}.'' Seminar at {KITP}, 2015.

\bibitem{Sachdev:1992fk}
S.~Sachdev and J.~Ye, ``{Gapless spin fluid ground state in a random, quantum
  Heisenberg magnet},''
  \href{http://dx.doi.org/10.1103/PhysRevLett.70.3339}{{\em Phys. Rev. Lett.}
  {\bfseries 70} (1993) 3339},
\href{http://arxiv.org/abs/cond-mat/9212030}{{\ttfamily arXiv:cond-mat/9212030
  [cond-mat]}}.

\bibitem{Sachdev:2015efa}
S.~Sachdev, ``{Bekenstein-Hawking Entropy and Strange Metals},''
  \href{http://dx.doi.org/10.1103/PhysRevX.5.041025}{{\em Phys. Rev.}
  {\bfseries X5} no.~4, (2015) 041025},
\href{http://arxiv.org/abs/1506.05111}{{\ttfamily arXiv:1506.05111 [hep-th]}}.

\bibitem{Garcia-Garcia:2016mno}
A.~M. Garc\'ia-Garc\'ia and J.~J.~M. Verbaarschot, ``{Spectral and
  thermodynamic properties of the Sachdev-Ye-Kitaev model},''
  \href{http://dx.doi.org/10.1103/PhysRevD.94.126010}{{\em Phys. Rev.}
  {\bfseries D94} no.~12, (2016) 126010},
\href{http://arxiv.org/abs/1610.03816}{{\ttfamily arXiv:1610.03816 [hep-th]}}.

\bibitem{Cotler:2016fpe}
J.~S. Cotler, G.~Gur-Ari, M.~Hanada, J.~Polchinski, P.~Saad, S.~H. Shenker,
  D.~Stanford, A.~Streicher, and M.~Tezuka, ``{Black Holes and Random
  Matrices},'' \href{http://dx.doi.org/10.1007/JHEP05(2017)118}{{\em JHEP}
  {\bfseries 05} (2017) 118},
\href{http://arxiv.org/abs/1611.04650}{{\ttfamily arXiv:1611.04650 [hep-th]}}.

\bibitem{Polchinski:1998rr}
J.~Polchinski, {\em {String theory. Vol. 2: Superstring theory and beyond}}.
\newblock Cambridge University Press,
2007.
\newblock

\bibitem{Gurau:2017qna}
R.~Gurau, ``{The $\imath \epsilon$ prescription in the SYK model},''
\href{http://arxiv.org/abs/1705.08581}{{\ttfamily arXiv:1705.08581 [hep-th]}}.

\bibitem{Jevicki:2016bwu}
A.~Jevicki, K.~Suzuki, and J.~Yoon, ``{Bi-Local Holography in the SYK Model},''
  \href{http://dx.doi.org/10.1007/JHEP07(2016)007}{{\em JHEP} {\bfseries 07}
  (2016) 007},
\href{http://arxiv.org/abs/1603.06246}{{\ttfamily arXiv:1603.06246 [hep-th]}}.

\bibitem{Jevicki:2016ito}
A.~Jevicki and K.~Suzuki, ``{Bi-Local Holography in the SYK Model:
  Perturbations},'' \href{http://dx.doi.org/10.1007/JHEP11(2016)046}{{\em JHEP}
  {\bfseries 11} (2016) 046},
\href{http://arxiv.org/abs/1608.07567}{{\ttfamily arXiv:1608.07567 [hep-th]}}.

\bibitem{Bagrets:2016cdf}
D.~Bagrets, A.~Altland, and A.~Kamenev, ``{Sachdev-Ye-Kitaev model as Liouville
  quantum mechanics},''
  \href{http://dx.doi.org/10.1016/j.nuclphysb.2016.08.002}{{\em Nucl. Phys.}
  {\bfseries B911} (2016) 191--205},
\href{http://arxiv.org/abs/1607.00694}{{\ttfamily arXiv:1607.00694
  [cond-mat.str-el]}}.

\bibitem{Kitaev:2017awl}
A.~Kitaev and S.~J. Suh, ``{The soft mode in the Sachdev-Ye-Kitaev model and
  its gravity dual},'' \href{http://dx.doi.org/10.1007/JHEP05(2018)183}{{\em
  JHEP} {\bfseries 05} (2018) 183},
\href{http://arxiv.org/abs/1711.08467}{{\ttfamily arXiv:1711.08467 [hep-th]}}.

\bibitem{Ginsparg:1988ui}
P.~H. Ginsparg, ``{APPLIED CONFORMAL FIELD THEORY},'' in {\em {Les Houches
  Summer School in Theoretical Physics, June 28-August 5, 1988}}.

\bibitem{Gross:2017hcz}
D.~J. Gross and V.~Rosenhaus, ``{The Bulk Dual of SYK: Cubic Couplings},''
  \href{http://dx.doi.org/10.1007/JHEP05(2017)092}{{\em JHEP} {\bfseries 05}
  (2017) 092},
\href{http://arxiv.org/abs/1702.08016}{{\ttfamily arXiv:1702.08016 [hep-th]}}.

\bibitem{Czech:2016xec}
B.~Czech, L.~Lamprou, S.~McCandlish, B.~Mosk, and J.~Sully, ``{A Stereoscopic
  Look into the Bulk},'' \href{http://dx.doi.org/10.1007/JHEP07(2016)129}{{\em
  JHEP} {\bfseries 07} (2016) 129},
\href{http://arxiv.org/abs/1604.03110}{{\ttfamily arXiv:1604.03110 [hep-th]}}.

\bibitem{Das:2017wae}
S.~R. Das, A.~Ghosh, A.~Jevicki, and K.~Suzuki, ``{Space-Time in the SYK
  Model},''
\href{http://arxiv.org/abs/1712.02725}{{\ttfamily arXiv:1712.02725 [hep-th]}}.

\bibitem{Murugan:2017eto}
J.~Murugan, D.~Stanford, and E.~Witten, ``{More on Supersymmetric and 2d
  Analogs of the SYK Model},''
  \href{http://dx.doi.org/10.1007/JHEP08(2017)146}{{\em JHEP} {\bfseries 08}
  (2017) 146},
\href{http://arxiv.org/abs/1706.05362}{{\ttfamily arXiv:1706.05362 [hep-th]}}.

\bibitem{Turiaci:2017zwd}
G.~Turiaci and H.~Verlinde, ``{Towards a 2d QFT Analog of the SYK Model},''
  \href{http://dx.doi.org/10.1007/JHEP10(2017)167}{{\em JHEP} {\bfseries 10}
  (2017) 167},
\href{http://arxiv.org/abs/1701.00528}{{\ttfamily arXiv:1701.00528 [hep-th]}}.

\bibitem{Fu:2016vas}
W.~Fu, D.~Gaiotto, J.~Maldacena, and S.~Sachdev, ``{Supersymmetric
  Sachdev-Ye-Kitaev models},''
  \href{http://dx.doi.org/10.1103/PhysRevD.95.069904,
  10.1103/PhysRevD.95.026009}{{\em Phys. Rev.} {\bfseries D95} no.~2, (2017)
  026009}, \href{http://arxiv.org/abs/1610.08917}{{\ttfamily arXiv:1610.08917
  [hep-th]}}.
[Addendum: Phys. Rev.D95,no.6,069904(2017)].

\bibitem{Witten:2016iux}
E.~Witten, ``{An SYK-Like Model Without Disorder},''
\href{http://arxiv.org/abs/1610.09758}{{\ttfamily arXiv:1610.09758 [hep-th]}}.

\bibitem{Klebanov:2016xxf}
I.~R. Klebanov and G.~Tarnopolsky, ``{Uncolored random tensors, melon diagrams,
  and the Sachdev-Ye-Kitaev models},''
  \href{http://dx.doi.org/10.1103/PhysRevD.95.046004}{{\em Phys. Rev.}
  {\bfseries D95} no.~4, (2017) 046004},
\href{http://arxiv.org/abs/1611.08915}{{\ttfamily arXiv:1611.08915 [hep-th]}}.

\bibitem{Krishnan:2016bvg}
C.~Krishnan, S.~Sanyal, and P.~N. Bala~Subramanian, ``{Quantum Chaos and
  Holographic Tensor Models},''
  \href{http://dx.doi.org/10.1007/JHEP03(2017)056}{{\em JHEP} {\bfseries 03}
  (2017) 056},
\href{http://arxiv.org/abs/1612.06330}{{\ttfamily arXiv:1612.06330 [hep-th]}}.

\bibitem{Krishnan:2017ztz}
C.~Krishnan, K.~V.~P. Kumar, and S.~Sanyal, ``{Random Matrices and Holographic
  Tensor Models},'' \href{http://dx.doi.org/10.1007/JHEP06(2017)036}{{\em JHEP}
  {\bfseries 06} (2017) 036},
\href{http://arxiv.org/abs/1703.08155}{{\ttfamily arXiv:1703.08155 [hep-th]}}.

\bibitem{Gross:2017aos}
D.~J. Gross and V.~Rosenhaus, ``{All point correlation functions in SYK},''
\href{http://arxiv.org/abs/1710.08113}{{\ttfamily arXiv:1710.08113 [hep-th]}}.

\bibitem{Ferrari:2017ryl}
F.~Ferrari, ``{The Large D Limit of Planar Diagrams},''
\href{http://arxiv.org/abs/1701.01171}{{\ttfamily arXiv:1701.01171 [hep-th]}}.

\bibitem{Azeyanagi:2017mre}
T.~Azeyanagi, F.~Ferrari, P.~Gregori, L.~Leduc, and G.~Valette, ``{More on the
  New Large $D$ Limit of Matrix Models},''
\href{http://arxiv.org/abs/1710.07263}{{\ttfamily arXiv:1710.07263 [hep-th]}}.

\end{thebibliography}\endgroup

\end{document}